\documentclass{jpp}

\usepackage{graphicx,amsmath}
\usepackage{natbib,xtab}
\usepackage[pdftex,colorlinks,citecolor=blue]{hyperref}

\ifCUPmtlplainloaded \else
  \checkfont{eurm10}
  \iffontfound
    \IfFileExists{upmath.sty}
      {\typeout{^^JFound AMS Euler Roman fonts on the system,
                   using the 'upmath' package.^^J}%
       \usepackage{upmath}}
      {\typeout{^^JFound AMS Euler Roman fonts on the system, but you
                   dont seem to have the}%
       \typeout{'upmath' package installed. JPP.cls can take advantage
                 of these fonts, if you use 'upmath' package.^^J}%
      }
  \else
  \fi
\fi

\ifCUPmtlplainloaded \else
  \checkfont{msam10}
  \iffontfound
    \IfFileExists{amssymb.sty}
      {\typeout{^^JFound AMS Symbol fonts on the system, using the
                'amssymb' package.^^J}%
       \usepackage{amssymb}%
         
       \let\ge=\geqslant  
      }{}
  \fi
\fi

\ifCUPmtlplainloaded \else
  \IfFileExists{amsbsy.sty}
    {\typeout{^^JFound the 'amsbsy' package on the system, using it.^^J}%
     \usepackage{amsbsy}}
    {\providecommand\boldsymbol[1]{\mbox{\boldmath $##1$}}}
\fi

\newcommand{\gas}{\!\bs{R}_s}
\newcommand{\gai}{\!\bs{R}_i}
\newcommand{\ggs}{{\negthickspace\!\bs{R}_s}}
\newcommand{\gga}{{\negthickspace\!\bs{R}_i}}
\newcommand{\pD}[2]{\frac{\partial #2}{\partial #1}}
\newcommand{\D}[2]{\frac{{\rm d} #2}{{\rm d} #1}}
\newcommand{\bigD}[2]{\frac{{\rm D}#2}{{\rm D}#1}}
\newcommand\bs[1]{\boldsymbol{#1}}
\newcommand\bb[1]{\mbox{\boldmath{$#1$}}}
\newcommand\grad{\bb{\nabla}}
\newcommand\bcdot{\,\bb{\cdot}\,}

\newcommand\btimes{\,\bb{\times}\,}
\newcommand{\mc}[1]{\mathcal{#1}}
\newcommand{\mf}[1]{\,\mathfrak{#1}\!}

\newcommand{\msb}[1]{\mathsfbi{#1}}
\def\order#1#2{\underbrace{\vphantom{\Biggl(}#2}_{\hspace{-0.27em}\lefteqn{\bigcirc}{\hspace{0.285em}\bf #1}}}

\newcommand{\imag}{{\rm i}}
\newcommand{\eb}{\hat{\bb{b}}}
\newcommand{\ez}{\hat{\bb{z}}}
\newcommand{\ex}{\hat{\bb{x}}}
\newcommand{\ey}{\hat{\bb{y}}}
\newcommand{\valf}{v_{{\rm A}}}
\newcommand{\vasq}{v^2_{{\rm A}}}
\newcommand{\valfeff}{v_{{\rm A}\ast}}

\newcommand{\vthprl}[1]{v_{{\rm th}\parallel #1}}
\newcommand{\vthprp}[1]{v_{{\rm th}\perp #1}}
\newcommand{\thetaprl}[1]{\theta_{\parallel #1}}
\newcommand{\thetaprp}[1]{\theta_{\perp #1}}
\newcommand{\betaprp}[1]{\beta_{\perp #1}}
\newcommand{\betaprl}[1]{\beta_{\parallel #1}}
\newcommand{\cell}[1]{C^\parallel_{\ell #1}}
\newcommand{\czero}[1]{C^\parallel_{0 #1}}
\newcommand{\cone}[1]{C^\parallel_{1 #1}}
\newcommand{\ctwo}[1]{C^\parallel_{2 #1}}

\newcommand{\kzero}[1]{C^\perp_{0 #1}}
\newcommand{\kone}[1]{C^\perp_{1 #1}}
\newcommand{\ktwo}[1]{C^\perp_{2 #1}}
\newcommand{\df}[1]{\delta f_{#1}}
\newcommand{\fip}{f_{0i}}
\newcommand{\fem}{f_{0e}}
\newcommand{\felli}{F^\parallel_{\ell i}}
\newcommand{\fzeroi}{F^\parallel_{0i}}
\newcommand{\fonei}{F^\parallel_{1i}}
\newcommand{\ftwoi}{F^\parallel_{2i}}
\newcommand{\dpprp}[1]{\delta p_{\perp #1}}
\newcommand{\pprp}[1]{p_{\perp 0 #1}}
\newcommand{\dpprl}[1]{\delta p_{\parallel #1}}
\newcommand{\pprl}[1]{p_{\parallel 0 #1}}
\newcommand{\dns}{\delta n_s}
\newcommand{\dne}{\delta n_e}
\newcommand{\dni}{\delta n_i}
\newcommand{\nsp}{n_{0s}}
\newcommand{\nem}{n_{0e}}
\newcommand{\nip}{n_{0i}}
\newcommand{\dtprl}[1]{\delta T_{\parallel #1}}
\newcommand{\dtprp}[1]{\delta T_{\perp #1}}
\newcommand{\tprl}[1]{T_{\parallel 0{#1}}}
\newcommand{\tprp}[1]{T_{\perp 0{#1}}}
\newcommand{\tauprl}[1]{\tau_{\parallel{#1}}}
\newcommand{\tauprp}[1]{\tau_{\perp{#1}}}
\newcommand{\ddupar}{\delta u'_{\parallel s}}
\newcommand{\ddupari}{\delta u'_{\parallel i}}
\newcommand{\dupar}{u'_{\parallel 0s}}
\newcommand{\duparsq}{u'^2_{\parallel 0s}}
\newcommand{\dupari}{u'_{\parallel 0i}}

\newcommand{\dupare}{u'_{\parallel 0e}}
\newcommand{\dupara}{u'_{\parallel 0\alpha}}
\newcommand{\dBprl}{\delta B_\parallel}
\newcommand{\dBprp}{\delta \bb{B}_\perp}
\newcommand{\dnek}{\delta n_{e\bs{k}}}
\newcommand{\dnik}{\delta n_{i\bs{k}}}
\newcommand{\dpprpik}{\delta p_{\perp i\bs{k}}}
\newcommand{\dBprlk}{\delta B_{\parallel\bs{k}}}

\newcommand{\Aprlk}{A_{\parallel\bs{k}}}

\newsavebox{\astrutbox}
\sbox{\astrutbox}{\rule[-5pt]{0pt}{20pt}}

\title[Kinetic Turbulence in Pressure-Anisotropic Plasmas]{Inertial-Range Kinetic Turbulence in Pressure-Anisotropic Astrophysical Plasmas}

\author[M.~W.~Kunz, A.~A.~Schekochihin, C.~H.~K.~Chen, I.~G.~Abel, \& S.~C.~Cowley]%
{M.\ns W.\ns K\ls U\ls N\ls Z\ls$^1$%
  \thanks{Email address for correspondence: mkunz@princeton.edu}, 
A.\ns A.\ns S\ls C\ls H\ls E\ls K\ls O\ls C\ls H\ls I\ls H\ls I\ls N\ls$^{2,3}$,\break
C.\ns H.\ns K.\ns C\ls H\ls E\ls N\ls$^4$, I.\ns G.\ns A\ls B\ls E\ls L\ls$^{5}$, \and S.\ns C.\ns C\ls O\ls W\ls L\ls E\ls Y\ls$^{4,6}$}

\affiliation{$^1$Department of Astrophysical Sciences, Princeton University, Peyton Hall, Princeton, NJ 08544, USA\\[\affilskip]
$^2$Rudolf Peierls Centre for Theoretical Physics, University of Oxford, 1 Keble Road, Oxford OX1 3NP, UK\\[\affilskip]
$^3$Merton College, Merton Street, Oxford OX1 4JD, UK\\[\affilskip]
$^4$Department of Physics, Imperial College London, London SW7 2AZ, UK\\[\affilskip]
$^5$Princeton Center for Theoretical Science, Princeton University, Jadwin Hall, Princeton, NJ 08544, USA\\[\affilskip]
$^6$EURATOM/CCFE Fusion Association, Culham Science Centre, Abingdon, OX14 3DB, UK}

\pubyear{2015}
\volume{}
\pagerange{}
\date{27 Jan 2015}
\begin{document}

\maketitle

\begin{abstract}
A theoretical framework for low-frequency electromagnetic (drift-)kinetic turbulence in a collisionless, multi-species plasma is presented. The result generalises reduced magnetohydrodynamics (RMHD) and kinetic RMHD \citep{schekochihin09} to the case where the mean distribution function of the plasma is pressure-anisotropic and different ion species are allowed to drift with respect to each other---a situation routinely encountered in the solar wind and presumably ubiquitous in hot dilute astrophysical plasmas such as the intracluster medium. Two main objectives are achieved. First, in a non-Maxwellian plasma, the relationships between fluctuating fields (e.g., the Alfv\'{e}n ratio) are order-unity modified compared to the more commonly considered Maxwellian case, and so a quantitative theory is developed to support quantitative measurements now possible in the solar wind. Beyond these order-unity corrections, the main physical feature of low-frequency plasma turbulence survives the generalisation to non-Maxwellian distributions: Alfv\'{e}nic and compressive fluctuations are energetically decoupled, with the latter passively advected by the former; the Alfv\'{e}nic cascade is fluid, satisfying RMHD equations (with the Alfv\'{e}n speed modified by pressure anisotropy and species drifts), whereas the compressive cascade is kinetic and subject to collisionless damping (and for a bi-Maxwellian plasma splits into three independent collisionless cascades). Secondly, the organising principle of this turbulence is elucidated in the form of a conservation law for the appropriately generalised kinetic free energy. It is shown that non-Maxwellian features in the distribution function reduce the rate of phase mixing and the efficacy of magnetic stresses, and that these changes influence the partitioning of free energy amongst the various cascade channels. As the firehose or mirror instability thresholds are approached, the dynamics of the plasma are modified so as to reduce the energetic cost of bending magnetic-field lines or of compressing/rarefying them. Finally, it is shown that this theory can be derived as a long-wavelength limit of non-Maxwellian slab gyrokinetics.
\end{abstract}

\section{Introduction}\label{sec:introduction}

Reduced magnetohydrodynamics (RMHD) is a non-linear system of fluid equations used to describe anisotropic fluctuations in magnetised plasmas at lengthscales $L$ much larger than the ion gyroradius $\rho_i$ and at frequencies $\omega$ much smaller than the ion gyrofrequency $\Omega_i$. It was initially used to model elongated structures in tokamaks \citep{kp74,strauss76,strauss77} but has since become a standard paradigm for astrophysical contexts such as solar-wind turbulence \citep{zm92a,zm92b,bns98} and the solar corona \citep{odm03,pc13}. 

Although RMHD was initially derived from incompressible ideal MHD, a collisional fluid theory, it can also be obtained without assuming the plasma to be collisional \citep[][hereafter S09]{schekochihin09}. The resulting set of fluid-kinetic equations describing both Alfv\'{e}nic (RMHD) and compressive, i.e., density and magnetic-field-strength, fluctuations is referred to as kinetic reduced magnetohydrodynamics (KRMHD). S09 argued that KRMHD is an appropriate description for small-scale solar-wind fluctuations, which are anisotropic \citep[e.g.,][]{hwc12} and weakly collisional \citep[e.g.,][]{bc05}, as well as for inertial-range turbulence in the hot ionised phase of the interstellar medium and in the intracluster medium of galaxy clusters.

Two of the assumptions of KRMHD are that the equilibrium distribution functions of all species are Maxwellian (and, therefore, that the equilibrium pressure is isotropic) and that there is only one ionic species. The former assumption works well for plasmas such as the interstellar medium, where collisions are weak ($\lambda_{\rm mfp} \gg \rho_i$ and $\nu_{ii} \ll \Omega_i$, where $\lambda_{\rm mfp}$ is the collisional mean free path and $\nu_{ii}$ is the ion--ion collision frequency) but non-negligible ($\lambda_{\rm mfp} \ll L$ and $\nu_{ii} \gg \omega$). However, the collisional mean free path in space plasmas is on the order of $1~{\rm au}$---the distance between the Sun and the Earth---and proton (H ions), alpha (He ions), and electron pressures in the solar wind are observed to be highly anisotropic with respect to the local magnetic-field direction \citep[e.g.,][]{hellinger06,stverak08,bale09,mkg12}. The observed distribution functions in the solar wind (especially the electron one) also exhibit non-Maxwellian suprathermal tails \citep[see][and references therein]{maksimovic05,marsch06} containing small ($\sim$5\% of the total density) populations of energetic particles. In the intracluster medium, where $\lambda_{\rm mfp} \sim 0.1$--$30~{\rm kpc}$ is many orders of magnitude larger than $\rho_i \sim 1~{\rm npc}$, conservation of particles' first adiabatic invariant during (macroscale) turbulent stretching of the magnetic field is expected to render the distribution function anisotropic \citep[e.g.,][]{schekochihin05,sc06,kunz11}. How such anisotropic distribution functions affect the turbulent cascade in these systems is presently unknown, and it seems dangerous to describe their dynamics with a set of equations built upon the assumption of isotropy.

The assumption of a single ionic species is equally unwarranted in the solar wind, where the abundances of alpha particles and heavy ions have been established observationally for nearly fifty years \citep[for a review, see][]{sgg97}. The protons and alphas (as well as many other ions) drift with respect to the centre-of-mass frame, often with parallel drift velocities on the order of the Alfv\'{e}n speed \citep{asbridge76,marsch81,marsch82,neugebauer94,goldstein00}.

In this Paper, we present a generalisation of KRMHD to account for non-Maxwellian distribution functions and multiple ionic species. Our theory of pressure-anisotropic KRMHD can be applied to a broad range of plasmas that exhibit these characteristics and satisfy the original assumptions of KRMHD: that the turbulent fluctuations are small compared to the mean field, are spatially anisotropic with respect to it, have frequencies small compared to the ion cyclotron frequency, and have lengthscales large compared to the ion gyroradius. The purpose of this Paper is twofold: first, {\em to explain what is the organising principle governing the turbulent cascade in a pressure-anisotropic plasma}; and second, {\em to provide a quantitatively correct set of equations that describe that turbulence}. A reader interested only in the former can proceed directly to Section \ref{sum:W}, where we present a generalised free-energy invariant describing a turbulent cascade from large to small scales (as well as into phase space) in a non-Maxwellian plasma. A reader interested only in the latter will find those equations summarised in Section \ref{sum:akrmhd}.

Before proceeding with the derivation, we caution that pressure-anisotropic plasmas are subject to a variety of kinetic microscale instabilities if their pressure anisotropy $p_\perp - p_\parallel$ becomes larger than the magnetic pressure (times a factor of order unity). Some of these instabilities (e.g., ion-cyclotron, whistler) are ordered out of KRMHD by its restriction to sub-Larmor frequencies. Others, namely firehose and mirror, are included (\S\S\ref{sec:firehose}, \ref{sec:mirror}), but have growth rates that increase without bound with wavenumber due to the exclusion of finite-Larmor-radius effects that would have regularised them. Plasmas exhibiting super-Alfv\'{e}nic inter-species drifts can also be subject to cyclotron- and Landau-resonant electromagnetic instabilities \citep[e.g.,][]{ml87,gary91,dg98,vbc13}. With these complications borne in mind, KRMHD as a quantitative theory is only suitable for kinetic turbulence residing within the microscale stability boundaries. We will show that pressure anisotropies and interspecies drifts, even those lying within the stability boundaries, lead to order-unity modifications of the relations between different fluctuating fields and of the kinetics of the phase-space turbulent cascade compared to what was deduced previously for a Maxwellian plasma.

The Paper is organised as follows. In Section \ref{sec:equations} we derive the generalised KRMHD equations starting from the drift kinetics. Several consequences of these equations are detailed both mathematically and physically there and in Sections \ref{sec:alfveniccascade} and \ref{sec:compressivecascade}. These include linear waves and stability (\S\S\ref{sec:firehose}, \ref{sec:mirror}, Appendix \ref{app:linear}), nonlinearly conserved quantities (\S\S\ref{sec:WAW}, \ref{sec:Wcompr}, \ref{sum:W}), and their combined implications for the efficacy of phase mixing (\S\S\ref{sec:collisionlessinvariants}, \ref{sec:collisionlesscascades}). In Appendix \ref{app:gk}, we re-derive KRMHD systematically from a generalisation of slab gyrokinetics (\citealt{howes06}; S09) to non-Maxwellian distribution functions and multiple ionic species. The latter approach enables analytical and numerical studies of fluctuations at and below the ion gyroradius, including the effects of pressure anisotropy on the nonlinear perpendicular phase mixing and the phase-space cascade of kinetic Alfv\'{e}n waves and entropy. These topics will be the subject of a separate publication. Finally, in Appendix \ref{app:nomenclature} we provide a list of frequently used symbols and their definitions.

\section{General Non-Linear Equations of KRMHD}\label{sec:equations}

\subsection{Kinetic MHD}\label{sec:KMHD}

We begin with the equations of kinetic MHD (KMHD), as derived by \citet{kulsrud64,kulsrud83}. KMHD is a hybrid fluid-kinetic theory, appropriate for scales $k^{-1}$ larger than the ion gyroradius ($k\rho_i \ll 1$) and frequencies $\omega$ smaller than the ion gyrofrequency ($\omega \ll \Omega_i$) in a weakly collisional ($\rho_i \ll \lambda_{\rm mfp}$) plasma. The equations for species $s$ are the continuity equation
\begin{equation}\label{eqn:continuity}
\left( \pD{t}{} + \bb{u}_s \bcdot \grad \right) n_s = - n_s \grad \bcdot \bb{u}_s ,
\end{equation}
the momentum equation
\begin{equation}\label{eqn:force}
m_s n_s \left( \pD{t}{} + \bb{u}_s \bcdot \grad \right) \bb{u}_s = - \grad \bcdot \msb{P}_s + q_s n_s \left( \bb{E} + \frac{\bb{u}_s}{c} \btimes \bb{B} \right) + \bb{F}_s ,
\end{equation}
and the drift-kinetic equation
\begin{equation}\label{eqn:kinetic}
\bigD{t}{f_s} + \bigD{t}{\ln B} \frac{w_\perp}{2} \pD{w_\perp}{f_s}  + \left( \frac{q_s E_\parallel }{m_s} + \frac{w^2_\perp}{2} \grad \bcdot \eb - \bigD{t}{\bb{u}_{\perp s}} \bcdot \eb \right) \pD{v_\parallel}{f_s} = \left( \pD{t}{f_s} \right)_{\!\rm c} ,
\end{equation}
where ${\rm D}/{\rm D}t \doteq \partial / \partial t + \bb{u}_{\perp s} \bcdot \grad + v_\parallel \eb \bcdot \grad $. Our notation is standard: $m_s$, $q_s$ ($=$$Z_s e$),  $n_s$, $\bb{u}_s$, and $\msb{P}_s$ are, respectively, the mass, charge, number density, mean velocity, and pressure tensor of species $s$; $\bb{B}$ is the magnetic field; and $\bb{E}$ is the electric field. The rate of change in the momentum of species $s$ due to inter-species collisions, $\bb{F}_s$, is obtained by taking the first velocity moment of the collision term $( \partial f_s / \partial t )_{\rm c}$. The distribution function of species $s$, $f_s = f_s( v_\parallel , w_\perp )$, is written in terms of the velocity-space variables $v_\parallel$ and $w_\perp$, which are measured parallel and perpendicular, respectively, to the magnetic field direction, $\eb \doteq \bb{B} / B$. As in Kulsrud's original formulation, we use as the perpendicular kinetic variable $\bb{w}_\perp = \bb{v}_\perp - \bb{u}_{\perp s}$, the perpendicular velocity peculiar to the mean perpendicular flow of species $s$. The parallel component of Equation (\ref{eqn:force}) is redundant, as it may be straightforwardly obtained by taking the first parallel velocity moment of the drift-kinetic equation (\ref{eqn:kinetic}).

With Larmor gyrations ordered out of these equations, $f_s$ is independent of gyrophase. As a result, the pressure tensor is diagonal in a coordinate system defined by the parallel and perpendicular directions with respect to the magnetic field:
\begin{equation}\label{eqn:ptensor}
\msb{P}_s = p_{\perp s} \bigl( \msb{I} - \eb \eb \bigr) + p_{\parallel s}  \eb \eb ,
\end{equation}
where $\msb{I}$ is the unit dyadic, and
\begin{align}
p_{\parallel s} \doteq n_s T_{\parallel s} &= \int {\rm d}^3 \bb{v} \, m_s \left( v_\parallel - u_{\parallel s} \right)^2 f_s , \label{eqn:pprls}\\*
p_{\perp s} \doteq n_s T_{\perp s} &= \int {\rm d}^3 \bb{v} \, m_s \frac{w^2_\perp}{2} f_s   \label{eqn:pprps}
\end{align}
are the parallel and perpendicular pressures, respectively, of species $s$. By expanding Equation (\ref{eqn:kinetic}) in powers of $(m_ e / m_i)^{1/2}$ and using the quasineutrality constraint,
\begin{equation}\label{eqn:quasineutrality}
\sum_s q_s n_s = 0 ,
\end{equation}
it is straightforward to show that the parallel component of the electric field satisfies
\begin{equation}\label{eqn:eprl}
E_\parallel = - \eb \bcdot \left( \frac{\grad \bcdot \msb{P}_e}{en_e} \right) = - \frac{1}{e n_e} \left[ \eb \bcdot \grad p_{\parallel e} -  \left( p_{\perp e} - p_{\parallel e} \right) \grad \bcdot  \eb \right]
\end{equation}
to leading order.\footnote{Equation (\ref{eqn:eprl}) may also be obtained directly from the electron force equation (Eq.~\ref{eqn:force} with $s=e$) after neglecting both the inertial terms on its left-hand side and the friction force on its right-hand side.} Equation (\ref{eqn:eprl}) expresses parallel pressure balance for the (effectively massless) electron fluid, taking into account the forcing-out of large-pitch-angle particles (those with $\alpha \doteq \cos^{-1} (v_\parallel / v) \sim \pi/2$) from regions of increased magnetic-field strength. Indeed, in the absence of collisions, Equation (\ref{eqn:kinetic}) guarantees that each particle's adiabatic invariant $\mu_s \doteq m_s w^2_\perp / 2B$ is identically preserved (i.e.~$f_s$ remains constant along the phase-space trajectory ${\rm D}\mu_s/{\rm D}t = 0$). The second term in parentheses on the left-hand side of Equation (\ref{eqn:kinetic}) is then straightforwardly interpreted as the mirror force, $(m_s w^2_\perp/2) \grad\bcdot\eb = - \mu_s \eb \bcdot \grad B$, which, by Equation (\ref{eqn:eprl}), is compensated in the electron fluid by the parallel electric force, $-e E_\parallel$, and by the divergence of the parallel electron pressure, $-(1/n_e) \grad\bcdot ( p_{\parallel e} \eb )$.

\subsection{Particle Drifts and Centre-of-Mass Variables}\label{sec:drifts}

The same ordering as one uses to obtain a gyrophase-independent $f_s$ also guarantees that all species drift perpendicularly to the magnetic field with identical velocities. Indeed, the lowest-order contribution to the perpendicular electric field is $\bb{E}_\perp = -(\bb{u}_s/c) \btimes \bb{B}$, so that $\bb{u}_{\perp s} = \bb{u}_\perp = c \bb{E} \btimes \bb{B} / B^2$. With this borne in mind, we interpret the ${\rm D}/{\rm D}t$ operator in Equation (\ref{eqn:kinetic}) as measuring the rate of change of a quantity in a Lagrangian frame that is transported parallel to the magnetic field at velocity $v_\parallel$ and drifts perpendicular to the magnetic field at the $\bb{E} \btimes \bb{B}$ velocity.

It then follows that the mean drift of any species relative to the centre-of-mass velocity $\bb{u} \doteq \sum_s m_s n_s \bb{u}_s / \sum_s m_s n_s$ must be in the parallel direction, viz.,~$\bb{u}_s = \bb{u} + u'_{\parallel s} \eb$ with
\begin{equation}\label{eqn:uprls}
u'_{\parallel s} = \frac{1}{n_s} \int {\rm d}^3\bb{v} \left( v_\parallel - u_\parallel \right) f_s .
\end{equation}
In centre-of-mass variables with $\rho \doteq \sum_s m_s n_s$, Equations (\ref{eqn:continuity}) and (\ref{eqn:force}) become
\begin{equation}\label{eqn:comcontinuity}
\left( \pD{t}{} + \bb{u} \bcdot \grad \right) \rho = - \rho \grad \bcdot \bb{u} ,
\end{equation}
\begin{equation}\label{eqn:comforce}
\rho \left( \pD{t}{} + \bb{u} \bcdot \grad \right) \bb{u} = - \grad \left( p_\perp + \frac{B^2}{8\pi} \right) + \grad \bcdot \left[ \eb \eb \left( p_\perp - p_\parallel - \sum_s m_s n_s u'^2_{\parallel s} + \frac{B^2}{4\pi} \right) \right] ,
\end{equation}
where $p_\perp \doteq \sum_s p_{\perp s}$ and $p_\parallel \doteq \sum_s p_{\parallel s}$ are the total perpendicular and parallel pressures, respectively. The parallel drifts contribute to the total parallel pressure in a straightforward way. The magnetic field satisfies the ideal induction equation
\begin{equation}\label{eqn:induction}
\pD{t}{\bb{B}} = -c \grad \btimes \bb{E} = \grad \btimes \bigl( \bb{u} \btimes \bb{B} \bigr) ,
\end{equation}
i.e., the magnetic flux is frozen into a frame moving perpendicular to the magnetic field at the velocity $\bb{u}_\perp$.

\subsection{Reduced Ordering and Dimensionless Parameters}\label{sec:ordering}

We proceed by separating all fields into equilibrium values plus fluctuations: $f_s = f_{0s} + \df{s}$, $\bb{B} = B_0 \ez + \delta \bb{B}$, $n_s = \nsp + \dns$, $p_{\perp s} = \pprp{s} + \dpprp{s}$, $p_{\parallel s} = \pprl{s} + \dpprl{s}$, and $u'_{\parallel  s} = \dupar + \ddupar$. The mean magnetic field $B_0\ez$ and the equilibrium distribution function $f_{0s}$ are both taken to be spatially uniform (the slab limit). The asymptotic ordering
\begin{equation}\label{eqn:ordering}
 \frac{k_\parallel}{k_\perp} \sim \frac{\df{s}}{f_{0s}} \sim \frac{u_\perp}{\valf} \sim \frac{ \delta B_\perp}{B_0} \sim \frac{u_\parallel}{\valf} \sim \frac{\dBprl}{B_0} \sim \frac{\dns}{\nsp} \sim \frac{\dpprp{s}}{\pprp{s}} \sim \frac{\dpprl{s}}{\pprl{s}} \sim \frac{\ddupar}{\dupar} \sim \epsilon,
\end{equation}
where
\begin{equation}\label{eqn:valf}
\valf \doteq \frac{B_0}{\sqrt{4\pi\rho_0}}
\end{equation}
is the Alfv\'{e}n speed, is then applied to ``reduce'' the equations so that they describe the evolution of anisotropic ($k_\parallel \ll k_\perp$) fluctuations whose parallel Alfv\'{e}n timescale and perpendicular non-linear timescale are of the same order, $k_\parallel \valf \sim k_\perp u_\perp$---the so-called critical-balance conjecture \citep{gs95}, used here as an ordering assumption (S09). Such spatial anisotropy is both measured directly in the solar wind \citep[e.g.,][]{bwm96,hfo08,podesta09,wicks10,chen11a} and observed in numerical simulations of Alfv\'{e}nic turbulence \citep[e.g.,][]{smm83,opm94,cv00,mg01}. The perpendicular perturbations are taken to be Alfv\'{e}nic ($\delta B_\perp / B_0 \sim u_\perp / \valf$) and the compressive perturbations ($\delta u_\parallel$, $\dBprl$, $\dns$, $\dpprp{s}$, $\dpprl{s}$, $\ddupar$) are ordered comparable to the Alfv\'{e}nic ones, with the parallel and perpendicular thermal speeds of species $s$,
\begin{equation}\label{eqn:vths}
\vthprl{s} \doteq \sqrt{\frac{2\tprl{s}}{m_s}} \qquad {\rm and} \qquad \vthprp{s} \doteq \sqrt{\frac{2\tprp{s}}{m_s}} ,
\end{equation}
respectively, ordered comparable to the Alfv\'{e}n speed (i.e., $\betaprl{s}$, $\betaprp{s}$ defined in Eq.~\ref{eqn:betas} ordered unity). We further assume that the characteristic frequency of the fluctuations $\omega \sim k_\parallel \valf \sim \vthprl{i} / L$. This means that fast magnetosonic modes, for which $\omega \sim k_\perp \valf$, are ordered out of our equations. Fast-wave fluctuations are rarely seen in the solar wind \citep{howes12}; observations of turbulence in the solar wind confirm that it is primarily Alfv\'{e}nic \citep[e.g.,][]{bd71} and that its compressive component is substantially pressure-balanced (\citealt{roberts90,burlaga90,mt93,bpb04}; see Eq.~\ref{eqn:pbalance}). We expect the same to hold true in the intracluster medium, where observationally inferred turbulent velocities are  convincingly subsonic \citep[e.g.,][]{sf13,zhuravleva14}. Indeed, our reduced ordering is consistent with a small sonic Mach number: ${\rm Ma} = u_\perp / v_{{\rm th}i} \sim \epsilon ( \valf / v_{{\rm th}i} ) \ll 1$. 

The density fluctuations of the various species are related to one another via quasineutrality (Eq.~\ref{eqn:quasineutrality}):
\begin{equation}\label{eqn:neutrality}
\sum_s c_s \frac{\dns}{\nsp} = \sum_s \frac{c_s}{\nsp} \int {\rm d}^3\bb{v} \, \df{s} = 0, 
\end{equation}
where $c_s \doteq Z_s \nsp / \nem$ is the charge-weighted ratio of number densities; note that $c_e = -1$ and $\sum_s c_s = 0$. The perturbed pressures are calculated via
\begin{equation}\label{eqn:dpprl-dpprp}
\dpprl{s} = \int {\rm d}^3\bb{v}\, m_s \bigl( v_\parallel - \dupar \bigr)^2 \, \df{s}  \qquad {\rm and} \qquad \dpprp{s} = \int {\rm d}^3\bb{v}\, m_s \frac{w^2_\perp}{2}  \,\df{s}  .
\end{equation}
Perturbed parallel drifts may be obtained directly from taking first moments of the perturbed distribution functions,
\begin{equation}\label{eqn:drifts}
\ddupar = \frac{1}{\nsp} \int {\rm d}^3\bb{v} \, ( v_\parallel - \dupar )\, \df{s} - u_\parallel \qquad {\rm with} \qquad u_\parallel = \frac{ \sum_s m_s \int {\rm d}^3\bb{v}\, v_\parallel \df{s}}{\sum_s m_s \nsp},
\end{equation}
rather than indirectly via the momentum equations (\ref{eqn:force}) and (\ref{eqn:comforce}). In other words, they are not independent quantities.

The resulting set of equations has a number of dimensionless free parameters:
\begin{equation}\label{eqn:taus}
\tauprl{s} \doteq \frac{\tprl{s}}{\tprl{e}} \qquad {\rm and} \qquad \tauprp{s} \doteq \frac{\tprp{s}}{\tprp{e}}
\end{equation}
are the ratios of the parallel and perpendicular temperatures of species $s$ to their respective electron temperatures (note that $\tauprl{e} = \tauprp{e} = 1$); 
\begin{equation}\label{eqn:paniso}
\Delta_s \doteq \frac{\pprp{s}}{\pprl{s}} - 1
\end{equation}
is the dimensionless pressure anisotropy of species $s$; and
\begin{equation}\label{eqn:betas}
\betaprl{s} \doteq \frac{8 \pi \pprl{s}}{B^2_0} \qquad {\rm and} \qquad \betaprp{s} \doteq \frac{8 \pi \pprp{s}}{B^2_0}
\end{equation}
are the ratios of the parallel and perpendicular pressures of species $s$ to the magnetic pressure. We use the shorthand $\beta_\parallel \doteq \sum_s \betaprl{s}$ and $\beta_\perp \doteq \sum_s \betaprp{s}$. All of these quantities, evaluated in the equilibrium state, are taken to be order unity in the $\epsilon$ expansion (subsidiary limits in, e.g., high and low $\beta_\parallel$ can be taken after the $\epsilon$ expansion is done; see \S\ref{sec:mirror}). Likewise, we order $\dupar \sim \vthprl{i}$ for all species $s$. This precludes equilibrium parallel drifts in the electron fluid from entering into our equations, because $v_\parallel \sim \vthprl{e} \sim \sqrt{m_i / m_e} \vthprl{i} \gg \vthprl{i}$ for the electrons (i.e., the random thermal motions of electrons are characterised by speeds much in excess of the ion thermal speed and, therefore, any parallel drifts in the equilibrium state). 

We will also make frequent use of the following compact notation:
\begin{equation}\label{eqn:fprlfprp}
f^\parallel_{0s} \doteq - \vthprl{s}^2 \,\pD{(v_\parallel - \dupar)^2}{f_{0s}} \qquad {\rm and} \qquad f^\perp_{0s} \doteq -\vthprp{s}^2 \, \pD{w^2_\perp}{f_{0s}} ,
\end{equation}
which are dimensionless derivatives of a species' equilibrium distribution with respect to the square of the parallel velocity (peculiar to the species equilibrium drift velocity) and perpendicular velocity (peculiar to the $\bb{E}\btimes\bb{B}$ drift velocity), respectively; and 
\begin{equation}\label{eqn:Df0s}
\mf{D} f_{0s} \doteq \frac{\pprp{s}}{\pprl{s}} f^\parallel_{0s} - f^\perp_{0s} ,
\end{equation}
which measures the velocity-space anisotropy of that distribution. These definitions are, of course, only useful insofar as particle collisions and scatterings are unable to maintain a Maxwellian distribution in the equilibrium state, for which $f^\parallel_{0s} = f^\perp_{0s} = f_{0s}$ and $\mf{D} f_{0s} = 0$. To allow for this to be the case, we order the collision frequency $\nu_{ii}\ll\epsilon\omega$. This means that collisional relaxation of $f_{0s}$ towards a Maxwellian distribution occurs at higher orders than will be treated in this Paper, as does the heating of the background plasma due to collisional smoothing of the (secularly increasing) fine-scale structure in velocity space. For our purposes, the background equilibrium is thus stationary in time.

While most applications to non-Maxwellian space and astrophysical plasmas make use of the bi-Maxwellian distribution function,
\begin{equation}\label{eqn:biMax}
f_{\textrm{bi-M},s}(v_\parallel,w_\perp) \doteq \frac{\nsp}{\sqrt{\pi} \vthprl{s}} \exp \left[ - \frac{( v_\parallel - \dupar )^2}{\vthprl{s}^2} \right] \frac{1}{\pi\vthprp{s}^2} \exp \left( - \frac{w^2_\perp}{\vthprp{s}^2} \right) ,
\end{equation}
to describe the equilibrium distribution function of the plasma (for which $f^\parallel_{0s} = f^\perp_{0s} = f_{0s}$ and $\mf{D} f_{0s} = \Delta_s f_{0s}$), we keep our derivation of KRMHD general with respect to the form of $f_{0s}(v_\parallel,w_\perp)$.\footnote{Some restrictions on $f_{0s}$ are necessary in order to make sense of the generalised free-energy invariant of KRMHD, derived and discussed in Sections \ref{sec:Wcompr} and \ref{sum:W}. Namely, $f^\parallel_{0s}$ must be strictly positive throughout all of the parallel velocity space; i.e., $f_{0s}$ must decay monotonically away from $v_\parallel = \dupar$ and not be too flat around that point. This restriction also eliminates the possibility that $f_{0s}$ is unstable to high-frequency bump-on-tail instabilities, which are outside the KRMHD ordering. This class of distribution functions covers all plausible distribution functions for the solar wind (Maxwellian, bi-Maxwellian, kappa, bi-kappa, etc.).} That being said, because of its widespread use by the space and astrophysical communities, we will often refer back to Equation (\ref{eqn:biMax}) to present useful particular cases of our more general results. To ease application of our theory to the solar wind, in Appendix \ref{app:bikappa} we also specialise our equations for the case of a bi-kappa equilibrium distribution function.

Finally, we caution that there are some situations in weakly collisional magnetised astrophysical plasmas for which the KRMHD ordering (\ref{eqn:ordering}) is inapplicable. One such situation is when the equilibrium pressure anisotropy takes on values beyond the firehose stability boundary (Eq.~\ref{eqn:firehose}), since $k_\parallel$ is not much smaller than $k_\perp$ for the fastest-growing firehose modes. Likewise, beyond the mirror stability boundary (Eq.~\ref{eqn:mirrorstability}), the nonlinear evolution of the mirror instability involves the trapping of particles, a feature not accounted for in the KRMHD ordering (note however that the expected wavevector anisotropy $k_\parallel/k_\perp \ll 1$ of these modes {\em is} captured). That being said, the majority of the solar wind does lie within these stability boundaries \citep[e.g.][]{bale09}, and so our theory should be appropriate for describing Alfv\'{e}nic turbulence in that part of parameter space. A second limitation of the KRMHD ordering is that there are times and places in both the solar wind and the magnetosheath where $\delta B / B_0$ is not small \citep[e.g.][]{alm08}, usually at the outer scale. However, as the free energy stored in the turbulent fluctuations cascades to smaller scales (in the inertial range and beyond), the fluctuations become smaller and more anisotropic. If anything, these fluctuations tend towards the KRMHD (or, more generally, gyrokinetic---see Appendix \ref{app:gk}) limit.

\subsection{Alfv\'{e}nic Fluctuations}\label{sec:alfven}

To zeroth order in $\epsilon$, Equation (\ref{eqn:comcontinuity}) becomes $\grad_{\!\perp} \bcdot  \bb{u}_\perp = 0$. Likewise, the divergence-free constraint on the magnetic field, $\grad \bcdot  \bb{B} = 0$, becomes $\grad_{\!\perp} \bcdot  \dBprp = 0$. These simplifications allow the Alfv\'{e}nic fluctuations to be expressed in terms of scalar stream (flux) functions:
\begin{equation}\label{eqn:streamflux}
\bb{u}_\perp = \ez \btimes \grad_{\!\perp} \Phi \qquad {\rm and} \qquad \frac{\dBprp}{\sqrt{4\pi\rho_0}} = \ez \btimes \grad_{\!\perp} \Psi .
\end{equation}
We substitute these expressions into Equations (\ref{eqn:comforce}) and (\ref{eqn:induction}) and examine the result order-by-order in $\epsilon$.

To lowest order, Equations (\ref{eqn:comforce}) and (\ref{eqn:induction}) become, respectively,
\begin{equation}\label{eqn:pbalance}
\grad_{\!\perp} \left( \sum_s \dpprp{s} + \frac{B^2_0}{4\pi} \frac{\dBprl}{B_0} \right) = 0 ,
\end{equation}
\begin{equation}\label{eqn:redinduction}
\pD{t}{}\Psi + \{ \Phi \, , \Psi \} = \valf \pD{z}{}\Phi ,
\end{equation}
where the Poisson bracket
\begin{equation}\label{eqn:rmhdbracket}
\{ \Phi \, , \Psi \} \doteq \ez \bcdot  ( \grad_{\!\perp} \Phi \btimes \grad_{\!\perp} \Psi ).
\end{equation}
The first of these equations expresses perpendicular force balance, a result which will aid our description of the compressive fluctuations in Section \ref{sec:compressive}. The second equation is identical to the induction equation in standard RMHD. At the next order, we derive an evolution equation for the stream function. Its simplest form is obtained by taking the $z$-component of the curl of the force equation (\ref{eqn:comforce})---the vorticity equation---which gives
\begin{eqnarray}\label{eqn:redforce}
\lefteqn{
\pD{t}{} \nabla^2_\perp \Phi + \left\{ \Phi \, , \nabla^2_\perp \Phi \right\} 
}
\nonumber\\*&&\mbox{}
= \left[ 1 + \sum_s \frac{\betaprl{s}}{2} \!\left( \Delta_s  - \frac{2\duparsq}{\vthprl{s}^2} \right) \right] \left( \valf \pD{z}{} \nabla^2_\perp \Psi + \left\{ \Psi \, , \nabla^2_\perp \Psi \right\} \right) .
\end{eqnarray}
For an isotropic equilibrium pressure ($\pprp{s} = \pprl{s}$, $\Delta_s = 0$) and no equilibrium parallel drifts ($\dupar = 0$), this reduces to the standard RMHD momentum equation.

The nonlinearities in Equations (\ref{eqn:redinduction}) and (\ref{eqn:redforce}) involving the magnetic field imply that Alfv\'{e}nic fluctuations propagate along the locally deformed magnetic field rather than the uniform equilibrium field, and so the parallel and perpendicular directions do not strictly lie along the Cartesian axes defined by the guide field. Indeed, by introducing the Lagrangian operators
\begin{equation}\label{eqn:dbydt}
\D{t}{} \doteq \pD{t}{} + \bb{u}_\perp  \bcdot \grad_{\!\perp} = \pD{t}{} + \{ \Phi \,, \dots \} ,
\end{equation}
\begin{equation}\label{eqn:bdotgrad}
\eb  \bcdot \grad \doteq \pD{z}{} + \frac{\dBprp}{B_0}  \bcdot \grad_{\!\perp} = \pD{z}{} + \frac{1}{\valf} \{ \Psi \, , \dots \} ,
\end{equation}
Equations (\ref{eqn:redinduction}) and (\ref{eqn:redforce}) may be written compactly as
\begin{align}\label{eqn:redinduction2}
\pD{t}{\Psi} &= \valf  \eb  \bcdot \grad \Phi , \\
\label{eqn:redforce2}
\D{t}{} \nabla^2_\perp \Phi &= \valf   \eb  \bcdot \grad \frac{\valfeff^2}{\valf^2}  \nabla^2_\perp  \Psi ,
\end{align}
where we have introduced the effective Alfv\'{e}n speed
\begin{equation}\label{eqn:vaeff}
\valfeff \doteq \valf \left[ 1 + \sum_s  \frac{\beta_{\parallel s}}{2} \!\left( \Delta_s - \frac{2\duparsq}{\vthprl{s}^2} \right)  \right]^{1/2} .
\end{equation}
In systems where $p_{\perp 0} - p_{\parallel 0} - \sum_s m_s \nsp \duparsq < 0$, the speed at which deformations in the magnetic field are propagated is effectively reduced by the excess parallel pressure, which undermines the restoring force exerted by the tension of the magnetic-field lines. When $p_{\perp 0} - p_{\parallel 0} - \sum_s m_s \nsp \duparsq = - B^2_0 / 4\pi$, the magnetic tension is exactly balanced by the anisotropy in the distribution function and the plasma does not respond to (perpendicular) magnetic perturbations. Parallel drifts in the equilibrium distribution functions of different species make this criterion easier to satisfy by supplementing the parallel thermal pressure. For values of $p_{\perp 0} - p_{\parallel 0}$ below this threshold, the effective Alfv\'{e}n speed becomes imaginary and the stream and flux functions acquire a $\pi/2$ relative phase shift. The plasma then becomes firehose unstable, an effect that we discuss in Section \ref{sec:firehose}. For now we caution that, if the equilibrium pressure anisotropy and parallel drifts make $\valfeff / \valf$ as small as $\epsilon$, the reduced ordering (\ref{eqn:ordering}) is broken and the KRMHD theory, as derived here, becomes inapplicable.

Equations (\ref{eqn:redinduction2}) and (\ref{eqn:redforce2}) form a closed set, and so the Alfv\'{e}n-wave inertial-range cascade is completely decoupled from all other (compressive) types of perturbations (the Alfv\'{e}nic cascade is further discussed in Section \ref{sec:alfveniccascade}). While this result is usually derived starting from the collisional MHD limit, we have shown that the same holds true even for a collisionless plasma (as in S09) with arbitrary gyrotropic equilibrium distribution function.

\subsection{Compressive Fluctuations}\label{sec:compressive}

\subsubsection{Parallel Electric Field}\label{sec:efield}

To obtain the equations describing the density ($\dne$) and magnetic-field-strength ($\dBprl$) fluctuations, we return to the drift-kinetic equation (\ref{eqn:kinetic}). Applying the reduced ordering (\ref{eqn:ordering}) and neglecting collisions, we have for the electron species
\begin{equation}\label{eqn:eDKE}
\left( \D{t}{} + v_\parallel \eb \bcdot \grad \right) \left( \df{e} - \frac{\dBprl}{B_0} \frac{w^2_\perp}{\vthprp{e}^2} f^\perp_{0e} \right) + \bigl( v_\parallel - \dupare \bigr) \left( \frac{eE_\parallel}{\tprl{e}} + \frac{w^2_\perp}{\vthprl{e}^2} \eb\bcdot\grad \frac{\dBprl}{B_0} \right) f^\parallel_{0e} = 0 .
\end{equation}
Further expanding Equation (\ref{eqn:eDKE}) in the small parameter $(m_e / m_i)^{1/2}$ removes the ${\rm d}/{\rm d}t$ term and the equilibrium electron drift ($\dupare$) to lowest order. Dividing the result by $v_\parallel$ and employing the $\mf{D}$ notation (Eq.~\ref{eqn:Df0s}) to group terms, we find
\begin{equation}\label{eqn:eleckinetic}
\eb \bcdot \grad \left( \df{e} + \frac{w^2_\perp}{\vthprp{e}^2} \frac{\dBprl}{B_0} \mf{D}\fem \right) + \frac{eE_\parallel }{\tprl{e}} \, f^\parallel_{0e} = 0 .
\end{equation}
Integrating over the velocity space leads to an expression for the parallel electric field,
\begin{equation}\label{eqn:efield}
E_\parallel = -\frac{1}{\czero{e}}\, \eb \bcdot \grad\,  \frac{\tprl{e}}{e}\! \left( \frac{\dne}{\nem} + \Delta_{1e} \frac{\dBprl}{B_0}  \right) ,
\end{equation}
where we use the following notation: for integer $\ell$,
\begin{align}\label{eqn:ecoeff}
\cell{e} &\doteq \frac{1}{\nem} \int {\rm d}^3 \bb{v} \,  \frac{1}{\ell !} \! \left( \frac{w_\perp}{\vthprp{e}} \right)^{\!2\ell} f^\parallel_{0e} \\*
\mbox{} &= 1 ~ \textrm{for a bi-Maxwellian} \nonumber
\end{align}
are dimensionless coefficients related to perpendicular moments of the parallel-differentiated equilibrium electron distribution function, and
\begin{equation}\label{eqn:epaniso}
\Delta_{\ell e} \doteq \cell{e} \frac{\pprp{e}}{\pprl{e}} - 1
\end{equation}
is the dimensionless pressure anisotropy (cf.~Eq.~\ref{eqn:paniso}) of the electrons weighted by those coefficients. For isotropic electrons, the parallel electric field is entirely related to fluctuations in the electron (and therefore ion) density; the corresponding (first) term in Equation (\ref{eqn:efield}) ultimately leads to the Landau damping of ion acoustic waves. When the equilibrium electron pressure is anisotropic, fluctuations in magnetic-field strength also contribute to the parallel electric field; this second term enforces quasineutrality in the face of preferential exclusion of large-pitch-angle electrons from regions of enhanced field strength.

\subsubsection{Pressure Perturbations}

With knowledge of the parallel electric field (Eq.~\ref{eqn:efield}), we can rewrite the perturbed electron distribution as\footnote{Technically, the perturbed electron distribution function can only be determined up to an additive unknown function whose parallel gradient vanishes. We have set this homogeneous solution to zero, a simplification which may be justified by assuming stochastic field lines (S09).}
\begin{equation}\label{eqn:dfe}
\df{e} = \frac{1}{\czero{e}} \left( \frac{\dne}{\nem} + \Delta_{1e} \frac{\dBprl}{B_0} \right) f^\parallel_{0e} - \frac{w^2_\perp}{\vthprp{e}^2} \frac{\dBprl}{B_0} \mf{D} \fem
\end{equation}
and compute the perturbed parallel and perpendicular electron pressures by taking the appropriate second moments (cf.~Eq.~\ref{eqn:dpprl-dpprp}):
\begin{align}\label{eqn:pprle}
\frac{\dpprl{e}}{\pprl{e}} &= \frac{1}{\czero{e}} \left( \frac{\dne}{\nem} + \Delta_{1e} \frac{\dBprl}{B_0} \right) - \Delta_e \frac{\dBprl}{B_0} ,
\\*\label{eqn:pprpe}
\frac{\dpprp{e}}{\pprp{e}} &= \frac{\cone{e}}{\czero{e}} \left( \frac{\dne}{\nem} + \Delta_{1e} \frac{\dBprl}{B_0} \right) - 2 \Delta_{2e} \frac{\dBprl}{B_0} .
\end{align}
Note that, if the equilibrium distribution function of electrons is isotropic, no electron pressure anisotropy can be generated by the fluctuations and the electron fluid remains isothermal along magnetic-field lines:
\[
\eb \bcdot \grad \delta T_e = 0.
\]
The latter occurs physically by rapid electron conduction along field lines. Deviations from isothermality in non-Maxwellian plasmas arise when electrons conserve their adiabatic invariant $\mu_e$ in the presence of field-line compressions and rarefactions. Indeed, by taking the third velocity moments of Equation (\ref{eqn:dfe}), we see that the parallel flows of parallel and perpendicular electron heat, respectively
\[
Q_{\parallel e} \doteq \int {\rm d}^3\bb{v} \, m_e v^3_\parallel \, \df{e} \qquad {\rm and} \qquad Q_{\perp e} \doteq \int {\rm d}^3\bb{v} \,  m_e v_\parallel \frac{w^2_\perp}{2} \, \df{e} ,
\]
satisfy $\eb \bcdot \grad Q_{\parallel e} = \eb \bcdot \grad Q_{\perp e} = 0$. For bi-Maxwellian electrons, this translates into
\begin{equation}\label{eqn:isothermalelectrons}
\eb \bcdot \grad \dtprl{e} = 0 \qquad {\rm and} \qquad \eb \bcdot \grad \Bigl( \dtprp{e} +   \Delta_e \langle \mu_e \rangle \dBprl  \Bigr) = 0, 
\end{equation}
where $\langle \mu_e \rangle =  m_e \vthprp{e}^2 / 2B_0$ is the lowest-order contribution to the mean adiabatic invariant of the electrons. Equation (\ref{eqn:isothermalelectrons}) states that, while the parallel temperature of the electrons remains constant along field lines, the perpendicular temperature cannot do so without violating $\mu_e$ conservation \citep[cf.~eqs 39--40 of][]{shd97}.

\subsubsection{Reduced Drift-Kinetic Equation}\label{sec:reduceddke}

It is often computationally convenient as well as physically illuminating to replace the perturbed distribution function $\df{s}$ by the function
\begin{equation}\label{eqn:gs}
g_s \doteq \df{s} - \frac{w^2_\perp}{\vthprp{s}^2} \frac{\dBprl}{B_0} f^\perp_{0s} ,
\end{equation}
which is the perturbed distribution function if $f_{0s}$ is taken to be a function of the exact adiabatic invariant $\mu_s = m w^2_\perp / 2B$ (rather than of $m w^2_\perp / 2B_0$); to wit,
\begin{align}
f_s - f_{0s} ( v_\parallel, \mu_s ) &= f_s - f_{0s} \biggl( v_\parallel , \frac{m_s w^2_\perp}{2B_0} - \frac{m_s w^2_\perp}{2B_0} \left( 1 - \frac{B_0}{B} \right) \biggr) \nonumber\\*
\mbox{}&\simeq f_s - f_{0s} ( v_\parallel, w_\perp ) + \frac{w_\perp}{2} \frac{\dBprl}{B_0} \pD{w_\perp}{f_{0s}} \nonumber\\*
\mbox{}&= \delta f_s -  \frac{w^2_\perp}{\vthprp{s}^2} \frac{\dBprl}{B_0} f^\perp_{0s} = g_s .
\end{align}
Indeed, using Equation (\ref{eqn:dfe}) in Equation (\ref{eqn:gs}), we find that
\begin{equation}\label{eqn:ge}
g_e = \left[ \frac{1}{\czero{e}} \left( \frac{\dne}{\nem} + \Delta_{1e} \frac{\dBprl}{B_0} \right) - \frac{w^2_\perp}{\vthprl{e}^2} \frac{\dBprl}{B_0} \right] f^\parallel_{0e}
\end{equation}
does not contain any derivatives of the equilibrium distribution function with respect to $w_\perp$. We will see that the same holds true for $g_i$, whose evolution equation we now derive.

The evolution equation for the perturbed ion distribution function is obtained by applying the reduced ordering (Eq.~\ref{eqn:ordering}) to Equation (\ref{eqn:kinetic}) with $s = i$. Using Equation (\ref{eqn:efield}) for the parallel electric field, the reduced kinetic equation for the ions may be written in a compact form analogous to equation (145) of S09:
\begin{equation}\label{eqn:ionkinetic}
\left( \D{t}{} + v_\parallel \eb  \bcdot \grad \right) g_i + \bigl( v_\parallel - \dupari \bigr) \eb \bcdot \grad \left[ \frac{1}{\czero{e}}\frac{Z_i}{\tauprl{i}}  \left( \frac{\dne}{\nem} + \Delta_{1e} \frac{\dBprl}{B_0} \right) + \frac{w^2_\perp}{\vthprl{i}^2} \!\frac{\dBprl}{B_0} \right] f^\parallel_{0i} = 0 .
\end{equation}
Note that this equation does not contain any derivatives of the equilibrium distribution function with respect to $w_\perp$. In terms of $g_i$, Equations (\ref{eqn:neutrality}), (\ref{eqn:drifts}), and (\ref{eqn:pbalance}) become
\begin{equation}\label{eqn:neutrality2}
\frac{\dne}{\nem} - \frac{\dBprl}{B_0} =  \sum_i \frac{c_i}{\nip} \int {\rm d}^3\bb{v}\, g_i ,
\end{equation}
\begin{equation}\label{eqn:drifts2}
u_\parallel + \sum_i c_i \ddupari = \sum_i \frac{c_i}{\nip} \int {\rm d}^3\bb{v} \, ( v_\parallel - \dupari ) \, g_i ,
\end{equation}
\begin{align}\label{eqn:pbalance2}
\frac{\cone{e}}{\czero{e}} \left( \frac{\dne}{\nem} + \Delta_{1e} \frac{\dBprl}{B_0} \right) + 2 &\left( \sum_i c_i \frac{\tauprp{i}}{Z_i} + \frac{1}{\betaprp{e}} - \Delta_{2e} \right) \frac{\dBprl}{B_0} 
\nonumber\\*
\mbox{} &= - \sum_i  \frac{\tauprp{i}}{Z_i} \frac{c_i}{\nip} \int {\rm d}^3\bb{v} \, \frac{w^2_\perp}{\vthprp{i}^2} \, g_i ,
\end{align}
which reduce to equations (146)--(148) of S09 for a single-ion-species Maxwellian plasma.

Equations (\ref{eqn:ionkinetic})--(\ref{eqn:pbalance2}) evolve the ion distribution function $g_i$; the ``slow-wave quantities'' $u_\parallel$, $\delta u'_{\parallel i}$, and $\dBprl$; and the density fluctuations $\dne$. All nonlinearities are contained in the ${\rm d}/{\rm d}t$ and $\eb\bcdot\grad$ Lagrangian operators, which include the Alfv\'{e}nic quantities $\Phi$ and $\Psi$; these are determined separately and independently by Equations (\ref{eqn:redinduction}) and (\ref{eqn:redforce}). Nonlinear scattering/mixing of slow waves and the entropy mode by the Alfv\'{e}nic perturbations takes the form of passive advection of the distribution function $g_i$. In other words, even when the equilibrium distribution function is non-Maxwellian and there are parallel drifts between the various species, the compressive fluctuations are passively transported by the Alfv\'{e}nic fluctuations, a result that we have, thus, generalised from MHD \citep{lg01} and Maxwellian KRMHD (S09). The passive cascades of compressive fluctuations, as well as their kinetic damping and susceptibility to mirror instability in a pressure-anisotropic plasma are further discussed in Section \ref{sec:compressivecascade}.

\subsection{Summary}\label{sum:akrmhd}

The reduced theory derived here evolves $4+N_{\rm ion}$ unknown functions: $\Phi$, $\Psi$, $\dBprl$, $\dne$, and $g_i$ for each of the $N_{\rm ion}$ different ionic species. The stream and flux functions, $\Phi$ and $\Psi$ respectively, are related to the fluid quantities (perpendicular velocity and magnetic-field perturbations) via Equation (\ref{eqn:streamflux}). They satisfy a closed set of equations, Equations (\ref{eqn:redinduction})--(\ref{eqn:redforce}), which describe the decoupled cascade of Alfv\'{e}nic fluctuations whose phase speed is modified by pressure anisotropy and inter-species parallel drifts. In the collisional limit, they revert to the standard equations of RMHD. The density and magnetic-field-strength fluctuations (the ``compressive'' fluctuations, or the slow waves and the entropy mode in the collisional limit) require a kinetic description in terms of the ion distribution function $g_i$, which is evolved by the kinetic Equation (\ref{eqn:ionkinetic}). This kinetic equation itself contains $\dne$ and $\dBprl$, which are, in turn, calculated by taking velocity-space integrals of $g_i$ via Equations (\ref{eqn:neutrality2}) and (\ref{eqn:pbalance2}). The nonlinear evolution of $g_i$, $\dBprl$, and $\dne$ is due solely to passive advection of $g_i$ by the Alfv\'{e}nic turbulence, which mixes $\dne$ and $\dBprl$ in the direction transverse to the magnetic field. In the Lagrangian frame associated with the Alfv\'{e}nic fluctuations, the compressive fluctuations obey a one-dimensional linear equation, which may be solved independently of the Alfv\'{e}nic turbulence.

Here we summarise our new set of equations:
\begin{subequations}\label{sum:eqns}
\begin{equation}\label{sum:psi}
\pD{t}{\Psi} = \valf  \eb  \bcdot \grad \Phi ,
\end{equation}
\begin{equation}\label{sum:phi}
\D{t}{} \nabla^2_\perp \Phi = \valf \eb  \bcdot \grad \frac{\valfeff^2}{\vasq} \nabla^2_\perp \Psi ,
\end{equation}
\begin{equation}\label{sum:ionkinetic}
\left( \D{t}{} + v_\parallel \eb  \bcdot \grad \right) g_i + \bigl( v_\parallel - \dupari \bigr) \eb \bcdot \grad \left[ \frac{1}{\czero{e}}\frac{Z_i}{\tauprl{i}}  \left( \frac{\dne}{\nem} + \Delta_{1e} \frac{\dBprl}{B_0} \right) + \frac{w^2_\perp}{\vthprl{i}^2} \!\frac{\dBprl}{B_0} \right] f^\parallel_{0i} = 0 .
\end{equation}
\begin{align}\label{sum:dne}
\frac{\dne}{\nem} =  - \sum_i \Bigl[ \,\dots \Bigr]^{-1} \frac{c_i}{\nip} \int {\rm d}^3\bb{v} \left[ \frac{w^2_\perp}{\vthprp{i}^2} \!\!\!\right. & \left. \mbox{} - 2 c_i \left( \sum_{i'} \frac{c_{i'}  \tauprp{i'} Z_i}{c_i \tauprp{i} Z_{i'}} + \frac{1}{\betaprp{i}} \right) \right. \nonumber\\*
\mbox{} & \left. \mbox{} + \frac{Z_i}{\tauprp{i}} \left( 2 \Delta_{2e} - \frac{\cone{e}}{\czero{e}}\Delta_{1e} \right) \right] g_i ,
\end{align}
\begin{equation}\label{sum:dBprl}
\frac{\dBprl}{B_0} = - \sum_i \Bigl[ \,\dots \Bigr]^{-1} \frac{c_i}{\nip} \int {\rm d}^3\bb{v} \left( \frac{w^2_\perp}{\vthprp{i}^2} + \frac{\cone{e}}{\czero{e}} \frac{Z_i}{\tauprp{i}} \right) g_i ,
\end{equation}
\end{subequations}
where
\begin{equation}\label{eqn:bracket}
 \Bigl[ \,\dots \Bigr] \doteq \frac{Z_i}{\tauprp{i}} \left( \frac{\cone{e}}{\czero{e}}   - 2 \Delta_{2e}  +  \frac{\cone{e}}{\czero{e}}  \Delta_{1e} \right) + 2 c_i \left( \sum_{i'} \frac{c_{i'} \tauprp{i'} Z_i}{c_i \tau_{\perp i} Z_{i'}} + \frac{1}{\betaprp{i}} \right) .
\end{equation}
These equations reduce to equations (155)--(159) of S09 when the equilibrium distribution function is Maxwellian and only one ionic species is present (for which $\valfeff = \valf$, $\dupari = 0$, $\cell{e} = 1$, $\tauprl{s} = \tauprp{s}$, $\Delta_{\ell e} = 0$, $\vthprl{s} = \vthprp{s}$, $\betaprl{s} = \betaprp{s}$, and $f^\parallel_{0s} = f^\perp_{0s} = f_{0s}$). 

It should be noted that Equations (\ref{sum:eqns}) are ideal, in that they are ignorant of any physics capable of dissipating the large gradients in phase space that will inevitably be produced as the turbulence cascades to smaller and smaller scales. While some of this fine-scale structure is regularised by finite-Larmor-radius effects, which are included in the non-Maxwellian gyrokinetic theory derived in Appendix \ref{app:gk}, numerical implementation of Equations (\ref{sum:eqns}) requires the addition of finite collisionality and resistivity.

\section{Alfv\'{e}nic Fluctuations in the Inertial Range}\label{sec:alfveniccascade}

Having constructed a theoretical framework for the evolution of anisotropic kinetic turbulence in collisionless magnetised astrophysical plasmas, we now investigate its implications for the behaviour of Alfv\'{e}nic fluctuations in the inertial range (dynamical equations for these fluctuations were derived in Section \ref{sec:alfven}). In this Section we demonstrate that pressure anisotropy and parallel drifts do not interfere with the nonlinear mixing of counter-propagating Alfv\'{e}nic fluctuations. In doing so, we derive the two Alfv\'{e}nic invariants that are independently conserved and cascaded by these interactions. The explanation of the physical content of these invariants is aided by the linear theory of Alfv\'{e}n waves, which we present in the next Section.

\subsection{Linear Theory of Alfv\'{e}nic Fluctuations: Alfv\'{e}n Waves and Firehose Instability}\label{sec:firehose}

The linear theory of Alfv\'{e}nic fluctuations in KRMHD can be readily obtained by dropping the nonlinear terms in Equations (\ref{sum:psi}) and (\ref{sum:phi}) and adopting the solutions $\Phi$, $\Psi \sim \exp(- \imag \omega t + \imag \bb{k}\bcdot \bb{r} )$. The resulting dispersion relation is simply
\begin{equation}\label{eqn:alfven}
\omega = \pm k_\parallel \valfeff ,
\end{equation}
with eigenvectors satisfying $\Phi = \mp \Psi ( \valfeff / \valf )$. When
\begin{equation}\label{eqn:firehose}
p_{\perp 0} - p_{\parallel 0} - \sum_s m_s \nsp \duparsq < - \frac{B^2_0}{4\pi} ,
\end{equation}
the phase speed of the Alfv\'{e}n wave becomes imaginary and the firehose instability results \citep{rosenbluth56,ckw58,parker58,vs58}. Physically, negative pressure anisotropies and/or parallel drifts reduce the elasticity of the magnetic-field lines, undermining the supplied restoring force necessary to propagate the wave. In KRMHD, the fastest growth occurs at arbitrarily small parallel scales, with no small-scale regularisation accessible within the long-wavelength approximation, $k\rho_i \ll 1$, in which Equation (\ref{eqn:force}) is derived. To obtain the fastest growing mode, finite-Larmor-radius effects must be taken into account \citep[cf.][]{schekochihin10}. Direct calculation from the hot-plasma dispersion relation yields $k_\parallel \rho_i \sim | \beta_\perp / \beta_\parallel -1 + 2/\beta_\parallel |^{1/2}$ for the parallel ($k_\perp = 0$) firehose \citep{ks67,dv68} and $k\rho_i \sim 1$ for the oblique firehose with $k_\perp \ne 0$ \citep{ywa93,hm00}.

\subsection{Elsasser Fields and Alfv\'{e}n Ratio}\label{sec:elsasser}

The effective Alfv\'{e}n speed can be used to cast Equations (\ref{eqn:redinduction2}) and (\ref{eqn:redforce2}) in a symmetric form via the introduction of the generalised {\it Elsasser potentials},
\begin{equation}\label{eqn:elsasserpotentials}
\zeta^\pm \doteq \Phi \pm \frac{\valfeff}{\valf}  \Psi ,
\end{equation}
and the corresponding {\it Elsasser fields}
\begin{equation}\label{eqn:elsasserfields}
\bb{z}^\pm \doteq \ez \btimes \grad_{\!\perp} \zeta^\pm = \bb{u}_\perp \pm \frac{\valfeff}{\valf} \frac{\dBprp}{\sqrt{4\pi\rho_0}}.
\end{equation}
The latter are a straightforward generalisation of the standard \citet{elsasser50} variables to non-Maxwellian equilibria, for which $\valfeff \ne \valf$. Combining Equations (\ref{eqn:redinduction}) and (\ref{eqn:redforce}), one can show that the Elsasser potentials satisfy
\begin{equation}\label{eqn:elsasser}
\left( \pD{t}{} \mp \valfeff \pD{z}{} \right) \nabla^2_\perp \zeta^\pm  = -  \frac{1}{2}  \Bigl( \left\{ \zeta^+ \,, \nabla^2_\perp \zeta^- \right\} + \left\{ \zeta^- \, , \nabla^2_\perp \zeta^+ \right\} \mp \nabla^2_\perp \left\{ \zeta^+ \, , \zeta^- \right\} \Bigr) .
\end{equation}
Thus, the standard result that nonlinear interactions (``scatterings'') of Alfv\'{e}nic fluctuations occur only between counter-propagating fluctuations \citep{kraichnan65} holds true for general (gyrotropic) distribution functions. What is modified by the non-Maxwellian nature of the distribution function is the amount of (perpendicular) magnetic fluctuations that comprise each of the Elsasser potentials. As $\valfeff \rightarrow 0$, the magnetic fluctuations fail to propagate and the distinction between $\zeta^+$ and $\zeta^-$ is no longer meaningful. Indeed, the very idea of critical balance that underpins the RMHD ordering $k_\parallel \valf \sim k_\perp u_\perp$ is based upon a causality argument: fluctuations cannot be correlated over a distance larger than that over which an Alfv\'{e}n wave propagates in a nonlinear interaction time. Significantly reducing the signal speed, with $\valfeff / \valf \sim \epsilon$ or smaller, interferes with this argument and breaks the reduced ordering used in this Paper. This is what will happen if the firehose threshold is approached. On the unstable side of the threshold, the firehose fluctuations that emerge are not anisotropic in the same way that Alfv\'{e}nic, or more generally gyrokinetic, fluctuations are: in fact, they have $k_\parallel \sim k_\perp$ \citep{ywa93,hm00,kss14}. This is why the considerations in this Paper do not describe the turbulence on the unstable side of the firehose threshold.

The fact that the Alfv\'{e}n ratio
\begin{equation}\label{eqn:alfvenratio}
r_{\rm A} \doteq \left| \frac{\Phi}{\Psi} \right|^2 = 1 + \sum_s \frac{\betaprl{s}}{2} \left( \Delta_s - \frac{2\duparsq}{\vthprl{s}^2} \right)
\end{equation}
depends on the anisotropy inherent to the distribution function \citep[cf.][]{barnes79} becomes testable in the solar wind, where measurements find that the energy in magnetic-field fluctuations exceeds the energy in the velocity fluctuations, $r_{\rm A} < 1$ \citep[e.g.,][]{bd71,mg82,bbv85,roberts87,tmt89,mt90,gvm91,gns95,bpb98,prg07,salem09,pb10,chen11a,borovsky12}. This result is often interpreted in terms of ``residual energy'', $\sigma_{\rm r} \doteq ( r_{\rm A} - 1 ) / ( r_{\rm A} + 1 )$, the difference between the energy in velocity and magnetic-field fluctuations that is believed to be an inherent feature of the turbulence itself (\citealt{pfl76}; see \citealt{chen13} and \citealt{wicks13} for brief reviews of the relevant literature and contemporary analyses). While the observed scale-dependent component of the residual energy is likely to be intrinsic---recent theory predicts a $\propto k^{-2}_\perp$ residual-energy spectrum for both balanced and moderately imbalanced strong turbulence \citep[e.g.,][]{mg05,boldyrev11,bpw12}---the constant component can (at least partially) be attributed to non-MHD corrections to the Alfv\'{e}n speed due to pressure anisotropies and parallel drifts. Indeed, this interpretation is supported by Equation (\ref{eqn:alfvenratio}), and one may thus construct a more appropriate Alfv\'{e}n ratio by weighting the flux function by $\valfeff/\valf$ \citep{bd71,mg82}. This was the route followed by \citet{chen13}, who found that such a ``kinetic normalisation'', equivalent to using
\begin{equation}
r_{{\rm A}\ast} \doteq \left| \frac{\valf}{\valfeff} \frac{\Phi}{\Psi} \right|^2
\end{equation}
instead of $r_{\rm A}$, yields fluctuations that are closer to equipartition, with a mean residual energy of $\sigma_{{\rm r}\ast} \doteq ( r_{{\rm A}\ast} - 1 ) / ( r_{{\rm A}\ast} + 1 ) = -0.19$ and a mean Alfv\'{e}n ratio of $r_{{\rm A}\ast} = 0.71$ (rather than $\sigma_{\rm r} = -0.43$ and $r_{\rm A} = 0.40$ using Eq.~\ref{eqn:alfvenratio}). Using the appropriate kinetic normalisation is thus essential when measuring quantities like the residual energy in plasmas with anisotropic distribution functions, such as the solar wind.

 \subsection{Alfv\'{e}n-Wave Invariants}\label{sec:WAW}
 
Another standard result---that interactions between ``$+$'' and ``$-$'' waves occur without exchanging energy---can also be shown to hold true in general. Multiplying Equation (\ref{eqn:elsasser}) by $\rho_0 \zeta^\pm$ and integrating the result over space, we find
\begin{equation}
\D{t}{W^\pm_{\rm AW}} = 0,
\end{equation}
where
\begin{equation}\label{eqn:WAWpm}
W^\pm_{\rm AW} \doteq \frac{1}{2} \int{\rm d}^3\bb{r} \, \rho_0 \bigl| \grad_{\!\perp} \zeta^\pm \bigr|^2
\end{equation}
are the independently conserved (free) energies of the forward- and backward-propagating Alfv\'{e}nic fluctuations, respectively. Their sum,
\begin{align}\label{eqn:WAW}
W_{\rm AW} &\doteq W^+_{\rm AW} + W^-_{\rm AW} \nonumber\\*
\mbox{} &= \frac{1}{2} \int{\rm d}^3\bb{r} \, \rho_0 \left( \bigl| \grad_{\!\perp} \zeta^+ \bigr|^2 + \bigl| \grad_{\!\perp} \zeta^- \bigr|^2 \right) \nonumber\\*
\mbox{} &= \frac{1}{2} \int{\rm d}^3\bb{r} \, \rho_0 \left( \bigl| \grad_{\!\perp} \Phi \bigr|^2 + \frac{\valfeff^2}{\vasq} \bigl| \grad_{\!\perp} \Psi \bigr|^2 \right) \nonumber\\*
\mbox{} &= \int{\rm d}^3\bb{r} \left\{ \frac{\rho_0 u^2_\perp}{2} + \left[ \, 1 + \sum_s \frac{\betaprl{s}}{2} \left( \Delta_s - \frac{2\duparsq}{\vthprl{s}^2}\right) \right] \frac{\delta B^2_\perp}{8\pi} \right\},
\end{align}
is, of course, also conserved. If $\valfeff^2$ (equivalently, the expression in the square brackets in Equation \ref{eqn:WAW}) is positive---i.e., if the plasma is firehose-stable---$W_{\rm AW}$ is a positive-definite quantity measuring the total kinetic and potential energy stored in the Alfv\'{e}nic fluctuations. To interpret Equation (\ref{eqn:WAW}) when $\valfeff^2$ is driven negative by the pressure anisotropy and parallel drifts, we separate the various terms in the conservation law for $W_{\rm AW}$ as follows:
\begin{equation}\label{eqn:FHenergy}
\D{t}{} \int{\rm d}^3\bb{r} \left( \frac{\rho_0 u^2_\perp}{2} + \frac{\delta B^2_\perp}{8\pi} \right) = \sum_s  \frac{\nsp \tprl{s}}{2} \int{\rm d}^3\bb{r} \, \left| \Delta_s - \frac{2\duparsq}{\vthprl{s}^2} \right| \pD{t}{} \frac{\delta B^2_\perp}{B^2_0} .
\end{equation}
We then see that the terms on the right-hand side of this equation constitute a source for the kinetic and magnetic fluctuations on the left-hand side. Indeed, recent work on the nonlinear evolution of the firehose instability \citep[e.g.,][]{schekochihin08a,rosin11,kss14} has shown that the rate of relaxation of the pressure anisotropy $\Delta_s$ is related to $\partial \delta B^2_\perp / \partial t$ and so one can, heuristically, interpret the right-hand side of Equation (\ref{eqn:FHenergy}) as a velocity-space source of free energy multiplied by the rate at which fluctuations act to remove that source of free energy. In this case, the Alfv\'{e}nic invariant (\ref{eqn:WAW}) is minimised by growing fluctuations.

Another way to interpret Equation (\ref{eqn:WAW}) for a non-Maxwellian plasma is as follows. How close the equilibrium pressure-anisotropic distribution is to the firehose threshold has the effect of weighting the (free) energy associated with perpendicular magnetic perturbations. As the threshold is approached, bending field lines becomes energetically less demanding (dynamically, the negative-pressure-anisotropy stress cancels the tension force; see Eq.~\ref{eqn:comforce}). As the threshold is crossed, $W_{\rm AW}$ is no longer positive-definite and so can be conserved even if perturbations grow---which is indeed what happens.

In Section \ref{sum:W}, we show that $W_{\rm AW}$ is part of a generalised free-energy invariant conserved and cascaded to small scales in phase space by the plasma turbulence.

%
%
\section{Compressive Fluctuations in the Inertial Range}\label{sec:compressivecascade}

We now turn our focus to the behaviour of compressive fluctuations in the inertial range (dynamical equations for these fluctuations were derived in Section \ref{sec:compressive}). In this Section, we show that the compressive fluctuations possess their own invariant, which has a natural interpretation when the equilibrium distribution function is cast in terms of the particle kinetic energy and adiabatic invariant. For a bi-Maxwellian plasma with a single ionic species, the inertial-range cascade of compressive fluctuations can be further split into three independent kinetic cascades. We derive the linear theory of compressive fluctuations in a bi-Maxwellian plasma (the general linear theory is given in Appendix \ref{app:linear}) and use it to demonstrate how pressure anisotropy affects the efficacy of linear parallel phase mixing and the partitioning of free energy amongst the various cascade channels.

\subsection{Compressive Invariant}\label{sec:Wcompr}

In Section \ref{sec:reduceddke}, we derived the evolution and constraint equations for the perturbed ion distribution function $g_i$, whose moments describe compressive fluctuations in a pressure-anisotropic plasma. The nonlinear evolution of these fluctuations is due solely to passive advection of $g_i$ by the Alfv\'{e}nic turbulence, which mixes $\dne$ and $\dBprl$ in the direction transverse to the local magnetic field. (Mathematically, this is a statement that, in the Lagrangian frame associated with the Alfv\'{e}nic component of the turbulence, Equation (\ref{sum:ionkinetic}) is linear.) During this mixing, the compressive fluctuations satisfy an important conservation law, which we now derive.

Multiplying Equation (\ref{eqn:ionkinetic}) by $(\tauprl{i} / Z_i) (c_i/\nip) (g_i / f^\parallel_{0i})$ and integrating over the phase space, we find that
\begin{align}\label{eqn:entropy}
\D{t}{} &\int {\rm d}^3\bb{r} \, \frac{\tauprl{i}}{Z_i} \frac{c_i}{\nip} \int{\rm d}^3\bb{v} \, \frac{g^2_i}{2f^\parallel_{0i}} 
\\*\mbox{}
&+ \int{\rm d}^3\bb{r} \,\frac{c_i}{\nip} \int{\rm d}^3\bb{v} \, \bigl( v_\parallel - \dupari \bigr) g_i \, \eb \bcdot \grad  \left[ \frac{1}{\czero{e}}  \left( \frac{\dne}{\nem} + \Delta_{1e} \frac{\dBprl}{B_0} \right) +  \frac{\tauprl{i}}{Z_i} \frac{w^2_\perp}{\vthprl{i}^2} \!\frac{\dBprl}{B_0} \right] = 0 .\nonumber
\end{align}
On the other hand, multiplying Equation (\ref{eqn:ionkinetic}) by the term in the square brackets in Equation (\ref{eqn:entropy}), integrating the result over phase space, and performing integration by parts gives
\begin{align}
\int{\rm d}^3\bb{r} \int {\rm d}^3\bb{v} \, \D{t}{g_i} & \left[ \frac{1}{\czero{e}}  \left( \frac{\dne}{\nem} + \Delta_{1e} \frac{\dBprl}{B_0} \right) +  \frac{\tauprl{i}}{Z_i} \frac{w^2_\perp}{\vthprl{i}^2} \!\frac{\dBprl}{B_0} \right] 
\\*\mbox{}
&= \int{\rm d}^3\bb{r} \int {\rm d}^3\bb{v} \, v_\parallel g_i \, \eb \bcdot \grad  \left[ \frac{1}{\czero{e}}  \left( \frac{\dne}{\nem} + \Delta_{1e} \frac{\dBprl}{B_0} \right) +  \frac{\tauprl{i}}{Z_i} \frac{w^2_\perp}{\vthprl{i}^2} \!\frac{\dBprl}{B_0} \right] . \nonumber
\end{align}
Using this expression in Equation (\ref{eqn:entropy}), summing over ion species, and using Equations (\ref{eqn:neutrality2}) and (\ref{eqn:pbalance2}) to eliminate the resulting velocity-space integrals of $g_i$ produces the following conservation law:
\begin{align}\label{eqn:Wcompr-conservation}
\D{t}{W_{\rm compr}} &= \int{\rm d}^3\bb{r} \, \sum_i \dupari \left[ \tprl{i} \dni \frac{Z_i}{\tauprl{i}}\eb \bcdot \grad   \frac{1}{\czero{e}} \left(\frac{\dne}{\nem} + \Delta_{1e} \frac{\dBprl}{B_0} \right) + \dpprp{i} \eb\bcdot\grad\frac{\dBprl}{B_0} \right] \nonumber\\*
\mbox{} &= - \int{\rm d}^3\bb{r} \, \dupari \left( Z_i e \dni E_\parallel - \dpprp{i} \eb\bcdot\grad\frac{\dBprl}{B_0} \right) ,
\end{align}
where
\begin{align}\label{eqn:Wcompr}
W_{\rm compr} \doteq \frac{\nem \tprl{e}}{2} &\int {\rm d}^3\bb{r} \left\{ \sum_i \frac{\tauprl{i}}{Z_i} \frac{c_i}{\nip} \int {\rm d}^3\bb{v} \,\frac{g^2_i}{f^\parallel_{0i}} + \frac{1}{\czero{e}} \left( \frac{\dne}{\nem} - \frac{\dBprl}{B_0} \right)^2   \right.
\\*\mbox{}
&- \left. \frac{\pprp{e}}{\pprl{e}} \left[ \frac{\cone{e}}{\czero{e}} - 2 \Delta_{2e} + \frac{\cone{e}}{\czero{e}} \Delta_{1e} + 2 \left( \sum_i c_i  \frac{\tauprp{i}}{Z_i} + \frac{1}{\betaprp{e}} \right) \right] \frac{\dBprl^2}{B^2_0} \right\} .\nonumber
\end{align}
In the absence of interspecies drifts, $W_{\rm compr}$ is an invariant conserved by Equations (\ref{eqn:ionkinetic})--(\ref{eqn:pbalance2}). The simpler version of $W_{\rm compr}$ that is conserved for the pressure-isotropic case (eq.~201 of S09) is related to the perturbed entropy $\delta S_s$ of the system (cf.~\citealt{kh94,sugama96,howes06,schekochihin08b}). With interspecies parallel drifts, the right-hand side of Equation (\ref{eqn:Wcompr-conservation}) constitutes a source or sink for this quantity. It is the work done by the fluctuating parallel electric field (Equation \ref{eqn:efield}) and by magnetic-mirror forces acting on the interspecies drifts, and represents the exchange of free energy between these drifts and the compressive fluctuations.\footnote{See Appendix \ref{app:landau} for a specific example of this physics, where we explicitly demonstrate that free energy can flow into or out of the interspecies drifts depending upon whether the system is unstable to an ion-acoustic (streaming) instability.} 

Adding and subtracting the phase-space integral of $\tprl{e} g^2_e / 2 f^\parallel_{0e}$ to the right-hand side of Equation (\ref{eqn:Wcompr}), with $g_e$ given by Equation (\ref{eqn:ge}), the compressive invariant may be re-written in the following compact form:
\begin{equation}\label{eqn:Wcompr2}
W_{\rm compr} = \int {\rm d}^3\bb{r} \left[ \sum_s \int{\rm d}^3\bb{v} \, \frac{\tprl{s} g^2_s}{2f^\parallel_{0s}} - ( 1 + \beta_\perp ) \frac{\dBprl^2}{8\pi} \right] .
\end{equation}
We now make a transformation analogous to that made in equation (149) of S09. Defining
\begin{equation}\label{eqn:dftilde}
\delta\widetilde{f}_s \doteq g_s + \frac{w^2_\perp}{\vthprl{s}^2} \frac{\dBprl}{B_0} f^\parallel_{0s} = \df{s} + \frac{w^2_\perp}{\vthprp{s}^2} \frac{\dBprl}{B_0} \mf{D} f_{0s}
\end{equation}
and using Equations (\ref{eqn:ge}) and (\ref{eqn:pbalance2}) to eliminate the resulting velocity-space integrals over $g_e$ and $g_i$, respectively, we can rewrite Equation (\ref{eqn:Wcompr2}) in a particularly useful form:
\begin{equation}\label{eqn:Wcompr3}
W_{\rm compr} = \int{\rm d}^3\bb{r} \left[ \sum_s \int{\rm d}^3\bb{v} \, \frac{\tprl{s} \delta \widetilde{f}^{\,2}_s}{2f^\parallel_{0s}} + \Bigl( 1 - \sum_s \betaprp{s} \Delta_{2s} \Bigr) \frac{\delta B^2_\parallel}{8\pi} \right] ,
\end{equation}
where
\begin{equation}\label{eqn:delta2s}
\Delta_{2s} = \left( \frac{1}{\nsp} \int{\rm d}^3\bb{v} \, \frac{1}{2} \frac{w^4_\perp}{\vthprp{s}^4} f^\parallel_{0s} \right) \frac{\pprp{s}}{\pprl{s}} -  1
\end{equation}
is the extension of $\Delta_{2e}$ to arbitrary species $s$ (cf.~Eq.~\ref{eqn:epaniso}). This form of $W_{\rm compr}$ parallels the final expression in Equation (\ref{eqn:WAW}) for the Alfv\'{e}nic invariant: we have a quantity that has one interpretation if $1 - \sum_s \betaprp{s} \Delta_{2s} > 0$, namely that it is the generalised energy of the compressive fluctuations, and another if $1 - \sum_s \betaprp{s} \Delta_{2s} < 0$, in which case the final term in Equation (\ref{eqn:Wcompr3}) becomes a free-energy source for the mirror instability, for which the expression multiplying $\dBprl^2/8\pi$ in Equation (\ref{eqn:Wcompr3}) is related to the stability parameter (see Section \ref{sec:highbetalimit} and Appendix \ref{app:mirror}); neglecting interspecies drifts,
\begin{equation}
\D{t}{} \int {\rm d}^3\bb{r} \left( \sum_s \int {\rm d}^3\bb{v} \, \frac{\tprl{s} \delta\widetilde{f}^{\,2}_s}{2f^\parallel_{0s}} + \frac{\delta B^2_\parallel}{8\pi} \right) = \sum_s \nsp \tprp{s} \int{\rm d}^3\bb{r} \, \Delta_{2s} \pD{t}{} \frac{\delta B^2_\parallel}{B^2_0} .
\end{equation}
In the unstable case, one can interpret $\partial \delta B^2_\parallel / \partial t$ as the rate of relaxation of the pressure anisotropy as the mirror fluctuations grow \citep[see, e.g.,][]{schekochihin08a,kss14,rqv14,rsc14}. In this case, $W_{\rm compr}$ is minimised by growing fluctuations.

Another interpretation of what happens when the stability threshold is approached and crossed is analogous to the one we offered at the end of Section \ref{sec:WAW} for a similar situation concerning the Alfv\'{e}nic fluctuations. Within the mirror stability boundary, $W_{\rm compr}$ is a positive-definite conserved free-energy-like quantity. As the system gets closer to the mirror threshold, it becomes energetically ``cheaper'' to produce magnetic compressions or rarefactions ($\dBprl$)---dynamically, this is due to the fact that the effect of positive $p_{\perp 0} - p_{\parallel 0}$ is to reduce the magnetic pressure response \citep[cf.][]{sk93}. Once the threshold is crossed, $W_{\rm compr}$ is no longer positive definite and its conservation is compatible with the growth of $\dBprl$ and $\delta\widetilde{f}_s$ (the mirror instability).

The astute reader will recognise that the factor multiplying $\dBprl^2$ in the free-energy invariant, namely $1 - \sum_s \betaprp{s} \Delta_{2s}$, is not the {\em exact} mirror stability parameter, Equation (\ref{eqn:mirrorstability}). While the latter reduces to the former in the case of very high $\beta$ or of cold electrons (for which the right-hand side of Eq.~\ref{eqn:mirrorstability} vanishes), in general there is a stabilising term due to the interaction of linearly resonant particles with the parallel electric field (i.e.~Landau damping). This physics is contained inside the first term in the compressive invariant, proportional to $\delta\widetilde{f}^2_s$. In order to see this, and to make better sense of the structure of $W_{\rm compr}$, we must understand the physical meaning of $\delta\widetilde{f}_s$.

\subsection{Meaning of $\delta\widetilde{f}_s$: $(v_\parallel,w_\perp)$ vs.~$(\varepsilon_s,\mu_s)$ Coordinates}\label{sec:eandmu}

Our decision to write $f_{0s}$ as a function of $v_\parallel$ and $w_\perp$, while analytically convenient, is not the most natural choice for interpreting the compressive invariant $W_{\rm compr}$. Instead, let us introduce the kinetic energy and adiabatic invariant of a particle of species $s$, given respectively by
\begin{align}\label{eqn:energy}
\varepsilon_s &\doteq  \frac{1}{2} m_s \bigl( v_\parallel - \dupar \bigr)^2 + \frac{1}{2} m_s w^2_\perp, \\*
\label{eqn:mu}
\mu_s &\doteq \frac{m_s w^2_\perp}{2B} ,
\end{align}
and rewrite the drift-kinetic equation (\ref{eqn:kinetic}) using $\varepsilon_s$ and $\mu_s$ as our velocity-space coordinates \citep[e.g.,][]{hazeltine73}:
\begin{equation}\label{eqn:kineticEMU}
\bigD{t}{f_s} + \left[ m_s \bigl( v_\parallel - \dupar \bigr) \left( \frac{q_s}{m_s} E_\parallel - \bigD{t}{\bb{u}_\perp}\bcdot \eb \right) + \mu_s \left( \D{t}{B} + \dupar \pD{z}{B} \right) \right] \pD{\varepsilon_s}{f_s} = \left( \pD{t}{f_s} \right)_{\rm c} ,
\end{equation}
where $v_\parallel - \dupar = \pm \sqrt{2 m_s ( \varepsilon_s - \mu_s B )}$. This is perhaps the most transparent form of KMHD: particles parallel-stream, $\bb{E}\btimes\bb{B}$-drift, conserve $\mu_s$, and change their kinetic energy by interacting with a parallel electric field $E_\parallel$ and/or a changing magnetic field (mirroring) in the frame of the equilibrium species drift; the term $-({\rm D}\bb{u}_\perp/{\rm D}t) \bcdot \eb = \bb{u}_\perp \bcdot ( {\rm D}\eb / {\rm D}t)$ is  an inertial term having to do with the fact that the direction of the magnetic-field line changes as the particle streams along it and so the plane of the $\bb{E}\btimes\bb{B}$ drift tilts. 

When compared to the formulation of KMHD in $(v_\parallel, w_\perp)$ coordinates (Eq.~\ref{eqn:kinetic}), this formulation makes clear that there is a more general class of equilibrium solutions in a collisionless drift-kinetic plasma than those satisfying $f_{s} = f_{0s}(v_\parallel , w_\perp)$, namely, $f_{s} = \widetilde{f}_{0s} ( \varepsilon_s, \mu_s)$. Thus, what is referred to as the ``equilibrium state'' may, in fact, contain inhomogeneous fluctuations (e.g., Alfv\'{e}n waves), so long as the energy of each particle is conserved. As a result, there are pieces of $\delta f_s$ in the $(v_\parallel, w_\perp)$ formulation that may be absorbed into the equilibrium distribution function when $f_{0s}$ is taken to be a function $\varepsilon_s$ and $\mu_s$ (e.g., fluctuations in magnetic-field strength that do not violate $\mu_s$-conservation). To see that this is the case, we relate the two formulations via
\begin{align}\label{eqn:femu}
f_{0s}\left( \bigl(v_\parallel-\dupar\bigr)^2,w^2_\perp \right) &= f_{0s} \left( \frac{2\varepsilon_s}{m_s} - \frac{2\mu_s B}{m_s} , \frac{2\mu_s B}{m_s} \right) \nonumber\\*
\mbox{} &\simeq f_{0s} \left( \frac{2\varepsilon_s}{m_s} - \frac{2\mu_s B_0}{m_s} , \frac{2\mu_s B_0}{m_s} \right) \nonumber\\*
\mbox{} &\qquad\quad+ \frac{2\mu_s B_0}{m_s}\frac{\dBprl}{B_0} \left[ - \pD{(v_\parallel-\dupar)^2}{f_{0s}} + \pD{w^2_\perp}{f_{0s}} \right] \nonumber\\*
\mbox{} &= \widetilde{f}_{0s} ( \varepsilon_s , \mu_s ) + \frac{w^2_\perp}{\vthprp{s}^2} \frac{\dBprl}{B_0} \mf{D}f_{0s} .
\end{align}
Comparing Equations (\ref{eqn:dftilde}) and (\ref{eqn:femu}), it then becomes clear that the perturbed distribution function appearing in the compressive invariant $W_{\rm compr}$ (see Eq.~\ref{eqn:Wcompr3}) satisfies
\begin{equation}\label{eqn:dftildedef}
\delta\widetilde{f}_s \doteq \delta f_s + \frac{w^2_\perp}{\vthprp{s}^2} \frac{\dBprl}{B_0} \mf{D}f_{0s} = \delta f_s + f_{0s} - \widetilde{f}_{0s} = f_s - \widetilde{f}_{0s},
\end{equation}
i.e., it is the perturbed distribution function if $f_{0s}$ is taken to be a function of $(\varepsilon_s,\mu_s)$ instead of $(v_\parallel,w_\perp)$. 

The meaning of the first entropy-like term in the compressive invariant $W_{\rm compr}$ (see Eq.~\ref{eqn:Wcompr3}) is then readily apparent: it is the non-Alfv\'{e}nic piece of the distribution function that represents changes in the kinetic energy of the particles due to interactions with the compressive fluctuations. In it are contributions from Landau-resonant particles, whose energy is changed by the parallel electric and magnetic-mirror forces in such a way as to facilitate Landau (Appendix \ref{app:landau}) and Barnes (Appendix \ref{app:mirror}) damping of ion-acoustic waves and slow modes. As long as the plasma stays within the mirror and streaming stability boundaries, the compressive invariant is positive-definite. 

In the next four sections (\S\S\ref{sec:parallelkinetics}--\ref{sec:collisionlesscascades}), we show that the conservation of $W_{\rm compr}$ represents a turbulent cascade of compressive fluctuations from large to small scales in phase space.

\subsection{Parallel Kinetics: Two Decoupled Collisionless Cascades}\label{sec:parallelkinetics}

We begin by noting that, under our collisionless ordering, the $w_\perp$ dependence in Equation (\ref{sum:ionkinetic}) can be integrated out. We introduce two auxiliary functions, $G_n(v_\parallel)$ and $G_B(v_\parallel)$, defined so that
\begin{equation}\label{eqn:GnBdef}
\frac{\dne}{\nem} = \int {\rm d} v_\parallel \, G_n \qquad {\rm and} \qquad \frac{\dBprl}{B_0} = \int {\rm d} v_\parallel \, G_B ,
\end{equation}
i.e., they contain all the $w_\perp$ integrals and species summations in the right linear combination as per Equations (\ref{sum:dne}) and (\ref{sum:dBprl}). Then Equation (\ref{sum:ionkinetic}) reduces to the following two coupled one-dimensional kinetic equations (cf.~eqs 179 and 180 of S09):
\begin{subequations}\label{eqn:GnB}
\begin{align}\label{eqn:Gn}
\D{t}{G_n} + v_\parallel \eb \bcdot \grad G_n &=  \sum_i \bigl( v_\parallel - \dupari \bigr) \fzeroi (v_\parallel) \,  \eb\bcdot \grad   \left[ \lambda^{nn}_{i}(v_\parallel) \frac{\dne}{\nem} + \lambda^{nB}_{i}(v_\parallel) \frac{\dBprl}{B_0} \right] ,
\\*\label{eqn:GB}
\D{t}{G_B} + v_\parallel \eb \bcdot \grad G_B &= \sum_i \bigl( v_\parallel - \dupari \bigr) \fzeroi (v_\parallel)  \, \eb\bcdot \grad  \left[ \lambda^{Bn}_{i}(v_\parallel) \frac{\dne}{\nem} + \lambda^{BB}_{i}(v_\parallel) \frac{\dBprl}{B_0} \right] ,
\end{align}
\end{subequations}
where the $v_\parallel$- and ion-species-dependent $\lambda_i$ coefficients, given in Appendix \ref{app:coefficients}, depend upon various perpendicular moments of the parallel-differentiated equilibrium ion distribution function:
\begin{equation}\label{eqn:Felli}
\felli (v_\parallel) \doteq \frac{2\pi}{\nip} \int_0^\infty {\rm d} w_\perp  w_\perp \, \frac{1}{\ell !} \left( \frac{w_\perp}{\vthprp{i}} \right)^{2\ell} f^\parallel_{0i} ( v_\parallel, w_\perp )
\end{equation}
for integer $\ell$. This coupled system of integral equations, compactly expressed as
\begin{align}\label{eqn:GnB2}
\left( \D{t}{} + v_\parallel \eb\bcdot \grad \right) \left[\!\!
\begin{array}{c}
G_n(v_\parallel) \\
G_B(v_\parallel)
\end{array}
\!\!\right] &= \sum_i \bigl( v_\parallel - \dupari \bigr) \fzeroi (v_\parallel) \\*
\mbox{} & \times \eb\bcdot\grad \int_{-\infty}^\infty {\rm d}v'_\parallel \, \Biggl[\!\!
\begin{array}{cc}
\lambda^{nn}_{i}(v_\parallel) & \lambda^{nB}_{i}(v_\parallel) \\
\lambda^{Bn}_{i}(v_\parallel) & \lambda^{BB}_{i}(v_\parallel)
\end{array}
\!\!\Biggr] \!\! \left[\!\!
\begin{array}{c}
G_n(v'_\parallel) \\
G_B(v'_\parallel)
\end{array}
\!\!\right] , \nonumber
\end{align}
has a simple physical interpretation. Kinetic fluctuations in each of the $i' = 1\dots N_{\rm ion}$ ionic species collectively excite compressive fluctuations in the magnetic-field strength (via perpendicular pressure balance), the electron density (via quasineutrality), and thereby the electric field (via parallel pressure balance in the massless electron fluid; see Eq.~\ref{eqn:efield}). These electromagnetic fields in turn feed back upon the kinetic fluctuations exhibited by each individual ionic species, in a way that is dictated by the various $\lambda_i$ coefficients. This system of equations describing the evolution of the compressive fluctuations and their interactions with the particles is closed, signaling the fact that the compressive dynamics proceeds independently from that of the Alfv\'{e}nic fluctuations.

Because each ion species responds to the compressive fluctuations in a different way ($\lambda_i \ne \lambda_{i'}$ for $i \ne i'$), Equation (\ref{eqn:GnB2}) cannot be diagonalised in general. Moreover, the mode structure excited in the parallel-velocity space by these fluctuations will generally be different than the mode structure present in the fluctuations themselves [in general, $\lambda(v_\parallel) \ne \lambda(v'_\parallel)$ for $v_\parallel \ne v'_\parallel$], and so this system can only be diagonalized if the $\lambda_i$ coefficients are independent of $v_\parallel$. Both of these requirements for diagonalizing Equation (\ref{eqn:GnB2}) are guaranteed only for a plasma consisting of a single species of bi-Maxwellian ions, for which $F^{\parallel}_{\ell i}(v_\parallel) = F_{\rm M}(v_\parallel) \doteq ( 1 / \sqrt{\pi} \vthprl{i} ) \exp( - v^2_\parallel / \vthprl{i}^2 )$ for all integer $\ell$. Then, assuming a bi-Maxwellian plasma and proceeding with the diagonalisation, we find that Equation (\ref{eqn:GnB2}) is equivalent to
\begin{equation}\label{eqn:cascades}
\D{t}{G^\pm} + v_\parallel \eb\bcdot \grad G^\pm = \frac{v_\parallel F_{\textrm{M}}(v_\parallel)}{\Lambda^\pm} \, \eb\bcdot\grad \int^\infty_{-\infty} {\rm d}v'_\parallel \, G^\pm(v'_\parallel) ,
\end{equation}
where
\begin{equation}\label{eqn:Gpm}
G^+ = G_{B} + \frac{\pprl{i}}{\pprp{i}}  \frac{1}{\sigma_i} \left( 1 + \frac{Z_i}{\tauprp{i}} \right) G_n \qquad {\rm and} \qquad G^- =  G_n +  \frac{\varpi_i}{\sigma_i} \frac{\tauprl{i}}{Z_i}   \frac{2}{\betaprp{i}} \, G_B
\end{equation}
are the eigenvectors; 
\begin{equation}\label{eqn:lambdapm}
\Lambda^\pm = - \frac{\tauprl{i}}{Z_i} + \frac{\pprl{i}}{\pprp{i}}\frac{\varsigma_i}{\betaprp{i}} \pm \sqrt{ \left( \frac{\tauprl{i}}{\tauprp{i}} + \frac{\tauprl{i}}{Z_i} \right)^2 + \left( \frac{\pprl{i}}{\pprp{i}} \frac{\varsigma_i}{\betaprp{i}} \right)^2 } 
\end{equation}
are the (inverses of) the corresponding eigenvalues; and we have defined
\begin{eqnarray}\label{eqn:sigma}
\sigma_i \doteq \frac{\tauprl{i}}{\tauprp{i}} + \frac{\tauprl{i}}{Z_i} +  \frac{\pprl{i}}{\pprp{i}} \frac{1}{\betaprp{i}} + \sqrt{ \left( \frac{\tauprl{i}}{\tauprp{i}} + \frac{\tauprl{i}}{Z_i} \right)^2 + \left( \frac{\pprl{i}}{\pprp{i}} \frac{\varsigma_i}{\betaprp{i}} \right)^2 } , 
\end{eqnarray}
\begin{equation}\label{eqn:varsigma}
\varsigma_i \doteq 1 - \beta_\perp \Delta_e ,
\end{equation}
\begin{equation}\label{eqn:varpi}
\varpi_i \doteq 1 +  \frac{\pprl{e}}{\pprp{i}} \Delta_e \left( 1 - \frac{1}{2} \beta_\perp \Delta_e \right) ,
\end{equation}
the latter two expressions equating to unity for isotropic electrons. 

Equation (\ref{eqn:cascades}), which reduces to equation (181) of S09 for a Maxwellian plasma, describes two decoupled kinetic cascades. These are the subject of Section \ref{sec:collisionlessinvariants}, in which we derive the two collisionless invariants associated with $G^+$ and $G^-$. But first, we specialise $G^\pm$ and $\Lambda^\pm$ for application to two different astrophysical systems representing two distinct parameter regimes and determine what their values imply for the evolution of compressive fluctuations in the linear regime.

\subsection{Linear Theory: Collisionless Damping and Mirror Instability}\label{sec:mirror}

To develop the linear theory for a two-component bi-Maxwellian plasma,\footnote{The linear theory for a plasma with an arbitrary gyrotropic equilibrium distribution function and multiple ionic species is presented in Appendix \ref{app:linear}.} we Fourier transform Equation (\ref{eqn:cascades}) in time ($\partial / \partial t \rightarrow -\imag \omega$) and space ($\eb \bcdot \grad \rightarrow \imag k_\parallel$), divide both sides by $-\imag ( \omega - k_\parallel v_\parallel )$, and integrate over the parallel velocity. Dividing both sides of the resulting equation by $\int {\rm d}v_\parallel G^\pm$ and using $\int{\rm d} v_\parallel \, F_{\rm M}(v_\parallel) = 1$, we obtain the following dispersion relation:
\begin{equation}\label{eqn:disprel}
\Lambda^\pm - 1 = \frac{\omega}{k_\parallel} \int {\rm d}v_\parallel \, \frac{F_{\rm M}(v_\parallel)}{v_\parallel - \omega / k_\parallel} = \xi_i Z_{\rm M}(\xi_i) ,
\end{equation}
where $\xi_i \doteq \omega / k_\parallel \vthprl{i}$ is the dimensionless phase speed and the (Maxwellian) plasma dispersion function \citep{fc61}
\begin{equation}\label{eqn:maxwZ}
Z_{\rm M}(\xi) \doteq \frac{1}{\sqrt{\pi}} \int^\infty_{-\infty} {\rm d}x\, \frac{{\rm e}^{-x^2}}{x-\xi} ,
\end{equation}
the integration being performed along the Landau contour. Formally, Equation (\ref{eqn:disprel}) has an infinite number of solutions, most of which are strongly damped with damping rates ${\rm Im}(\xi_i) \sim 1$. A few of the more interesting solutions may be obtained analytically in the low- and high-beta limits.

\subsubsection{Low-Beta Limit: Solar Corona}\label{sec:lowbetalimit}

When $\beta_s \ll 1 \sim \Delta_s$, Equations (\ref{eqn:Gpm}) and (\ref{eqn:lambdapm}) give\footnote{In some regions of the solar atmosphere, this subsidiary expansion in low $\beta_s$ may conflict with the prior mass-ratio expansion if $\beta_i \sim m_e / m_i$.}
\begin{align}\label{eqn:lowbetalimit}
\Lambda^- - 1 &\simeq - \left( 1 + \frac{\tauprl{i}}{Z_i} \right) , & G^- &\simeq G_n + \frac{\tauprl{i}}{Z_i} \left( 1 + \Delta_i + \Delta_e \, \frac{Z_i}{\tauprl{i}} \right) G_B , \\*
\Lambda^+ -1 &\simeq \frac{\pprl{i}}{\pprp{i}} \frac{2}{\betaprp{i}}, & G^+ &\simeq G_B .
\end{align}
For the ``$-$'' branch, we have ${\rm Im}(\xi_i) \sim 1$, and so
\begin{equation}
\omega \sim -\imag |k_\parallel| \valf \sqrt{\betaprl{i}} ,
\end{equation}
which is much smaller than the Alfv\'{e}nic cascade rate $k_\parallel \valf$. For the ``$+$'' branch, consisting predominantly of fluctuations in magnetic-field strength,
\begin{equation}
\omega \sim -\imag |k_\parallel| \valf \sqrt{ \betaprl{i} \left| \, \ln \frac{\pprl{i}}{\pprp{i}} \frac{2}{\betaprp{i}} \, \right| } ,
\end{equation}
up to logarithmically small corrections. This damping rate is slightly greater than that of the ``$-$'' branch, though still much smaller than the Alfv\'{e}nic cascade rate. Compressive fluctuations in a low-beta plasma are therefore weakly damped. Pressure anisotropies do not affect this conclusion.

\subsubsection{High-Beta Limit: Intracluster Medium}\label{sec:highbetalimit}

When $\beta_s \sim 1/\Delta_s \gg 1$, we have
\begin{align}
\Lambda^- - 1 &\simeq -2 \left( \frac{\tauprl{i}}{\tauprp{i}} + \frac{\tauprl{i}}{Z_i} \right) , & G^- &\simeq G_n , \\*
\Lambda^+ -1 &\simeq \frac{\pprl{i}}{\pprp{i}} \frac{1}{\betaprp{i}} \Bigl( 1 - \sum_s \betaprp{s} \Delta_s \Bigr) , & G^+ &\simeq G_B + \frac{1}{2} \frac{Z_i}{\tauprl{i}} \frac{\pprl{i}}{\pprp{i}} \, G_n .\label{eqn:highbetalimit}
\end{align}
The ``$-$'' branch corresponds to the density fluctuations and is strongly damped with ${\rm Im}(\xi_i) \sim 1$. In contrast, the damping rate of the ``$+$'' branch is small: it can be obtained by expanding $Z_{\rm M}(\xi_i) = \imag \sqrt{\pi} + \mc{O}(\xi_i)$, which gives
\begin{equation}\label{eqn:barnes}
\gamma \doteq -\imag \omega = - \frac{|k_\parallel|\valf}{\sqrt{\pi\betaprl{i}}}  \frac{\pprl{i}^2}{\pprp{i}^2} \Bigl( 1 - \sum_s \betaprp{s} \Delta_s \Bigr) .
\end{equation}
This expression generalises (in the limit $k_\parallel / k_\perp \ll 1$) for bi-Maxwellian distribution functions what is known in astrophysics as the \citet{barnes66} damping and in plasma physics as transit-time damping \citep{stix62}. Particles that are almost at rest with respect to the slow wave (i.e.~``Landau resonant'' with $\omega \sim k_\parallel v_\parallel$) are subject to the action of the mirror force associated with the magnetic compressions in the wave. Since, for a monotonically decreasing distribution function ($f^\parallel_{0i} > 0$), there are more particles with $v_\parallel < \omega / k_\parallel$ than with $v_\parallel > \omega / k_\parallel$, the energy exchange between resonant particles and the wave leads to a net gain (loss) of energy by the particles (wave). Put differently, the only way to maintain perpendicular pressure balance for a slow wave is to increase the energy of the resonant particles (referred to as ``betatron acceleration'', due to the last term in square brackets in Eq.~\ref{eqn:kineticEMU}) at the expense of the wave energy. The result is wave damping.

Because $G_n$ is strongly damped, the fluctuations that are damped at the rate (\ref{eqn:barnes}) are predominantly of the magnetic-field strength. For large ion beta, the damping rate is a small fraction $\sim$$1/\sqrt{\betaprl{i}}$ of the Alfv\'{e}nic cascade rate, becoming even smaller in a plasma exhibiting positive pressure anisotropy. This reduction is due to the proportional increase (for $\Delta_s > 0$) in the number of large-pitch-angle particles in the magnetic troughs ($\dBprl < 0$). This inflates the field lines (in order to maintain perpendicular pressure balance), thereby (partially) offsetting the damping of the field-strength fluctuations. If the concentration of these particles leads to more perpendicular pressure than can be stably balanced by the magnetic pressure, the troughs must grow deeper to compensate. This process runs away as the resonant particles in the deepening troughs lose energy at a rate $\mu\partial B/\partial t \sim \mu \gamma \dBprl$ (``betatron deceleration''). This is the {\em mirror instability} \citep[e.g.][]{sk93}. It is the inevitable outcome of trying to maintain perpendicular pressure balance in the midst of an effectively negative magnetic pressure (see Appendix \ref{app:linear} and, in particular, Eq.~\ref{eqn:dpprpi}).

\subsection{Collisionless Invariants and Phase Mixing}\label{sec:collisionlessinvariants}

The linear theory elucidated, we now return to the principal result of Section \ref{sec:parallelkinetics}: that, for a plasma with a single species of bi-Maxwellian ions, the compressive fluctuations can be decomposed into two decoupled kinetic cascades. Here we show that these two cascades independently obey their own conservation laws. 

If we multiply Equation (\ref{eqn:cascades}) by $G^\pm(v_\parallel) / F_{\rm M}(v_\parallel)$, integrate over space and parallel velocity, and perform integration by parts on the right-hand side, we find that
\begin{equation}\label{eqn:invariants1}
\D{t}{} \int {\rm d}^3\bb{r} \int {\rm d}v_\parallel \, \frac{ ( G^\pm )^2 }{2 F_{\rm M}(v_\parallel)} = - \frac{1}{\Lambda^\pm} \int {\rm d}^3\bb{r} \left( \int {\rm d}v_\parallel \, G^\pm \right) \eb\bcdot\grad \int {\rm d}v_\parallel \, v_\parallel G^\pm .
\end{equation}
On the other hand, integrating Equation (\ref{eqn:cascades}) over parallel velocity gives
\begin{equation}
\D{t}{} \int {\rm d}v_\parallel \, G^\pm = - \eb\bcdot\grad \int {\rm d}v_\parallel \, v_\parallel G^\pm ,
\end{equation}
because $\int {\rm d}v_\parallel \, v_\parallel F_{\rm M}(v_\parallel) = 0$. Using this to replace the final term on the right-hand side of Equation (\ref{eqn:invariants1}), we find
\begin{equation}
\D{t}{W^\pm_{\rm compr}} = 0 ,
\end{equation}
where the two invariants are
\begin{equation}\label{eqn:invariants2}
W^\pm_{\rm compr} = \frac{\nip \tprl{i}}{2}  \int {\rm d}^3\bb{r} \left[ \int {\rm d}v_\parallel \, \frac{(G^\pm)^2}{F_{\rm M}(v_\parallel)} - \frac{1}{\Lambda^\pm} \left( \int {\rm d}v_\parallel \, G^\pm \right)^2 \right] .
\end{equation}
For these invariants to be conserved, the first term in the square bracket must grow to compensate for the decay of the second due to collisionless damping. To see how this arrangement proceeds, it is useful to split
\begin{equation}
G^\pm = \widetilde{G}^\pm + F_{\rm M}(v_\parallel) \int{\rm d}v_\parallel \, G^\pm ,
\end{equation}
noting that $\int {\rm d}v_\parallel \, \widetilde{G}^\pm = 0$ since $\int {\rm d}v_\parallel \, F_{\rm M}(v_\parallel) = 1$. Then
\begin{equation}\label{eqn:invariants3}
W^\pm_{\rm compr} = \frac{\nip\tprl{i}}{2} \int {\rm d}^3\bb{r} \left[ \int {\rm d}v_\parallel \, \frac{(\widetilde{G}^\pm)^2}{F_{\rm M}(v_\parallel)} + \left( 1 - \frac{1}{\Lambda^\pm} \right) \left( \int {\rm d}v_\parallel \, G^\pm \right)^2 \right] .
\end{equation}
For a Maxwellian equilibrium distribution function, the two invariants $W^\pm_{\rm compr}$ are guaranteed to be positive definite because $\Lambda^+ > 1$ and $\Lambda^- < 0$ (see Eq.~\ref{eqn:lambdapm}). Collisionless damping leads to exponential decay of the density and magnetic-field-strength fluctuations, or, equivalently, of $\int {\rm d}v_\parallel\, G^\pm$, while conserving $W^\pm_{\rm compr}$. This means that damping is a redistribution of the conserved quantity $W^\pm_{\rm compr}$: the first term grows to compensate for the decay of the second. This is a manifestation of linear parallel phase mixing \citep{landau46,hdp92,kh94,krommes99,ws04}: free energy passes from the low (density) moment of the distribution function to higher moments (contained in $\widetilde{G}$). As time goes on, the latter part of the solution becomes increasingly oscillatory in $v_\parallel$ ($\widetilde{G}^\pm \propto {\rm e}^{-\imag k_\parallel v_\parallel t}$, the so-called ballistic response), representing the development of finer structure in the parallel-velocity space (see \S 6.2.4 of S09 for further discussion). 

What is novel for anisotropic distribution functions is what happens near the mirror threshold $\Lambda^+ = 1$ (cf.~Eq.~\ref{eqn:highbetalimit}). When the perpendicular pressure is greater than the parallel pressure, the Barnes damping rate is reduced (Eq.~\ref{eqn:barnes}), and so the free energy is transferred from the electromagnetic fluctuations to the ballistic kinetic fluctuations at a slower rate. As we explained in our discussion Equation (\ref{eqn:Wcompr3}), it is easier to compress magnetic-field lines ($\dBprl \ne 0$) when $p_{\perp 0} > p_{\parallel 0}$, so the energetic cost of perturbing the magnetic-field strength is relatively small; this is the physical origin of the small factor $(1 - 1/\Lambda^+)$ multiplying the second term in Equation (\ref{eqn:invariants3}). In other words, as the mirror threshold is approached, the generation of fine-scale structure in velocity space by phase mixing is achieved by damping increasingly large magnetic-field-strength fluctuations at an ever decreasing rate. When $\Lambda^+ < 1$ and the plasma is driven mirror-unstable, the second term in Equation (\ref{eqn:invariants3}) becomes increasingly negative as the magnetic-field-strength fluctuations grow. Consequently, the first term in Equation (\ref{eqn:invariants3}) must grow increasingly positive to compensate, which corresponds to the production of fine-scale structure in velocity space. This structure is caused by the fraction of particles that are linearly resonant with (and nonlinearly trapped by) the unstable mirror mode \citep[e.g.][]{sk93}.

\subsection{$W_{\rm compr}$ Revisited: Compressive Phase-Space Cascade}\label{sec:collisionlesscascades}

We now show that, for a bi-Maxwellian plasma with a single ion species, the compressive invariant $W_{\rm compr}$ obtained in Section \ref{sec:Wcompr} incorporates the two invariants derived in Section \ref{sec:collisionlessinvariants}. We begin by expressing the density and magnetic-field-strength fluctuations in terms of $G^\pm$:
\begin{equation}\label{eqn:dnemG}
\frac{\dne}{\nem} = \frac{1}{\kappa_i} \left( \sigma_i \int {\rm d} v_\parallel \, G^-  - \varpi_i \frac{\tauprl{i}}{Z_i} \frac{2}{\betaprp{i}} \int {\rm d} v_\parallel \, G^+ \right) ,
\end{equation}
\begin{equation}\label{eqn:dBprlG}
\frac{\dBprl}{B_0} = \frac{1}{\kappa_i} \left[ \, \sigma_i \int {\rm d} v_\parallel \, G^+  - \frac{\pprl{i}}{\pprp{i}} \left( 1 + \frac{Z_i}{\tauprp{i}} \right) \int {\rm d} v_\parallel \, G^- \, \right] ,
\end{equation}
where $\sigma_i$ and $\varpi_i$ are defined in Equations (\ref{eqn:sigma}) and (\ref{eqn:varpi}) and\footnote{There is a factor of $2$ missing in the definition of $\kappa$ in S09 (their eq.~204). This error affects their equation (213).}
\begin{equation}\label{eqn:kappa}
\kappa_i \doteq 2 \, \sqrt{ \left( \frac{\tauprl{i}}{\tauprp{i}} + \frac{\tauprl{i}}{Z_i} \right)^2 + \left( \frac{\pprl{i}}{\pprp{i}} \frac{\varsigma_i}{\betaprp{i}} \right)^2 } .
\end{equation}
To express $g_i$ in terms of $G^\pm$, we decompose $g_i$ as follows (cf.~\S6.2.5 of S09):
\begin{equation}\label{eqn:ghat}
g_i = \frac{\nip}{\pi\vthprp{i}^2} {\rm e}^{-x } \, \hat{g}( x , v_\parallel ) , \quad \hat{g}(x,v_\parallel) \doteq \sum_{\ell=0}^\infty L_\ell(x) G_\ell(v_\parallel) ,
\end{equation}
where $x = w^2_\perp / \vthprp{i}^2$ and the Laguerre polynomials $L_\ell(x) = ({\rm e}^x / \ell !)({\rm d}^\ell / {\rm d}x^\ell ) \, x^\ell {\rm e}^{-x}$. Because Laguerre polynomials are orthogonal, we have
\begin{equation}\label{eqn:LaguerreW}
\frac{1}{\nip} \int {\rm d}^3\bb{v} \, \frac{g^2_i}{2\fip} = \sum_{\ell=0}^\infty \int {\rm d}v_\parallel \, \frac{G^2_\ell}{2F_{\rm M}(v_\parallel)} ,
\end{equation}
where the expansion coefficients are determined via the Laguerre transform:
\begin{equation}
G_\ell(v_\parallel) = \int_0^\infty {\rm d}x \, {\rm e}^{-x} \, L_\ell(x) \hat{g}(x,v_\parallel) .
\end{equation}
Since $L_0 = 1$ and $L_1 = 1 - x$, Equations (\ref{eqn:neutrality2}) and (\ref{eqn:pbalance2}) can be used to write the $\ell=0$ and $\ell=1$ expansion coefficients as linear combinations of $G_n$ and $G_B$. Using Equations (\ref{eqn:Gpm}) and (\ref{eqn:lambdapm}) to replace $G_n$ and $G_B$ by suitable combinations of $\Lambda^+ G^+$ and $\Lambda^- G^-$, we find after some straightforward but tedious algebra that
\begin{align}\label{eqn:G0}
G_0 = -\frac{1}{\kappa} &\left\{ \left[ \,\sigma \, \frac{\pprp{i}}{\pprl{i}} \left( 1 + \frac{\pprl{e}}{\pprp{i}} \Delta_e \right) - \varpi_i \frac{2}{\betaprp{i}} \right] \Lambda^+ G^+ \right. 
\nonumber\\*
&\left. \mbox{} + \frac{Z_i}{\tauprl{i}} \left[ \, \sigma - \left( \frac{\tauprl{i}}{\tauprp{i}} + \frac{\tauprl{i}}{Z_i} \right) \left( 1 +  \frac{\pprl{e}}{\pprp{i}} \Delta_e \right) \right] \Lambda^- G^-  \right\} ,
\end{align}
\begin{equation}\label{eqn:G1}
G_1 = \frac{1}{\kappa} \left[ \, \sigma \, \frac{\pprp{i}}{\pprl{i}} \, \Lambda^+ G^+ - \left( 1 + \frac{Z_i}{\tauprp{i}} \right) \Lambda^- G^- \, \right] .
\end{equation}
Substituting Equation (\ref{eqn:ghat}) into the ion kinetic equation (\ref{sum:ionkinetic}), we see that all higher-order expansion coefficients satisfy a simple homogeneous equation:
\begin{equation}
\D{t}{G_\ell} + v_\parallel \eb \bcdot \grad G_\ell = 0, \quad \ell > 1 .
\end{equation}
Thus, the distribution function can be explicitly written in terms of $G^\pm$:
\begin{equation}\label{eqn:giseries}
g_i = \left[ G_0(v_\parallel) + \left( 1 - \frac{w^2_\perp}{\vthprp{i}^2} \right) G_1(v_\parallel) \right] \frac{\nip}{\pi\vthprp{i}^2} {\rm e}^{-w^2_\perp / \vthprp{i}^2} + \widetilde{g}_i ,
\end{equation}
where $\widetilde{g}_i$ comprises all $G_\ell$ with $\ell > 1$. In other words, $\widetilde{g}_i$ is a passively mixed, undamped, ballistic-type mode that contributes to neither density nor magnetic-field strength:
\begin{equation}\label{eqn:gtilde}
\D{t}{\widetilde{g}_i} + v_\parallel \eb \bcdot \grad \widetilde{g}_i = 0 , \quad \int {\rm d}^3\bb{v} \, \widetilde{g}_i = 0, \quad \int{\rm d}^3\bb{v} \, \frac{w^2_\perp}{\vthprp{i}^2} \widetilde{g}_i = 0 ;
\end{equation}
it is the homogeneous solution of Equation (\ref{sum:ionkinetic}).

Equipped with expressions for $G_0$ (Eq.~\ref{eqn:G0}), $G_1$ (Eq.~\ref{eqn:G1}), $\dne/\nem$ (Eq.~\ref{eqn:dnemG}), and $\dBprl/B_0$ (Eq.~\ref{eqn:dBprlG}) written in terms of $G^\pm$, we now substitute these into our expression for the compressive invariant (Eq.~\ref{eqn:Wcompr2} with Eq.~\ref{eqn:LaguerreW}) to find that
\begin{align}\label{eqn:Wcomprdecomposed}
W_{\rm compr} &= \int {\rm d}^3\bb{r} \int {\rm d}^3\bb{v} \, \frac{\tprl{i} \widetilde{g}^{\,2}_i}{2\fip} 
\nonumber\\*\mbox{}
&+ \frac{\pprp{i}^2}{\pprl{i}^2} \left\{ 1 + \frac{2}{\kappa_i} \left( \frac{\tauprl{i}}{\tauprp{i}} + \frac{\tauprl{i}}{Z_i} \right) \left( 1 + \frac{\pprl{e}}{\pprp{i}} \Delta_e \right)  \right.
\nonumber\\*\mbox{}
&\qquad\qquad + \left.  \frac{1}{2} \left[ \left( 1 + \frac{\pprl{e}}{\pprp{i}} \Delta_e \right)^2 -1 \right]  \left( 1 - \frac{2}{\kappa_i} \frac{\pprl{i}}{\pprp{i}} \frac{\varsigma_i}{\betaprp{i}} \right) \right\} ( \Lambda^+ )^2 \, W^+_{\rm compr}
\nonumber\\*\mbox{}
&+ \frac{1}{2} \frac{Z_i^2}{\tauprl{i}^2} \left( 1 + \frac{2}{\kappa_i} \frac{\pprl{i}}{\pprp{i}}\frac{\varsigma_i}{\betaprp{i}} \right) ( \Lambda^- )^2 \, W^-_{\rm compr} .
\end{align}
Thus, for a bi-Maxwellian plasma, the generalised invariant for compressive fluctuations splits into three independently cascading parts: $W^\pm_{\rm compr}$ associated with the density and magnetic-field-strength fluctuations and a purely kinetic part given by the first term in Equation (\ref{eqn:Wcomprdecomposed}):
\begin{equation}\label{eqn:Wg}
W_{\widetilde{g}_i} \doteq \int {\rm d}^3\bb{r} \int {\rm d}^3\bb{v} \, \frac{\tprl{i}\widetilde{g}^{\,2}_i}{2f_{0i}} .
\end{equation}
All three cascade channels lead to small perpendicular spatial scales via passive mixing by the Alfv\'{e}nic turbulence and to small scales in $v_\parallel$ via the linear parallel phase mixing, the rates of mixing being functions of the velocity-space anisotropy of the equilibrium distribution function.

\section{Conclusions}

\subsection{Generalised Free-Energy Cascade in a Pressure-Anisotropic Plasma}\label{sum:W}

Assembling the results of Sections \ref{sec:alfveniccascade} and \ref{sec:compressivecascade}, we now arrive at the central unifying concept of this Paper. The Alfv\'{e}nic invariants (Eq.~\ref{eqn:WAW}) and the compressive invariant (Eq.~\ref{eqn:Wcompr3}) together make up the {\em generalised free energy},
\begin{align}\label{eqn:W}
W &\doteq W^+_{\rm AW} + W^-_{\rm AW} + W_{\rm compr} \nonumber\\*
\mbox{} &= \int{\rm d}^3\bb{r} \left\{ \sum_s \int {\rm d}^3\bb{v} \, \frac{\tprl{s} \delta\widetilde{f}^{\,2}_s}{2f^\parallel_{0s}} + \frac{\rho_0 u^2_\perp}{2}  \right. \nonumber\\*
\mbox{} &\qquad\qquad+ \left. \left[ 1 + \sum_s \frac{\betaprl{s}}{2} \left( \Delta_s - \frac{2\duparsq}{\vthprl{s}^2}\right) \right] \frac{\delta B^2_\perp}{8\pi} + \Bigl( 1 - \sum_s \betaprp{s} \Delta_{2s} \Bigr) \frac{\delta B^2_\parallel}{8\pi} \right\} ,
\end{align}
which is the quantity conserved by Equations (\ref{sum:eqns}) in the absence of equilibrium interspecies drifts and cascaded to small scales in phase space across the inertial range of KRMHD turbulence (analogous to the energy cascade in fluid or MHD turbulence). It contains (in the order of appearance in Eq.~\ref{eqn:W}) the perturbed entropy of the system in the frame of the Alfv\'{e}nic fluctuations, the energy associated with the $\bb{E}\btimes\bb{B}$ motion, the energy carried by the magnetic fluctuations, and terms arising from the exchange of free energy between the magnetic fluctuations and the equilibrium pressure anisotropy and parallel interspecies drifts. Just as shown by S09 for a two-species Maxwellian plasma, the inertial-range kinetic cascade can generally be split into three independent cascades of the generalised Alfv\'{e}nic and compressive-fluctuation energies: $W^+_{\rm AW}$, $W^-_{\rm AW}$, and $W_{\rm compr}$. In Section \ref{sec:collisionlesscascades}, we showed that, for a single-ion-species bi-Maxwellian plasma, $W_{\rm compr}$ can be further decomposed into three independently cascading parts: $W^+_{\rm compr}$, $W^-_{\rm compr}$, and $W_{\widetilde{g}_i}$. In all of these cascades, what is affected by pressure anisotropies and interspecies drifts are the amount of magnetic-field fluctuations associated with each of the invariants (\S\S\ref{sec:WAW}, \ref{sec:parallelkinetics}) and the rate at which linear parallel phase mixing generates small-scale structure in velocity space (\S\ref{sec:mirror}).

Because we have ordered collisions out of our equations, the collisionless invariant given by Equation (\ref{eqn:W}) is just one of an infinite number of invariants of the system. We place special emphasis on this particular invariant, $W$, for two reasons. 

First, it neatly reduces to {\em the} generalised invariant for a (collisional) Maxwellian plasma (cf.~eqs.~74 and 153 of S09):
\[
W \rightarrow \int{\rm d}^3\bb{r} \left( \sum_s \int{\rm d}^3\bb{v} \, \frac{T_{0s} \delta f^2_s}{2 f_{0s}} + \frac{\rho_0 u^2_\perp}{2} + \frac{|\delta\bb{B}|^2}{8\pi} \right) ,
\]
which is variously referred to as the generalised grand canonical potential \citep{hallatschek04} or free energy \citep{fowler68,scott10} because of its similarity to the Helmholtz free energy $A \doteq -\sum_s T_{0s} \delta S_s + \delta U $, where $\delta U$ is the potential energy stored in the fluctuations and $\delta S_s$ is the entropy associated with the perturbed distribution function. 

Secondly, it encodes rather neatly in a thermodynamical context the main (linear and nonlinear) physical effect associated with the presence of firehose (\S\ref{sec:firehose}) and mirror (\S\ref{sec:mirror}) instabilities. The firehose and mirror (in)stability parameters appear in Equation (\ref{eqn:W}) as prefactors of the perpendicular and parallel magnetic energies, respectively. As either of these thresholds is approached, these pre-factors get smaller and thus the energetic weight associated with the corresponding type of magnetic fluctuation gets smaller. At the firehose threshold, bending magnetic-field lines is free; at the mirror threshold, compressing them is free. Beyond these thresholds, it becomes energetically profitable to grow magnetic fluctuations.

As long as the plasma stays within the firehose and mirror stability boundaries, the generalised free-energy invariant $W$ is positive-definite and so describes a turbulent cascade from large to small scales (as well as into phase space). In this case, linear stability implies nonlinear stability. As these thresholds are crossed, the reduced ordering underpinning our equations breaks down, since there are no restoring forces to keep such fluctuations small. Thus, Equation (\ref{eqn:W}) can only be meaningfully interpreted as a free-energy invariant when the plasma is stable.

In the presence of equilibrium interspecies drifts, the conservation law for the generalised free energy acquires source/sink terms:
\begin{equation}\label{eqn:dWdt}
\D{t}{W} = - \int {\rm d}^3\bb{r} \, \sum_i \dupari \left( Z_i e \dni E_\parallel - \dpprp{i} \eb\bcdot\grad\frac{\dBprl}{B_0} \right),
\end{equation}
which correspond to the change in the free energy due to the work done on the system by the fluctuating parallel electric (Eq.~\ref{eqn:efield}) and by magnetic-mirror forces acting on the equilibrium parallel drifts.

\subsection{Quantitative Details Matter}\label{sum:details}

Within the stability boundaries imposed by the firehose, mirror, and streaming instabilities, the theory presented here is analogous to that presented in S09 for a single-ion-species Maxwellian plasma, with many of the differences amounting to a somewhat mundane change of multiplicative coefficients. But this in itself is interesting, for it demonstrates that many of the salient qualitative features of the gyrokinetic and KRMHD theories of astrophysical turbulence are robust with respect to deviations from velocity-space isotropy. On a more practical quantitative level, these otherwise benign coefficients do affect, at the order-unity level, the relationships between various fluctuating fields and partitioning of free energy that lie at the heart of predictive theories of solar-wind turbulence. In light of the ever-increasing scrutiny placed upon theories of Alfv\'{e}nic turbulence by the wealth of data from the solar wind, as well as the astrophysical importance of knowing the proportion of turbulent energy that is distributed between the ion and electron populations, such details matter.

The equations derived and discussed herein may be readily incorporated into existing (Maxwellian) KRMHD and gyrokinetic numerical codes. Such an advancement is now underway.
 
\acknowledgements

Support for M.~W.~K.~was provided by NSF through Max-Planck/Princeton Center for Plasma Physics Grant No.~PHY-1144374 and by NASA through Einstein Postdoctoral Fellowship Grant No.~PF1-120084, the latter issued by the Chandra X-ray Observatory Center, which is operated by the Smithsonian Astrophysical Observatory for and on behalf of NASA under Contract No.~NAS8-03060. Support for C.~H.~K.~Chen was provided by an Imperial College Junior Research Fellowship. The authors thank Michael Barnes, Ron Davidson, Greg Hammett, Anjor Kanekar, John Krommes, Felix Parra, Jason TenBarge, and George Wilkie for useful conversations. The completion of this work was facilitated by the generous hospitality and material support provided by the International Space Science Institute (ISSI) in Bern to ISSI/ISSI-Beijing Team 304 and by Merton College, Oxford.

\appendix

\section{Definitions of $\lambda_i$-coefficients in Equation (\ref{eqn:GnB})}\label{app:coefficients}

In Section \ref{sec:parallelkinetics}, we derived Equation (\ref{eqn:GnB2}) describing the parallel kinetics of the compressive fluctuations, which explicitly shows how kinetic fluctuations evolve under the influence of the density and magnetic-field-strength fluctuations they excite. This Equation involves $4N_{\rm ion}$ coupling coefficients, which we give here in terms of the usual plasma parameters and various perpendicular moments of the parallel-differentiated equilibrium ion distribution function, denoted $F^\parallel_{\ell i}$ (see Eq.~\ref{eqn:Felli}):
\begin{align}
\lambda^{nn}_i &\doteq  \Bigl[ \,\dots \Bigr]^{-1} \frac{c_i}{\czero{e}} \frac{Z_i}{\tauprl{i}} &&\hspace{-1em} \left[ \frac{\fonei}{\fzeroi} + \frac{Z_i}{\tauprp{i}} \left( 2 \Delta_{2e} - \frac{\cone{e}}{\czero{e}} \Delta_{1e} \right) - 2 c_i \left( \sum_{i'} \frac{c_{i'} \tauprp{i'} Z_i}{c_i \tauprp{i} Z_{i'}} + \frac{1}{\betaprp{i}} \right) \right] \,,
\\*
\lambda^{nB}_i &\doteq  \Bigl[ \,\dots \Bigr]^{-1} \frac{c_i}{\czero{e}} \frac{Z_i}{\tauprl{i}} &&\hspace{-1em} \left\{ \Delta_{1e} \frac{\fonei}{\fzeroi} + 2\czero{e} \frac{\tauprl{i}}{Z_i} \frac{\pprp{i}}{\pprl{i}} \frac{\ftwoi}{\fzeroi} + \left( \Delta_{1e} + \czero{e} \frac{\tauprl{i}}{Z_i} \frac{\pprp{i}}{\pprl{i}} \frac{\fonei}{\fzeroi} \right) \right.
\nonumber \\*
\mbox{} & \mbox{} &&\hspace{-1em} \left.\times \left[  \frac{Z_i}{\tauprp{i}} \left( 2 \Delta_{2e} - \frac{\cone{e}}{\czero{e}} \Delta_{1e} \right) - 2 c_i \left( \sum_{i'} \frac{c_{i'} \tauprp{i'} Z_i}{c_i \tauprp{i} Z_{i'}} + \frac{1}{\betaprp{i}} \right)\right] \right\} \,,
\\*
\lambda^{Bn}_i &\doteq  \Bigl[ \,\dots \Bigr]^{-1} \frac{c_i}{\czero{e}} \frac{Z_i}{\tauprl{i}} &&\hspace{-1em} \left( \frac{\fonei}{\fzeroi} + \frac{\cone{e}}{\czero{e}} \frac{Z_i}{\tauprp{i}} \right) \,,
\\*
\lambda^{BB}_i &\doteq  \Bigl[ \,\dots \Bigr]^{-1} \frac{c_i}{\czero{e}} \frac{Z_i}{\tauprl{i}}  &&\hspace{-1em} \left[ \Delta_{1e} \frac{\fonei}{\fzeroi} + 2\czero{e} \frac{\tauprl{i}}{Z_i} \frac{\pprp{i}}{\pprl{i}} \frac{\ftwoi}{\fzeroi}  + \frac{\cone{e}}{\czero{e}} \frac{Z_i}{\tauprp{i}} \left( \Delta_{1e} + \czero{e} \frac{\tauprl{i}}{Z_i} \frac{\pprp{i}}{\pprl{i}} \frac{\fonei}{\fzeroi} \right) \right] \,, 
\end{align}
where the bracket $[\, \dots ]$ is given by Equation (\ref{eqn:bracket}).

\section{Linear KRMHD Theory for Arbitrary $f_{0s}$}\label{app:linear}

In Section \ref{sec:mirror} we derived the linear theory of compressive fluctuations under the assumption that the plasma contained only a single species of bi-Maxwellian ions. This simplification was necessary in order to diagonalize the ion kinetic Equation (\ref{eqn:GnB}) and thereby decompose the compressive invariant $W_{\rm compr}$ into its three independently cascading parts. In this Appendix, the linear theory for the compressive fluctuations is derived for a plasma containing multiple ionic species with arbitrary $f_{0s}$.

\subsection{Linear Dispersion Relation}\label{app:disprel}

To obtain the linear dispersion relation governing the compressive fluctuations, we return to Equation (\ref{sum:ionkinetic}). Dropping the nonlinear terms and Fourier transforming in time ($\partial / \partial t \rightarrow -\imag \omega$) and space ($\eb \bcdot \grad \rightarrow \imag k_\parallel$), we find
\begin{equation}\label{eqn:glinear}
g_{i\bs{k}} = - \left[ \frac{1}{\czero{e}} \frac{Z_i}{\tauprl{i}}  \left( \frac{\dnek}{\nem} + \Delta_{1e} \frac{\dBprlk}{B_0} \right) + \frac{w^2_\perp}{\vthprl{i}^2} \frac{\dBprlk}{B_0} \right] \frac{ v_\parallel - \dupari}{ v_\parallel - \omega / k_\parallel} \, f^\parallel_{0i}.
\end{equation}
Computing the required moments of this equation introduces poles where $\omega = k_\parallel v_\parallel$, which correspond to wave-particle resonances and generically result in the growth or decay of compressive fluctuations. Anticipating this, we define the ion coefficients
\begin{align}\label{eqn:icoeff}
\cell{i} &\doteq \frac{1}{\nip} \int {\rm d}^3 \bb{v} \,  \frac{1}{\ell !} \! \left( \frac{w_\perp}{\vthprp{i}} \right)^{\!2\ell} \frac{ v_\parallel - \dupari }{ v_\parallel - \omega / k_\parallel} \, f^\parallel_{0i} \\*
\mbox{} &= \int{\rm d}v_\parallel \, F^\parallel_{\ell i}(v_\parallel) + \frac{\omega - k_\parallel\dupari}{|k_\parallel|} \int {\rm d}v_\parallel \,\frac{F^\parallel_{\ell i}(v_\parallel)}{v_\parallel - \omega / |k_\parallel| },\nonumber
\end{align}
which engender suitable generalisations of the plasma dispersion function for non-Maxwellian distributions: e.g.,
\begin{equation}\label{eqn:cellmaxwell}
\cell{i} = 1 + \xi_i Z_{\rm M} ( \xi_i ) \quad \textrm{for a bi-Maxwellian, where } \xi_i \doteq \frac{\omega - k_\parallel \dupari}{|k_\parallel| \vthprl{i}} .
\end{equation}
Note that these ion coefficients resemble the corresponding electron ones (cf.~Eq.~\ref{eqn:ecoeff}) in the limit $\omega$, $k_\parallel \dupare \ll k_\parallel v_\parallel$. By analogy with Equation (\ref{eqn:epaniso}) for the electron fluid, we also define the dimensionless pressure anisotropy of the ions appropriately weighted by $\cell{i}$:
\begin{equation}\label{eqn:ipaniso}
\Delta_{\ell i} \doteq \cell{i} \frac{\pprp{i}}{\pprl{i}} - 1 .
\end{equation}
Note that the effect of equilibrium parallel drifts is implicit in the modified pressure anisotropy $\Delta_{\ell i}$, as they enter through the $\xi_i$ dependence of the $\cell{i}$ coefficients.

With these definitions in hand, we proceed as follows. Taking the zeroth and $m_i w^2_\perp / 2 $ moments of Equation (\ref{eqn:glinear}) gives, respectively,
\begin{equation}\label{eqn:dni}
\frac{\dnik}{\nip} = - \frac{\czero{i}}{\czero{e}}  \frac{Z_i}{\tauprl{i}} \left( \frac{\dnek}{\nem} + \Delta_{1e} \frac{\dBprlk}{B_0} \right) - \Delta_{1i} \frac{\dBprlk}{B_0} ,
\end{equation}
\begin{equation}\label{eqn:dpprpi}
\frac{\dpprpik}{\pprp{i}} = - \frac{\cone{i}}{\czero{e}}  \frac{Z_i}{\tauprl{i}}\left( \frac{\dnek}{\nem} + \Delta_{1e} \frac{\dBprlk}{B_0} \right) - 2 \Delta_{2i} \frac{\dBprlk}{B_0} .
\end{equation}
Equation (\ref{eqn:dpprpi}), along with Equation (\ref{eqn:pprpe}) for the perturbed perpendicular electron pressure, allows the pressure-balance relation (Eq.~\ref{eqn:pbalance}) in the linear regime to be written as
\begin{equation}
\sum_s  \frac{\cone{s}}{\czero{e}} \frac{Z_s}{\tauprl{s}} \frac{\betaprp{s}}{2} \left( \frac{\dnek}{\nem} + \Delta_{1e} \frac{\dBprlk}{B_0} \right) + \left( \sum_s \betaprp{s} \Delta_{2s}  - 1 \right)  \frac{\dBprlk}{B_0} = 0 .
\end{equation}
Combining this with the linearised quasineutrality equation (Eq.~\ref{eqn:neutrality} with Eq.~\ref{eqn:dni}), 
\begin{equation}
\sum_s c_s \frac{\czero{s}}{\czero{e}} \frac{Z_s}{\tauprl{s}} \left( \frac{\dnek}{\nem} + \Delta_{1e} \frac{\dBprlk}{B_0} \right) - \sum_s c_s \Delta_{1s} \frac{\dBprlk}{B_0} = 0,
\end{equation}
we obtain the {\em KRMHD dispersion relation}
\begin{equation}\label{eqn:akrmhd_disprel}
\left( \sum_s c^2_s \czero{s} \frac{2}{\betaprl{s}} \right) \left( \sum_s \betaprp{s} \Delta_{2s} - 1 \right) = \left( \sum_s c_s \Delta_{1s} \right)^2 .
\end{equation}
The left-hand side of this equation exhibits the usual modification of the plasma beta parameter by the pressure anisotropy of each species. For a single-ion-species bi-Maxwellian plasma, Equation (\ref{eqn:akrmhd_disprel}) reduces to Equation (\ref{eqn:disprel}), which most notably contains the pressure-anisotropic version of the Barnes damping (Eq.~\ref{eqn:barnes}). We now specialise Equation (\ref{eqn:akrmhd_disprel}) for two interesting cases.

\subsection{Landau Damping and Ion Acoustic Instability}\label{app:landau}

Let us first consider the $\beta_\parallel \rightarrow 0$ electrostatic limit of Equation (\ref{eqn:akrmhd_disprel}):
\begin{equation}\label{eqn:unmagnetized_disprel}
1 + \sum_i c_i \frac{\czero{i}}{\czero{e}} \frac{Z_i}{\tauprl{i}} = 0 .
\end{equation}
We specialise Equation (\ref{eqn:unmagnetized_disprel}) for a Maxwellian plasma (cf.~Eqs \ref{eqn:ecoeff} and \ref{eqn:cellmaxwell}) consisting of massless electrons, cold ions ($|\xi_i| \gg 1$), and a small population ($c_\alpha \ll 1$) of drifting ($\dupara \ll \vthprl{\alpha}$) hot alpha particles ($|\xi_\alpha| \ll 1$). Expanding $Z_{\rm M}(\xi_i) \simeq -1/\xi_i - 1/2\xi^3_i$ and $Z_{\rm M}(\xi_\alpha) \simeq \imag \sqrt{\pi}$, and assuming that the decay/growth rate $\gamma$ is much smaller than the real part of the frequency $\omega_{\rm r}$, we find
\begin{equation}\label{eqn:iai}
\omega_{\rm r} = \pm k_\parallel \sqrt{\frac{Z_i T_{0e}}{m_i}} \doteq \pm k_\parallel v_{\rm s} \qquad {\rm and} \qquad \gamma \simeq -  |k_\parallel| v_s \, c_\alpha\frac{Z_\alpha}{\tau_\alpha} \left( \frac{\pi}{8} \frac{Z_i}{\tau_\alpha} \frac{m_\alpha}{m_i} \right)^{1/2} \left( 1 \mp \frac{\dupara}{v_{\rm s}} \right) ,
\end{equation}
where $v_{\rm s}$ is the ion sound speed. For $\omega_{\rm r}> k_\parallel \dupara$, this is simply Landau damping of an ion-acoustic wave, occurring at a reduced rate due to the flattening of the total distribution function in the vicinity of the alpha-particle drift velocity. For $\omega_{\rm r} < k_\parallel\dupara$, Equation (\ref{eqn:iai}) represents a sort of ion-acoustic instability, driven by the differential streaming of alpha particles (rather than of electrons, which is the standard case; e.g.~\citealt{davidson83}). (Note that the assumption $\gamma \ll \omega_{\rm r}$ is satisfied because $c_\alpha \ll 1$.) Outside of the limits taken, this branch of the dispersion relation is heavily damped (\S\ref{sec:lowbetalimit}).

In Section \ref{sec:Wcompr}, we stated that the right-hand side of the evolution equation for the compressive invariant (Eq.~\ref{eqn:Wcompr}) represents the work done by the fluctuating parallel electric field and by magnetic-mirror forces acting on the interspecies drifts, and thus is related to the rate at which free energy is exchanged between these drifts and the compressive fluctuations. Here we demonstrate this explicitly for the electrostatic case investigated in this section. Using Equation (\ref{eqn:dni}) to express the density perturbation of the alphas in terms of the density perturbation of the electrons, expanding $\czero{\alpha} \simeq 1 + \imag \sqrt{\pi} \xi_\alpha$, and using Equation (\ref{eqn:iai}) for the real and imaginary parts of $\xi_\alpha$, we find that our free-energy equation (Eq.~\ref{eqn:dWdt}) becomes
\begin{equation}
\D{t}{W} = - p_{0\alpha}  \sum_{\bs{k}}  |k_\parallel | v_{{\rm th}\alpha}   \left| \frac{Z_\alpha}{\tau_\alpha} \frac{\dupara}{v_{{\rm th}\alpha}} \frac{\dnek}{\nem} \right|^2 \left( \frac{\omega_{\rm r}}{k_\parallel \dupara} - 1 \right).
\end{equation}
When the plasma is streaming-unstable, namely $\omega_{\rm r} < k_\parallel\dupara$, the right-hand side of this equation becomes positive, and free energy is extracted from the interspecies drifts and put into the compressive fluctuations.

\subsection{Barnes Damping and Mirror Instability}\label{app:mirror}

Next, we treat the linear theory for Barnes damping and the mirror instability. Two results are sought: ({\it i}\,) the general mirror stability threshold for a single-ion-species non-Maxwellian plasma; and ({\it ii}\,) the decay/growth rate for a bi-Maxwellian plasma consisting of massless electrons, hot ions ($|\xi_i| \ll 1$), and a small ($c_\alpha \ll 1$) population of drifting ($\dupara \ll \vthprl{\alpha}$) hot alpha particles ($|\xi_\alpha| \ll 1$).

For ({\it i}\,), we note that the transition from stability to instability proceeds through $\omega \rightarrow 0$ (this is not generally true in the case with particle drifts, as we show below). In this limit, the coefficients
\[
\cell{s} \rightarrow \frac{1}{\nsp} \int {\rm d}^3\bb{v} \, \frac{1}{\ell !} \left( \frac{w_\perp}{\vthprp{s}} \right)^{2\ell} f^\parallel_{0s} 
\]
for both the ion and electron species, and the stability criterion may be read off directly from Equation (\ref{eqn:akrmhd_disprel}):
\begin{equation}\label{eqn:mirrorstability}
1 - \sum_s \betaprp{s} \left[\left( \frac{1}{\nsp} \int{\rm d}^3\bb{v} \, \frac{1}{2} \frac{w^4_\perp}{\vthprp{s}^4} f^\parallel_{0s} \right) \frac{\pprp{s}}{\pprl{s}} -  1 \right] \ge - { {\displaystyle \left( \sum_s \frac{c_s}{\nsp} \int{\rm d}^3\bb{v}\, \frac{w^2_\perp}{\vthprl{s}^2} f^\parallel_{0s} \right)^2 } \over {\displaystyle \sum_s \frac{2}{\betaprl{s}} \frac{c^2_s}{\nsp} \int{\rm d}^3\bb{v} \, f^\parallel_{0s} } } .
\end{equation}
This is in agreement with existing expressions in the literature \citep[e.g.][eq.~34]{pokhotelov02,hellinger07}.

For ({\it ii}\,), we expand $Z_{\rm M}(\xi_i) \simeq Z_{\rm M}(\xi_\alpha) \simeq \imag \sqrt{\pi}$ in Equation (\ref{eqn:akrmhd_disprel}) and note that $\xi_\alpha$ can be expressed in terms of $\xi_i$:
\[
\xi_\alpha = \frac{\vthprl{i}}{\vthprl{\alpha}} \left( \xi_i - \frac{k_\parallel}{|k_\parallel|}\frac{\dupara}{\vthprl{i}} \right) .
\]
With $\Delta_s \sim 1/\betaprp{s} \ll 1$, the leading-order terms in the dispersion relation are 
\begin{equation}
\imag \sqrt{\pi} \betaprp{i} \left[ \xi_i \left(  \frac{\tprp{i}}{\tprl{i}} + \frac{\betaprp{\alpha}}{\betaprp{i}} \frac{\tprp{\alpha}}{\tprl{\alpha}} \frac{\vthprl{i}}{\vthprl{\alpha}} \right) - \frac{k_\parallel}{|k_\parallel|}\frac{\betaprp{\alpha}}{\betaprp{i}} \frac{\tprp{\alpha}}{\tprl{\alpha}} \frac{\dupara}{\vthprl{\alpha}} \right] = 1 - \sum_s \betaprp{s} \Delta_s  .
\end{equation}
Solving for the real and imaginary parts of the frequency, we find
\begin{subequations}
\begin{align}
\omega_{\rm r} &=- k_\parallel \dupara ~ \frac{\betaprp{\alpha}}{\betaprp{i}} \frac{\tprl{i}}{\tprl{\alpha}} \frac{\tprp{\alpha}}{\tprp{i}} \frac{\vthprl{i}}{\vthprl{\alpha}}\left( 1 + \frac{\betaprp{\alpha}}{\betaprp{i}} \frac{\tprl{i}}{\tprl{\alpha}} \frac{\tprp{\alpha}}{\tprp{i}} \frac{\vthprl{i}}{\vthprl{\alpha}} \right)^{-1} ,\\*
\gamma &= - \frac{|k_\parallel| \valf}{\sqrt{\pi\betaprl{i}}} \frac{\pprl{i}^2}{\pprp{i}^2} \Bigl( 1 - \sum_s \betaprp{s} \Delta_s \Bigr) \left( 1 + \frac{\betaprp{\alpha}}{\betaprp{i}} \frac{\tprl{i}}{\tprl{\alpha}} \frac{\tprp{\alpha}}{\tprp{i}} \frac{\vthprl{i}}{\vthprl{\alpha}} \right)^{-1} .
\end{align}
\end{subequations}
Thus, the usual decay/growth rate (Eq.~\ref{eqn:barnes}) is modified by the pressure anisotropy of the alpha particles and reduced by the effect of the hot alphas on the Landau resonance. The mode acquires a real part proportional to the alpha-particle drift so that, even at marginal stability, $\omega \ne 0$.

\section{Derivation of KRMHD from Pressure-Anisotropic Gyrokinetics}\label{app:gk}

Heretofore, we have worked under the assumption that both $\omega / \Omega_s$ and $k \rho_s$ are asymptotically small to all orders in $\epsilon$, where $\Omega_s \doteq q_s B_0 / m_s c$ is the gyrofrequency and $\rho_s \doteq \vthprp{s} / \Omega_s$ is the gyroradius of species $s$. In this Appendix, we relax this assumption and allow for fluctuations with $k_\perp \rho_s \sim 1$ and $\omega \sim \epsilon \Omega_s$. The resulting set of nonlinear equations generalises Maxwellian slab gyrokinetics (\citealt{howes06}; S09) to arbitrary equilibrium distribution function $f_{0s}$. Some of these results have been obtained before---notably, the nonlinear gyrokinetic equation for arbitrary $f_{0s}$ was derived already by \citet{fc82}. Our restriction to slab geometry removes many of the complications introduced by those authors' applications to axisymmetric tokamaks and makes the theory rather easier to grasp, which hopefully gives this Appendix some pedagogical value. We will also use these equations to show that the generalisation of KRMHD derived in this paper can be rigorously obtained from a gyrokinetic description by taking the $k_\perp \rho_s \ll 1$ limit.

\subsection{Basic Equations and Gyrokinetic Ordering}\label{app:gkordering}

We begin with the kinetic Vlasov-Landau (or Boltzmann) equation
\begin{equation}\label{eqn:vlasov}
\dot{f}_s \doteq \pD{t}{f_s} + \bb{v} \bcdot \grad f_s + \frac{q_s}{m_s} \left( \bb{E} + \frac{\bb{v} \btimes \bb{B}}{c} \right) \bcdot \pD{\bb{v}}{f_s} = \left( \pD{t}{f_s} \right)_{\rm c}.
\end{equation}
The electric and magnetic fields are expressed in terms of scalar and vector potentials:
\begin{equation}\label{eqn:BcurlA}
\bb{E} = - \grad \varphi - \frac{1}{c} \pD{t}{\bb{A}} \qquad {\rm and} \qquad \bb{B} = B_0 \ez + \grad \btimes \bb{A} ,
\end{equation}
where $\grad \bcdot \bb{A} = 0$ (the Coulomb gauge). In the non-relativistic limit, these fields satisfy the plasma quasineutrality constraint (which follows from the Poisson equation to lowest order in $k^2 \lambda^2_{\rm D}$, where $\lambda_{\rm D}$ is the Debye length),
\begin{equation}\label{eqn:poisson}
0 = \sum_s q_s n_s = \sum_s q_s \int {\rm d}^3\bb{v} \, f_s,
\end{equation}
and the pre-Maxwell version of Amp\`{e}re's law,
\begin{equation}\label{eqn:ampere}
- \nabla^2 \bb{A} = \frac{4\pi}{c} \bb{j} = \frac{4\pi}{c} \sum_s q_s \int {\rm d}^3\bb{v} \, \bb{v} f_s .
\end{equation}
In this Appendix, {\em and in contrast with the main text}, we work with the velocity-space coordinate $\bb{v}_\perp$, i.e., the full velocity perpendicular to the magnetic-field direction, rather than with the perpendicular velocity $\bb{w}_\perp = \bb{v}_\perp - \bb{u}_\perp$ peculiar to the $\bb{E}\btimes\bb{B}$ flow $\bb{u}_\perp$.

Equations (\ref{eqn:vlasov})--(\ref{eqn:ampere}) are reduced by expanding
\begin{equation}
f_s = f_{0s} + \delta f_{1s} + \delta f_{2s} + \dots, 
\end{equation}
where the subscript indicates the order in $\epsilon$, and by adopting the ordering of Section \ref{sec:ordering} along with $k_\perp \rho_s \sim 1$ and $\omega \sim \epsilon \Omega_s$:
\begin{gather}
\frac{\omega}{\Omega_s}  \sim \frac{\rho_s}{L} \sim \frac{k_\parallel}{k_\perp} \sim \frac{u_\perp}{\valf} \sim \frac{\delta B_\perp}{B_0} \sim \frac{u_\parallel}{\valf} \sim \frac{\dBprl}{B_0} \sim \frac{\delta f_{1s}}{f_{0s}} \sim \epsilon , \nonumber\\*
k_\perp \rho_s \sim \betaprl{s} \sim \betaprp{s} \sim \Delta_s \sim \tauprl{s} \sim \tauprp{s} \sim 1 .
\end{gather}
We remind the reader that the collision frequency $\nu_{ii} \ll \epsilon^2 \Omega_s$, allowing for non-Maxwellian $f_{0s}$ (cf.~\S A2.2 of \citealt{howes06}). Also, because the electron-ion mass-ratio expansion is not performed at the outset of the gyrokinetic derivation ($m_e / m_i \ll 1$ is taken as a subsidiary expansion in \S\ref{sec:gkelectrons}), we must allow for the possibility of parallel electron drifts in the equilibrium state ($\dupare \sim \vthprl{e}$). 

The formal expansion of Equations (\ref{eqn:vlasov})--(\ref{eqn:ampere}) that results is worked out order by order. We begin with Equation (\ref{eqn:vlasov}), ordered relative to $\omega f_{0s}$.

\subsection{Gyrokinetic Equation}\label{sec:gkeqn}

\subsubsection{Minus-First Order, $\mc{O}(1/\epsilon)$}

The largest term in Equation (\ref{eqn:vlasov}) corresponds to Larmor motion of the equilibrium distribution about the uniform guide field:
\begin{equation}\label{eqn:vlasovm1}
- \Omega_s \ez \bcdot \left( \bb{v} \btimes \pD{\bb{v}}{f_{0s}} \right) = 0 .
\end{equation}
Decomposing the particle velocity in terms of the parallel velocity $v_\parallel$, the perpendicular velocity $v_\perp$, and the gyrophase angle $\vartheta$,
\begin{equation}\label{eqn:velocity}
\bb{v} = v_\parallel \ez + v_\perp \bigl( \cos \vartheta \, \ex + \sin \vartheta \, \ey \bigr) ,
\end{equation}
we find that Equation (\ref{eqn:vlasovm1}) takes on the simple form
\begin{equation}\label{eqn:gyroindependence}
- \Omega_s \pD{\vartheta}{f_{0s}} = 0 .
\end{equation}
The equilibrium distribution function is thus independent of gyrophase (gyrotropic): $f_{0s} = f_{0s}(v_\parallel, v_\perp, t)$.

\subsubsection{Zeroth Order, $\mc{O}(1)$}\label{sec:boltz}

Proceeding to next order and decomposing the velocity into its parallel and perpendicular parts (Eq.~\ref{eqn:velocity}), Equation (\ref{eqn:vlasov}) becomes
\begin{equation}\label{eqn:vlasov0}
\bb{v}_\perp \bcdot \grad_{\!\perp} \delta f_{1s} + \frac{q_s}{m_s} \left( - \grad_{\!\perp} \varphi + \frac{\bb{v} \btimes \delta \bb{B}}{c} \right) \bcdot \pD{\bb{v}}{f_{0s}} - \Omega_s \pD{\vartheta}{\delta f_{1s}} = 0 .
\end{equation}
This equation is simplified by employing the shorthand (cf.~Eqns \ref{eqn:fprlfprp} and \ref{eqn:Df0s})
\begin{equation}\label{eqn:gkDf0s}
\mf{D} f_{0s} = \frac{\pprp{s}}{\pprl{s}} f^\parallel_{0s} - f^\perp_{0s} , ~ {\rm where}~ f^\parallel_{0s} = - \vthprl{s}^2 \pD{(v_\parallel - \dupar)^2}{f_{0s}} ~ {\rm and} ~ f^\perp_{0s} = -\vthprp{s}^2 \pD{v^2_\perp}{f_{0s}} ,
\end{equation}
and noting that
\begin{align}\label{eqn:vxdB}
\frac{\tprp{s}}{m_s} \frac{\bb{v} \btimes \delta \bb{B}}{c} \bcdot \pD{\bb{v}}{f_{0s}} &= \order{0}{\bb{v}_\perp \bcdot \grad_{\!\perp} \left( \frac{v_\parallel A_\parallel}{c} \mf{D}f_{0s} -  \frac{\dupar A_\parallel}{c} \frac{\tprp{s}}{\tprl{s}} f^\parallel_{0s}\right)} \nonumber\\*
 \mbox{} & - \order{1}{\bb{v}_\perp \bcdot \pD{z}{} \left( \frac{v_\parallel\bb{A}_\perp}{c} \mf{D} f_{0s} - \frac{\dupar\bb{A}_\perp}{c} \frac{\tprp{s}}{\tprl{s}} f^\parallel_{0s} \right) } ,
\end{align}
where the order in $\epsilon$ of each term is indicated. The resulting lowest-order equation for $\delta f_{1s}$,
\begin{align}\label{eqn:vlasov0sim}
 \bb{v}_\perp \bcdot \grad_{\!\perp} \delta f_{1s} - \Omega_s \pD{\vartheta}{\delta f_{1s}} =  &- \bb{v}_\perp \bcdot \grad_{\!\perp} \, \frac{q_s}{\tprl{s}} \left( \varphi - \frac{\dupar A_\parallel}{c} \right) f^\parallel_{0s} \nonumber\\*\mbox{} &+ \bb{v}_\perp \bcdot \grad_{\!\perp} \, \frac{q_s}{\tprp{s}} \left( \varphi - \frac{v_\parallel A_\parallel}{c} \right) \mf{D}f_{0s} ,
\end{align}
admits a homogeneous solution and a particular solution.

The homogeneous solution $h_s$ satisfies
\begin{equation}\label{eqn:homogeneous}
\bb{v}_\perp \bcdot \grad_{\!\perp} h_s - \left. \Omega_s  \pD{\vartheta}{h_s} \right|_{\bs{r}} = \left. - \Omega_s \pD{\vartheta}{h_s} \right|_{\bs{R}_s} = 0 ,
\end{equation}
where we have transformed the $\vartheta$ derivative taken at constant position $\bb{r}$ to one taken at constant guiding centre---the centre of the ring orbit that the particle follows in a strong guide field:
\begin{equation}\label{eqn:guidingcentre}
\bb{R}_s = \bb{r} + \frac{\bb{v} \btimes \ez}{\Omega_s}.
\end{equation}
Thus, $h_s$ is independent of the gyrophase angle at constant guiding centre $\bb{R}_s$ (but not at constant position $\bb{r}$):
\begin{equation}\label{eqn:hs}
h_s = h_s(t,\bb{R}_s,v_\parallel,v_\perp).
\end{equation}
It represents the response of charged rings to the perturbed fields, and is thus referred to as the {\em gyrokinetic response}. 

The particular solution of Equation (\ref{eqn:vlasov0sim})---the so-called adiabatic, or ``Boltzmann'', response---is given by 
\begin{equation}\label{eqn:boltzmann}
\delta f_{1s,{\rm Boltz}} = - \frac{q_s}{\tprl{s}} \left( \varphi - \frac{\dupar A_\parallel}{c} \right) f^\parallel_{0s} + \frac{q_s}{\tprp{s}} \left( \varphi - \frac{v_\parallel A_\parallel}{c} \right) \mf{D} f_{0s} .
\end{equation}
It arises from the evolution of $f_{0s}$ under the influence of the perturbed electromagnetic fields, a fact that is most clearly demonstrated by transforming Equation (\ref{eqn:boltzmann}) to $(\varepsilon_s,\mu_s$) coordinates (see \S\ref{sec:gkeandmu}). It is instructive to note that the combination
\begin{equation}\label{eqn:varphiprime}
\varphi'_s \doteq \varphi - \dupar A_\parallel /c
\end{equation}
in the first parentheses of Equation (\ref{eqn:boltzmann}) is the fluctuating electrostatic potential in the frame of the parallel-drifting species $s$.

Decomposing $\delta f_{1s}$ into its adiabatic (Eq.~\ref{eqn:boltzmann}) and non-adiabatic (Eq.~\ref{eqn:hs}) parts \citep[cf.][]{al80,ctb81}, the complete solution for $f_s$ may be written
\begin{equation}\label{eqn:df1s}
f_s = f_{0s}(v_\parallel,v_\perp,t) + \delta f_{1s,{\rm Boltz}} + h_s(t,\bb{R}_s,v_\parallel,v_\perp) + \delta f_{2s} + \dots
\end{equation}
Next we derive an evolution equation for the the gyrokinetic response $h_s$.

\subsubsection{First Order, $\mc{O}(\epsilon)$}\label{sec:gkfirstorder}

Introducing the {\em gyrokinetic potential}
\begin{equation}\label{eqn:gkpotential}
\chi \doteq \varphi - \frac{v_\parallel A_\parallel}{c} - \frac{\bb{v}_\perp \bcdot \bb{A}_\perp}{c}
\end{equation}
and using Equations (\ref{eqn:vxdB}) and (\ref{eqn:df1s}), at this order Equation (\ref{eqn:vlasov}) becomes
\begin{align}\label{eqn:vlasovp1}
\pD{t}{h_s} + \dot{\bb{R}}_s \bcdot \pD{\bb{R}_s}{h_s} &- \frac{q_s}{\tprl{s}} \left(  \pD{t}{\chi} + \dupar \pD{z}{\chi} \right) f^\parallel_{0s} + \frac{q_s}{\tprp{s}} \left( \pD{t}{\chi} + v_\parallel \pD{z}{\chi} \right) \mf{D}f_{0s}  \nonumber\\*
\mbox{} &=  \left. \Omega_s \pD{\vartheta}{\delta f_{2s}} \right|_{\bs{R}_s} -  \frac{q_s}{m_s} \left( - \grad_{\!\perp} \varphi + \frac{\bb{v}\btimes\delta\bb{B}}{c} \right) \bcdot \pD{\bb{v}}{\delta f_{1s}} ,
\end{align}
where
\begin{equation}\label{eqn:dRdt}
\dot{\bb{R}}_s = v_\parallel \ez + \frac{c}{B_0} \left( - \grad \varphi - \frac{1}{c} \pD{t}{\bb{A}} + \frac{\bb{v}\btimes\delta\bb{B}}{c} \right) \btimes \ez
\end{equation}
is the velocity of the guiding centre. Upon performing a ring average of Equation (\ref{eqn:vlasovp1}) over $\vartheta$ at fixed $\bb{R}_s$, defined, for any function $a(t,\bb{r},\bb{v})$, as
\begin{equation}\label{eqn:ringaverage}
\langle a(t, \bb{r} , \bb{v} )\rangle_{\gas} \doteq \frac{1}{2\pi} \oint {\rm d}\vartheta \, a \left( t, \bb{R}_s - \frac{\bb{v}\btimes\ez}{\Omega_s}, \bb{v} \right) ,
\end{equation}
we find that the entire right-hand side of Equation (\ref{eqn:vlasovp1}) vanishes. This follows from the periodicity of $\delta f_{2s}$ in $\vartheta$ and from the fact that, for any arbitrary function $a(\bb{r})$, the ring average $\langle \bb{v}_\perp \bcdot \grad a \rangle_{\gas} = 0$ \citep[see eq.~A21 of][]{howes06}. Thus, the ring-averaged Equation (\ref{eqn:vlasovp1}) is
\begin{align}\label{eqn:gavlasovp1}
\pD{t}{h_s} + \langle \dot{\bb{R}}_s \rangle_{\gas} \bcdot \pD{\bb{R}_s}{h_s} &= \frac{q_s}{\tprl{s}} \left(  \pD{t}{\langle\chi\rangle_{\gas} } + \dupar \pD{z}{\langle\chi\rangle_{\gas} } \right) f^\parallel_{0s} \nonumber\\*\mbox{} &- \frac{q_s}{\tprp{s}} \left( \pD{t}{\langle\chi\rangle_{\gas} } + v_\parallel \pD{z}{\langle\chi\rangle_{\gas} } \right) \mf{D}f_{0s} .
\end{align}
Using the decomposition $\delta\bb{B} = \grad A_\parallel \btimes \ez + \dBprl \ez$ (Eq.~\ref{eqn:BcurlA}) in Equation (\ref{eqn:dRdt}) and retaining only first-order contributions, the ring-averaged guiding-centre velocity is
\begin{align}\label{eqn:gadRdt}
\langle \dot{\bb{R}}_s \rangle_{\gas} &= v_\parallel \ez - \frac{c}{B_0} \left\langle \grad_{\!\perp} \varphi \right\rangle_{\gas} \btimes \ez + \frac{v_\parallel}{B_0} \left\langle \grad_{\!\perp} A_\parallel \right\rangle_{\gas} \btimes \ez - \frac{1}{B_0} \left\langle \bb{v}_\perp \dBprl \right\rangle_{\gas} \nonumber\\*
&= v_\parallel \ez - \frac{c}{B_0} \pD{\bb{R}_s}{\langle \chi \rangle_{\gas}} \btimes \ez ,
\end{align}
where we have used the identity $\langle \bb{v}_\perp \dBprl \rangle_{\gas} = - \langle \grad_{\!\perp} ( \bb{v}_\perp \bcdot \bb{A}_\perp ) \rangle_{\gas}$. 

Substituting Equation (\ref{eqn:gadRdt}) into Equation (\ref{eqn:gavlasovp1}), we obtain the {\em gyrokinetic equation}
\begin{align}\label{eqn:gyrokinetic}
\pD{t}{h_s} + v_\parallel \pD{z}{h_s} + \frac{c}{B_0}  \{ \langle \chi \rangle_{\gas} , h_s \} &= \frac{q_s f^\parallel_{0s}}{\tprl{s}} \left( \pD{t}{} + \dupar \pD{z}{} \right) \langle\chi\rangle_{\gas} 
\nonumber\\*&\mbox{}- \frac{q_s \mf{D}f_{0s}}{\tprp{s}}  \left( \pD{t}{} + v_\parallel \pD{z}{} \right) \langle\chi\rangle_{\gas},
\end{align}
where the Poisson bracket is defined in the usual way:
\begin{equation}\label{eqn:gkbracket}
\{ \langle \chi \rangle_{\gas} \, , h_s \} \doteq \ez \bcdot \left( \pD{\bb{R}_s}{\langle \chi \rangle_{\gas}} \btimes \pD{\bb{R}_s}{h_s} \right) .
\end{equation}
The differences between Equation (\ref{eqn:gyrokinetic}) and the usual (Maxwellian) slab gyrokinetic equation (cf.~eq.~25 of \citealt{howes06}) lie entirely on the right-hand side, which arises from (ring-averaged) changes in the kinetic energy and magnetic moment of the particles. In particular, the final term in Equation (\ref{eqn:gyrokinetic}), absent in Maxwellian gyrokinetics, ensures that $h_s$ evolves in such a way as to preserve adiabatic invariance. Indeed, if we augment the gyrokinetic response in the following way,\footnote{By Equation (\ref{eqn:homogeneous}), we can add any gyrophase-independent (at constant guiding centre $\bb{R}_s$) function to the gyrokinetic response $h_s$ and still satisfy the zeroth-order kinetic equation.}
\begin{equation}\label{eqn:hstilde}
\widetilde{h}_s \doteq h_s + \frac{q_s\mf{D}f_{0s}}{\tprp{s}} \langle \chi \rangle_{\gas} ,
\end{equation}
the gyrokinetic equation (\ref{eqn:gyrokinetic}) can be expressed more compactly as
\begin{equation}
\pD{t}{\widetilde{h}_s} + v_\parallel \pD{z}{\widetilde{h}_s} + \frac{c}{B_0}  \{ \langle \chi \rangle_{\gas} , \widetilde{h}_s \} = \frac{q_s f^\parallel_{0s}}{\tprl{s}} \left( \pD{t}{} + \dupar \pD{z}{} \right) \langle\chi\rangle_{\gas} .
\end{equation}
These matters are discussed in detail in Section \ref{sec:gkeandmu}.

\subsection{Field Equations}\label{sec:gkmaxwell}

The equations governing the electromagnetic potentials are obtained by substituting Equation (\ref{eqn:df1s}) into the leading-order expansions of the quasineutrality constraint (Eq.~\ref{eqn:poisson}) and Amp\`{e}re's law (Eq.~\ref{eqn:ampere}). This procedure requires computing parallel moments of the perpendicular-differentiated equilibrium distribution function: we denote
\begin{subequations}\label{eqn:kcoeffs}
\begin{align}
\kzero{s} &\doteq \frac{1}{\nsp} \int {\rm d}^3\bb{v} \, f^\perp_{0s} , \\*
\kone{s} &\doteq \frac{1}{\nsp} \int {\rm d}^3\bb{v} \, \frac{v_\parallel}{\vthprl{s}} f^\perp_{0s} \times \left(\frac{\dupar}{v_{{\rm th}\parallel s}}\right)^{-1} , \\*
\ktwo{s} &\doteq \frac{1}{\nsp} \int {\rm d}^3\bb{v} \, \frac{v^2_\parallel}{\vthprl{s}^2} f^\perp_{0s} \times \left( \frac{1}{2} + \frac{\duparsq}{\vthprl{s}^2} \right)^{-1} ,
\end{align}
\end{subequations}
all of which equate to unity for a drifting bi-Maxwellian distribution (Eq.~\ref{eqn:biMax}).\footnote{If there are no interspecies parallel drifts and if $f^\perp_{0s}$ is symmetric about $v_\parallel = \dupar$, then $\dupar\kone{s} = 0$.}

To first order, $\mc{O}(\epsilon)$, Equation (\ref{eqn:poisson}) becomes
\begin{align}\label{eqn:gkqn1}
0 &= \sum_s q_s \dns \nonumber\\*
\mbox{} &= \sum_s q_s \left[ \int {\rm d}^3\bb{v} \, h_s \left( t, \bb{r} + \frac{\bb{v} \btimes \ez}{\Omega_s} , v_\parallel, v_\perp \right) -  \frac{q_s \nsp}{\tprp{s}} \left( \kzero{s} \, \varphi - \kone{s} \frac{\dupar A_\parallel}{c} \right)  \right] .
\end{align}
Since the field variables $\varphi$ and $A_\parallel$ are functions of the spatial variable $\bb{r}$, the velocity integral of $h_s$ must be performed at constant location $\bb{r}$ of the charges rather than at constant guiding centre $\bb{R}_s$. This introduces a gyro-averaging operation dual to the ring average defined in Equation (\ref{eqn:ringaverage}):
\begin{equation}\label{eqn:gyroaverage}
\langle h_s ( t , \bb{R}_s , v_\parallel , v_\perp ) \rangle_{\bs{r}} \doteq \frac{1}{2\pi} \oint {\rm d} \vartheta \, h_s \left( t , \bb{r} + \frac{\bb{v} \btimes \ez}{\Omega_s} , v_\parallel , v_\perp \right) .
\end{equation}
Equation (\ref{eqn:gkqn1}) may then be written as
\begin{equation}\label{eqn:gkqn}
0 = \sum_s q_s \left[ \int {\rm d}^3\bb{v} \, \langle h_s \rangle_{\bs{r}} - \frac{q_s \nsp}{\tprp{s}} \left( \kzero{s} \, \varphi - \kone{s} \frac{\dupar A_\parallel}{c} \right)  \right] .
\end{equation}
Likewise, the parallel and perpendicular components of Amp\`{e}re's law become, respectively,
\begin{align}\label{eqn:gkprlamp}
\nabla^2_\perp A_\parallel &= - \frac{4\pi}{c} j_\parallel \nonumber\\*
\mbox{} &= - \frac{4\pi}{c} \sum_s q_s \left[ \int {\rm d}^3\bb{v}\, v_\parallel \langle h_s \rangle_{\bs{r}} - \frac{q_s \nsp \vthprl{s}}{2T_{\perp0s}} \left( \kone{s}\,\varphi \, \frac{2\dupar}{\vthprl{s}}  +  \widetilde{\Delta}_s\frac{\vthprl{s} A_\parallel}{c} \right) \right] ,
\end{align}
\begin{equation}
\label{eqn:gkprpamp}
\nabla^2_\perp \dBprl = -\frac{4\pi}{c} \ez \bcdot ( \grad_{\!\perp} \btimes \bb{j}_\perp ) = -\frac{4\pi}{c} \ez \bcdot \left[ \grad_{\!\perp} \btimes \sum_s q_s \int {\rm d}^3\bb{v}\, \langle \bb{v}_\perp h_s \rangle_{\bs{r}} \right] ,
\end{equation}
where
\begin{equation}\label{eqn:tildedelta}
\widetilde{\Delta}_s \doteq \frac{\pprp{s}}{\pprl{s}} - \ktwo{s} \left( 1 + \frac{2\duparsq}{\vthprl{s}^2} \right)
\end{equation}
is the pressure anisotropy of species $s$ augmented by the parallel ram pressure from equilibrium parallel drifts. Upon integrating by parts with respect to the gyroangle, Equation (\ref{eqn:gkprpamp}) can also be written as
\begin{equation}\label{eqn:gkpbalance}
\grad_{\!\perp} \grad_{\!\perp} \, \bb{:} \, \left( \delta\msb{P}_\perp + \msb{I} \, \frac{B_0\dBprl}{4\pi} \right) = 0,
\end{equation}
where
\begin{equation}
\delta\msb{P}_\perp = \sum_s \int{\rm d}^3\bb{v} \, m_s \langle \bb{v}_\perp\bb{v}_\perp h_s \rangle_{\bs{r}}
\end{equation}
is the perpendicular pressure tensor. The perpendicular component of Amp\'{e}re's law is therefore a statement of perpendicular pressure balance for the compressive fluctuations.

Together with the gyrokinetic equation (\ref{eqn:gyrokinetic}), the field equations (\ref{eqn:gkqn})--(\ref{eqn:gkprpamp}) provide a closed system that describes the evolution of a gyrokinetic plasma with non-Maxwellian $f_{0s}$ and parallel species drifts. It remains to show that, in the $k \rho_i$, $m_e/m_i \ll 1$ limit, these equations reduce to those of KRMHD.

\subsection{Meaning of $\delta f_{1s,{\rm Boltz}}$ and $h_s$: $(v_\parallel,v_\perp)$ vs.~$(\varepsilon_s,\mu_s)$ Coordinates}\label{sec:gkeandmu}

We pause the derivation here to offer a few comments regarding our choice of velocity-space coordinates, which will hopefully aid the reader's grasp of the physical content encapsulated in the Boltzmann response (Eq.~\ref{eqn:boltzmann}) and the gyrokinetic equation (\ref{eqn:gyrokinetic}). Just as in the main text (see \S\ref{sec:eandmu}), we have opted for analytical convenience by choosing to work with $v_\parallel$ and $v_\perp$ as our velocity variables. If, instead, our focus were to be on physical insight, then arguably better variables would be the total particle energy in the parallel-drifting frame,
\begin{equation}\label{eqn:ebar}
\overline{\varepsilon}_s = \varepsilon_{0s} + \varepsilon_{1s} \doteq \frac{1}{2} m_s \bigl| \bb{v} - \dupar \ez \bigr|^2 + q_s \varphi'_s ,
\end{equation}
and the gyrophase-dependent part of the first adiabatic invariant,
\begin{equation}\label{eqn:mubar}
\overline{\mu}_s = \mu_{0s} + \mu_{1s} \doteq \frac{m_s v^2_\perp}{2B_0} + \frac{q_s}{B_0} \left( \varphi - \frac{v_\parallel A_\parallel}{c} \right) ,
\end{equation}
written out to first order in the fluctuation amplitudes \citep[e.g.][]{kruskal58,taylor67,hth67,ctb81,parrathesis}. The ring-averaged time derivatives of both of these quantities, taken along a particle orbit, are first-order in $\epsilon$, namely
\begin{align}\label{eqn:edot}
\langle \dot{\overline{\varepsilon}}_s \rangle_{\gas} &= q_s \left( \pD{t}{} + \dupar \pD{z}{} \right) \langle\chi\rangle_{\gas} \sim \mc{O}(\epsilon\omega\varepsilon_{0s}), \\*
\label{eqn:mudot}
\langle \dot{\overline{\mu}}_s \rangle_{\gas} &= \frac{q_s}{B_0} \left( \pD{t}{} + v_\parallel \pD{z}{} \right) \langle\chi\rangle_{\gas} \sim \mc{O}(\epsilon\omega\mu_{0s}),
\end{align}
and thus are the same order as the gyrokinetic equation (see \S\ref{sec:gkfirstorder}).

In Section \ref{sec:boltz}, we stated without proof that the Boltzmann response (Eq.~\ref{eqn:boltzmann}) arises from the evolution of $f_{0s}$ under the influence of the perturbed electromagnetic fields. Using 
\begin{equation}\label{eqn:jacobian}
f^{\parallel}_{0s} = - \tprl{s} \pD{\varepsilon_{0s}}{f_{0s}} \qquad {\rm and} \qquad f^\perp_{0s} = - \tprp{s} \left( \frac{1}{B_0} \pD{\mu_{0s}}{f_{0s}} + \pD{\varepsilon_{0s}}{f_{0s}} \right) 
\end{equation}
to transform the $(v_\parallel,v_\perp)$-derivatives into $(\varepsilon_{0s},\mu_{0s})$-derivatives, the sum of the equilibrium distribution function and the Boltzmann response may be written in the following suggestive form:
\begin{align}\label{eqn:f0pf1B}
f_{0s}(\varepsilon_{0s},\mu_{0s}) + \delta f_{1s,{\rm Boltz}} &= f_{0s} + q_s \varphi'_s \,\pD{\varepsilon_{0s}}{f_{0s}} + \frac{q_s}{B_0} \left( \varphi - \frac{v_\parallel A_\parallel}{c} \right) \pD{\mu_{0s}}{f_{0s}} \nonumber\\*
\mbox{} &= f_{0s} + \varepsilon_{1s} \,\pD{\varepsilon_{0s}}{f_{0s}} + \mu_{1s} \, \pD{\mu_{0s}}{f_{0s}} \nonumber\\*
\mbox{} &\simeq f_{0s} ( \varepsilon_{0s} + \varepsilon_{1s} , \mu_{0s} + \mu_{1s} ) \equiv f_{0s} ( \overline{\varepsilon}_s , \overline{\mu}_s ),
\end{align}
i.e., it is the electromagnetically perturbed distribution function $f_{0s}(\overline{\varepsilon}_s,\overline{\mu}_s)$ Taylor expanded about $f_{0s}(\varepsilon_{0s},\mu_{0s})$ and written out to first order in the change in particle energy and the (gyrophase-dependent) change in the magnetic moment. Thus, the Boltzmann response does not change the form of the equilibrium distribution function if the latter is written as a function of sufficiently precisely conserved particle invariants.

Using Equation (\ref{eqn:f0pf1B}), we can absorb the Boltzmann response into $f_{0s}$ and write the total distribution function of species $s$ (Eq.~\ref{eqn:df1s}) as
\begin{equation}
f_s = f_{0s}(\overline{\varepsilon}_s ,\overline{\mu}_s) + h_s(t,\bb{R}_s,\overline{\varepsilon}_s,\overline{\mu}_s) + \delta f_{2s} + \dots .
\end{equation}
The gyro-averaged Vlasov equation at $\mc{O}(\epsilon \omega f_{0s})$ then becomes
\begin{equation}
\pD{t}{h_s} + \langle \dot{\bb{R}}_s \rangle_{\gas} \bcdot \pD{\bb{R}_s}{h_s} = - \langle \dot{\overline{\varepsilon}}_s \rangle_{\gas}  \pD{\overline{\varepsilon}_s}{f_{0s}} - \langle \dot{\overline{\mu}}_s \rangle_{\gas} \pD{\overline{\mu}_s}{f_{0s}} .
\end{equation}
Using Equations (\ref{eqn:edot}) and (\ref{eqn:mudot}) for the ring-averaged rates of change of $\overline{\varepsilon}_s$ and $\overline{\mu}_s$, respectively, it then becomes clear that our gyrokinetic Equation (\ref{eqn:gyrokinetic}) is simply the ring-averaged Vlasov equation, $\langle \dot{f}_s \rangle_{\gas} = 0$, written to lowest order in $\epsilon$.

It is often convenient and physically intuitive to subtract from $\overline{\mu}_s$ a first-order gyrophase-independent piece, namely
\begin{equation}\label{eqn:gkmu}
\mu_s \doteq \overline{\mu}_s - \frac{q_s}{B_0} \langle \chi \rangle_{\gas} .
\end{equation}
In doing so, we gain an order in the ring-averaged conservation property, $\langle \dot{\mu}_s \rangle_{\gas} \sim \mc{O} ( \epsilon^2 \omega \mu_{0s} )$, making $\mu_s$ the asymptotically conserved adiabatic invariant. The additional contribution to $\overline{\mu}_s$ in Equation (\ref{eqn:gkmu}) also ensures that, at long wavelengths, $\mu_s$ reduces to the adiabatic invariant defined by Equation (\ref{eqn:mu}):
\begin{align}
\mu_s &= \frac{m_s v^2_\perp}{2B_0} + \frac{q_s}{B_0} \left( \varphi - \langle\varphi\rangle_{\gas} - \frac{v_\parallel A_\parallel}{c} + \left\langle \frac{v_\parallel A_\parallel}{c}\right\rangle_{\ggs} \right) + \frac{q_s}{B_0} \left\langle \frac{\bb{v}_\perp\bcdot\bb{A}_\perp}{c} \right\rangle_{\ggs} \nonumber\\*
\mbox{} &\simeq \frac{m_s v^2_\perp}{2B_0} - \frac{q_s}{B_0} \frac{\bb{v}\btimes\ez}{\Omega_s} \bcdot \grad \left[ \varphi(\bb{r}) - \frac{v_\parallel A_\parallel(\bb{r})}{c} \right] - \frac{m_s v^2_\perp}{2B_0} \frac{\dBprl}{B_0} \nonumber\\*
\mbox{} &= \frac{m_s v^2_\perp}{2B_0} - \frac{m_s \bb{v}_\perp\bcdot\bb{u}_\perp}{B_0} - \frac{m_s v_\parallel \bb{v}_\perp\bcdot\dBprp}{B^2_0} - \frac{m_s v^2_\perp}{2B_0} \frac{\dBprl}{B_0} \nonumber\\*
\mbox{} &\simeq \frac{m_s \bigl| \bb{v} - \bb{u}_\perp - \bb{v} \bcdot \eb \eb \bigr|^2}{2B} \nonumber\\*
\mbox{} &\equiv \frac{m_s w^2_\perp}{2B} .
\end{align}
Similarly, we define the energy variable
\begin{equation}\label{eqn:gkenergy}
\varepsilon_s \doteq \overline{\varepsilon}_s - q_s \langle\varphi'_s\rangle_{\gas},
\end{equation}
which reduces in the long-wavelength limit to the energy variable defined by Equation (\ref{eqn:energy}):
\begin{align}
\varepsilon_s &= \frac{1}{2} m_s | \bb{v} - \dupar \ez |^2 + q_s \, \bigl( \varphi'_s - \langle \varphi'_s \rangle_{\gas} \bigr) \nonumber\\*
\mbox{} &\simeq \frac{1}{2} m_s v^2 - m_s \dupar v_z + \frac{1}{2} m_s \duparsq - q_s \frac{\bb{v}\btimes\ez}{\Omega_s} \bcdot \grad \left[ \varphi(\bb{r}) - \frac{\dupar A_\parallel(\bb{r})}{c} \right] \nonumber\\*
\mbox{} &=\frac{1}{2} m_s v^2 - m_s \dupar \bb{v} \bcdot \biggl( \eb - \frac{\delta\bb{B}_\perp}{B_0} \biggr) + \frac{1}{2} m_s \duparsq - m_s \bb{v}_\perp \bcdot \bb{u}_\perp - m_s \dupar \bb{v}_\perp\bcdot\frac{\delta\bb{B}_\perp}{B_0} \nonumber\\*
\mbox{} &\simeq \frac{1}{2} m_s \bigl( \bb{v}\bcdot\eb - \dupar \bigr)^2 + \frac{1}{2} m_s \bigl| \bb{v} - \bb{u}_\perp -\bb{v}\bcdot\eb\eb \bigr|^2 \nonumber\\*
\mbox{} &\equiv \frac{1}{2} m_s (v_\parallel - \dupar )^2 + \frac{1}{2} m_s w^2_\perp ,
\end{align}
i.e., it is the kinetic energy of the particle as measured in the frame moving with the $\dupar$ and $\bb{E}\btimes\bb{B}$ drifts. 

If we then write $f_s = \widetilde{f}_{0s}(\varepsilon_s,\mu_s) + \delta\widetilde{f}_s$, it is straightforward to show by using Equations (\ref{eqn:hstilde}), (\ref{eqn:f0pf1B}), (\ref{eqn:gkmu}), and (\ref{eqn:gkenergy}) that the following expressions are equivalent to leading order:
\begin{subequations}\label{eqn:gkdftilde}
\begin{align}
\delta \widetilde{f}_s &= \delta f_s + f_{0s}(\varepsilon_{0s},\mu_{0s}) - \widetilde{f}_{0s}(\varepsilon_s,\mu_s) \\*
\mbox{} &\simeq \delta f_s - q_s \,\bigl( \varphi'_s - \langle\varphi'_s\rangle_{\gas} \bigr) \, \pD{\varepsilon_{0s}}{f_{0s}} - \frac{q_s}{B_0} \left( \varphi - \frac{v_\parallel A_\parallel}{c} - \langle\chi\rangle_{\gas} \right) \pD{\mu_{0s}}{f_{0s}} \\*
\mbox{} &= \delta f_s - \bigl( \delta f_{1s,{\rm Boltz}} - \langle\delta f_{1s,{\rm Boltz}}\rangle_{\gas} \bigr) - \frac{q_s}{B_0} \left\langle \frac{\bb{v}_\perp\bcdot\bb{A}_\perp}{c} \right\rangle_{\ggs} \pD{\mu_{0s}}{f_{0s}} \label{eqn:gkdftilde3} \\*
\mbox{} &= h_s + \langle \delta f_{1s,{\rm Boltz}} \rangle_{\gas} - \frac{q_s}{B_0} \left\langle \frac{\bb{v}_\perp\bcdot\bb{A}_\perp}{c} \right\rangle_{\ggs} \pD{\mu_{0s}}{f_{0s}} \label{eqn:gkdftilde4} \\*
\mbox{} &= \widetilde{h}_s + q_s \langle \varphi'_s \rangle_{\gas} \pD{\varepsilon_{0s}}{f_{0s}} \label{eqn:gkdftilde5}.
\end{align}
\end{subequations}
Equation (\ref{eqn:gkdftilde}) is the $k_\perp \rho_i \sim 1$ generalisation of the perturbed distribution function $\delta\widetilde{f}_s$ defined by Equation (\ref{eqn:dftilde}), which prominently features in the generalised free energy (Eq.~\ref{eqn:W}) of KRMHD; i.e., it is the perturbed distribution function if $f_{0s}$ is taken to be a function of $(\varepsilon_s,\mu_s)$ instead of $(v_\parallel,w_\perp)$. 

At long wavelengths, the difference between the Boltzmann response and its ring average that features in Equation (\ref{eqn:gkdftilde3}) is
\begin{align}\label{eqn:gkboltzmann}
\delta f_{1s,{\rm Boltz}} - \langle \delta f_{1s,{\rm Boltz}} \rangle_{\gas} &\simeq \frac{q_s}{\tprp{s}} \frac{\bb{v}_\perp\btimes\ez}{\Omega_s}\bcdot\grad \left[ \varphi(\bb{r}) - \frac{v_\parallel A_\parallel(\bb{r})}{c} \right] f^\perp_{0s} \nonumber\\*
\mbox{}&\qquad\quad+ \frac{q_s}{\tprl{s}} \frac{\bb{v}_\perp\btimes\ez}{\Omega_s} \bcdot\grad \left[ \frac{v_\parallel A_\parallel(\bb{r})}{c} - \frac{\dupar A_\parallel(\bb{r})}{c} \right] f^\parallel_{0s} \nonumber\\*
\mbox{} &= - \bb{u}_\perp \bcdot \pD{\bb{v}_\perp}{f_{0s}} - \frac{\dBprp}{B_0}\bcdot\left[ v_\parallel \pD{\bb{v}_\perp}{f_{0s}} - \bb{v}_\perp \biggl( 1 - \frac{\dupar}{v_\parallel} \biggr) \pD{v_\parallel}{f_{0s}} \right],
\end{align}
where in the third line we have identified (cf.~Eq.~\ref{eqn:streamflux}) 
\begin{subequations}\label{eqn:gkstreamflux}
\begin{align}
\bb{u}_\perp &=  \frac{c}{B_0}  \ez \btimes \grad_{\!\perp} \varphi(\bb{r}) \doteq \ez \btimes \grad_{\!\perp} \Phi(\bb{r}) , \\*
\frac{\dBprp}{\sqrt{4\pi\rho_0}} &= - \frac{\valf}{B_0} \ez \btimes \grad_{\!\perp} A_\parallel (\bb{r}) \doteq \ez \btimes \grad_{\!\perp} \Psi (\bb{r}) .
\end{align}
\end{subequations}
Comparing Equations (\ref{eqn:gkdftilde3}) and (\ref{eqn:gkdftilde4}), we see that the Alfv\'{e}nic fluctuations comprise the piece of the gyrokinetic response $h_s$ that is cancelled at long wavelengths by the Boltzmann response. Indeed, by substituting Equations (\ref{eqn:gkboltzmann}) and (\ref{eqn:gkstreamflux}) into the expression for the full distribution function, $f_s = f_{0s} + \delta f_{1s,{\rm Boltz}} + h_s + \dots$, we find that we may absorb the Alfv\'{e}nic fluctuations into the equilibrium distribution:
\begin{equation}\label{eqn:gkalfven}
f_{0s} \Biggl(\frac{1}{2} m_s \bigl| \bb{v} - \dupar \ez \bigr|^2 , \frac{m_sv^2_\perp}{2B_0} \Biggr) \longrightarrow f_{0s} \Biggl( \frac{1}{2} m_s \bigl(v_\parallel - \dupar\bigr)^2 + \frac{1}{2} m_s w^2_\perp , \frac{m_s w^2_\perp}{2B_0} \Biggr) .
\end{equation}
In other words, Alfv\'{e}nic fluctuations do not change the form of the distribution function, but rather define the moving frame in which any changes to it are to be measured. Physically, this is because particles in a magnetised plasma adjust on a cyclotron timescale to take on the local $\bb{E}\btimes\bb{B}$ velocity. This principle is what underlies Kulsrud's formulation of KMHD, in which the perpendicular particle velocities are measured relative to the $\bb{E}\btimes\bb{B}$ drift, the latter being governed by a set of MHD-like fluid equations rather than a kinetic equation. The implication---that discussed at length in the main text of this Paper (\S\ref{sec:reduceddke}, in particular)---is that the compressive component of the turbulence is passively advected by the Alfv\'{e}nic fluctuations in the inertial range.

\subsection{A Tactical Step: From Rings to Gyrocentres}\label{app:htog}

Considering the content of Equation (\ref{eqn:gkalfven})---that the Alfv\'{e}nic fluctuations have a gyrokinetic response that is largely cancelled at long wavelengths by the Boltzmann response---it is often advantageous to not work directly with $h_s$. This choice is made by most (Maxwellian) $\delta f$ gyrokinetic codes (e.g., {\it Astro-GK}; \citealt{numata10}), which avoid the numerical error arising from this near-cancellation by working instead with the perturbed distribution function
\begin{align*}
g_s &= h_s - \frac{q_s}{T_{0s}} \left\langle \varphi - \frac{\bb{v}_\perp\bcdot\bb{A}_\perp}{c} \right\rangle_{\ggs} f_{0s} \\*
\mbox{} &= h_s + \langle \delta f_{1s,{\rm Boltz}} \rangle_{\gas} + \frac{q}{T_{0s}} \left\langle \frac{\bb{v}_\perp\bcdot\bb{A}_\perp}{c}\right\rangle_{\ggs} f_{0s} \\*
\mbox{} & = \langle \delta f_{1s} \rangle_{\gas} + \frac{q}{T_{0s}} \left\langle \frac{\bb{v}_\perp\bcdot\bb{A}_\perp}{c}\right\rangle_{\ggs} f_{0s} .
\end{align*}
In fact, many standard treatments of gyrokinetics use $g_s$ instead of $h_s$. In the electrostatic limit, the use of $g_s$ (which, in this limit, equals $\langle \delta f_{1s}\rangle_{\gas}$) aids in the interpretation of polarisation effects within gyrokinetics \citep{krommes12}, places the gyrokinetic equation in a numerically convenient characteristic form \citep{lee83}, and arises naturally from the Hamiltonian formulation of gyrokinetics \citep{dubin83,brizard00}.

We follow these practices and introduce, for non-Maxwellian $f_{0s}$,
\begin{subequations}\label{eqn:gkgs}
\begin{align}
g_s &\doteq  \widetilde{h}_s - \frac{q_s}{\tprl{s}} \left\langle \varphi'_s - \frac{\bb{v}_\perp\bcdot\bb{A}_\perp}{c} \right\rangle_{\ggs} f^\parallel_{0s}  \\*
\mbox{}& = h_s + \langle \delta f_{1s,{\rm Boltz}} \rangle_{\gas} + \frac{q_s}{\tprp{s}} \left\langle \frac{\bb{v}_\perp\bcdot\bb{A}_\perp}{c} \right\rangle_{\ggs} f^\perp_{0s}, \\*
\mbox{}& = \langle \delta f_{1s} \rangle_{\gas} + \frac{q_s}{\tprp{s}} \left\langle \frac{\bb{v}_\perp \bcdot \bb{A}_\perp}{c} \right\rangle_\ggs f^\perp_{0s} ,
\end{align}
\end{subequations}
which is the $k_\perp \rho_s \sim 1$ generalisation of the perturbed distribution function $g_s$ defined in KRMHD (cf.~Eq.~\ref{eqn:gs}). 
The physical distinction between the two formulations of gyrokinetics---one expressed in terms of $h_s$, the other in terms of $g_s$---is as follows. Written in terms of $h_s$, the set of gyrokinetic-Maxwell equations describes the dynamics of physically extended rings of charge as they move in a vacuum. If the fluctuating plasma is instead described by $g_s$, these equations describe a gas of point-particle-like gyrocentres moving in a polarizable medium \citep[see][]{krommes93a,krommes12,abel13}. Hence, we refer to $g_s$ as the {\em gyrocentre distribution function}. 

Using Equation (\ref{eqn:gkgs}) to replace $h_s$ in the gyrokinetic equation (\ref{eqn:gyrokinetic}), we find that $g_s$ evolves according to
\begin{eqnarray}\label{eqn:ggyrokinetic}
\pD{t}{g_s} + v_\parallel \pD{z}{g_s} + \frac{c}{B_0} \{ \langle \chi \rangle_{\gas} \, , g_s \} = -\frac{q_s}{\tprl{s}}  \bigl( v_\parallel - \dupar \bigr)  \left\langle \frac{1}{B_0} \{ A_\parallel , \varphi - \langle \varphi \rangle_{\gas} \} \right.
\nonumber\\*
\mbox{} + \left. \frac{1}{c} \pD{t}{A_\parallel} + \eb \bcdot \grad \varphi - \eb \bcdot \grad \left\langle \frac{\bb{v}_\perp\bcdot\bb{A}_\perp}{c} \right\rangle_{\ggs} \right\rangle_{\ggs} f^\parallel_{0s} ,
\end{eqnarray}
where
\begin{equation}\label{eqn:gkbdotgrad}
\eb \bcdot \grad = \pD{z}{} + \frac{\dBprp}{B_0} \bcdot \grad_{\!\perp} = \pD{z}{} - \frac{1}{B_0} \left\{ A_\parallel \, , \dots \right\}
\end{equation}
is the spatial derivative along the perturbed magnetic field (cf.~Eq.~\ref{eqn:bdotgrad}).\footnote{When taking the gradient of a ring-averaged function, e.g. the  term $\eb\bcdot\grad \langle \bb{v}_\perp\bcdot\bb{A}_\perp \rangle_{\gas}$ in Equation (\ref{eqn:ggyrokinetic}), the velocity variable $\bb{v}$ is held fixed.} As in S09, we have used compact notation in writing out the nonlinear terms: $\langle \{ A_\parallel \, , \varphi - \langle \varphi \rangle_{\gas} \} \rangle_{\gas} = \langle \{ A_\parallel (\bb{r}) \, , \varphi (\bb{r}) \} \rangle_{\gas} - \{ \langle A_\parallel \rangle_{\gas} \, , \langle \varphi \rangle_{\gas} \}$, where the first Poisson bracket involves derivatives with respect to $\bb{r}$ and the second with respect to $\bb{R}_s$.

The field equations (\ref{eqn:gkqn}), (\ref{eqn:gkprlamp}), and (\ref{eqn:gkpbalance}) are best written in Fourier space.

\subsection{Fourier Space}\label{sec:fourier}

The ring-averaging (Eq.~\ref{eqn:ringaverage}) and gyro-averaging (Eq.~\ref{eqn:gyroaverage}) procedures take on a rather compact form when expressed in Fourier space. First consider the gyrokinetic potential, decomposed into plane waves: $\chi(t,\bb{r},\bb{v}) = \sum_{\bs{k}} \chi_{\bs{k}}(t,\bb{v}) \exp( \imag \bb{k} \bcdot \bb{r} )$. The ring average of its Fourier coefficient is
\begin{align}\label{eqn:rachik}
\langle \chi_{\bs{k}}(t,\bb{v}) \rangle_{\gas} &= \frac{1}{2\pi} \oint {\rm d}\vartheta \left( \varphi_{\bs{k}} - \frac{v_\parallel \Aprlk}{c} - \frac{\bb{v}_\perp \bcdot \bb{A}_{\perp\bs{k}}}{c} \right) \exp\left( -\imag \bb{k}\bcdot \frac{\bb{v}_\perp\btimes\ez}{\Omega_s} \right) \nonumber\\*
\mbox{} &=  J_0(a_s) \left( \varphi_{\bs{k}} - \frac{v_\parallel \Aprlk}{c} \right) + \frac{T_{\perp 0s}}{q_s} \frac{2v^2_\perp}{\vthprp{s}^2} \frac{J_1(a_s)}{a_s} \frac{\dBprlk}{B_0} ,
\end{align}
where $a_s \doteq k_\perp v_\perp / \Omega_s$ and we have used the definition $\dBprlk = \ez \bcdot ( \imag \bb{k} \btimes \bb{A}_{\perp\bs{k}} )$. Simply put, ring averaging amounts to multiplication by either the zeroth- ($J_0$) or first-order ($J_1$) Bessel function, depending on whether the additional angular dependence of $\bb{v}_\perp$ appears in the integrand. Similarly, we can Fourier decompose $g_s(t,\bb{R}_s,v_\parallel,v_\perp) = \sum_{\bs{k}} g_{s\bs{k}}(t,v_\parallel,v_\perp) \exp( \imag \bb{k} \bcdot \bb{R}_s )$ and perform its gyro-averages at constant $\bb{r}$, viz.
\begin{equation}\label{eqn:gah}
\langle g_{s\bs{k}}(t,v_\parallel,v_\perp) \rangle_{\bs{r}} =  \frac{1}{2\pi} \oint {\rm d}\vartheta \, g_{s\bs{k}} \exp\left( \imag \bb{k}\bcdot \frac{\bb{v}_\perp\btimes\ez}{\Omega_s} \right) = J_0(a_s) g_{s\bs{k}} ,
\end{equation}
\begin{equation}\label{eqn:gavh}
\langle \bb{v}_\perp g_{s\bs{k}} (t,v_\parallel,v_\perp) \rangle_{\bs{r}} = \frac{1}{2\pi} \oint {\rm d}\vartheta \, \bb{v}_\perp g_{s\bs{k}} \exp\left( \imag \bb{k}\bcdot \frac{\bb{v}_\perp \btimes\ez}{\Omega_s} \right) = -  \imag \bb{k} \btimes \ez \, \frac{v^2_\perp}{\Omega_s} \frac{J_1(a_s)}{a_s} g_{s\bs{k}} ,
\end{equation}
\begin{align}\label{eqn:gavvh}
\langle \bb{v}_\perp \bb{v}_\perp g_{s\bs{k}} (t,v_\parallel,v_\perp) \rangle_{\bs{r}} &= \frac{1}{2\pi} \oint {\rm d}\vartheta \, \bb{v}_\perp \bb{v}_\perp g_{s\bs{k}} \exp\left( \imag \bb{k}\bcdot\frac{\bb{v}_\perp\btimes\ez}{\Omega_s}\right) \nonumber\\*
\mbox{} &= v^2_\perp \left[ \frac{\bb{k}_\perp\bb{k}_\perp}{k^2_\perp} \frac{J_1(a_s)}{a_s} + \frac{(\bb{k}\btimes\ez)(\bb{k}\btimes\ez)}{k^2_\perp} \D{a_s}{J_1(a_s)} \right] g_{s\bs{k}} .
\end{align}
These operations transform integro-differential equations in real space into integro-algebraic equations in Fourier space.

\subsection{Field Equations}

In order to write the field equations in terms of $g_s$ concisely, we first define several $v_\parallel$-, $v_\perp$-, and Bessel-function--weighted integrals over the equilibrium distribution function, suitably normalised:
\begin{subequations}\label{eqn:gammas}
\begin{align}
\Gamma_{00} (\alpha_s) & \doteq \frac{1}{\nsp} \int {\rm d}^3\bb{v} \, \bigl[ J_0(a_s) \bigr]^2 \, f_{0s} = 1 - \alpha_s + \dots \\*
\Gamma^\perp_{00} (\alpha_s) & \doteq \frac{1}{\nsp} \int {\rm d}^3\bb{v} \, \bigl[ J_0(a_s) \bigr]^2 \, f^\perp_{0s} = \kzero{s} - \alpha_s + \dots \\*
\Gamma^\perp_{01} (\alpha_s) & \doteq \frac{1}{\nsp} \int {\rm d}^3\bb{v} \, \bigl[ J_0(a_s) \bigr]^2 \left(\frac{v_\parallel}{\vthprl{s}}\right)  f^\perp_{0s} \times \left( \frac{\dupar}{\vthprl{s}} \right)^{-1}= \kone{s} - \alpha_s + \dots \\*
\Gamma^\perp_{02} (\alpha_s) & \doteq \frac{1}{\nsp} \int {\rm d}^3\bb{v} \, \bigl[ J_0(a_s) \bigr]^2 \left(\frac{v_\parallel}{\vthprl{s}}\right)^2 f^\perp_{0s} \times \left( \frac{1}{2} + \frac{\duparsq}{\vthprl{s}^2} \right)^{-1} = \ktwo{s} - \alpha_s + \dots \\*
\Gamma^\perp_{10} (\alpha_s) & \doteq \frac{1}{\nsp} \int {\rm d}^3\bb{v} \, \frac{v^2_\perp}{\vthprp{s}^2} \frac{2J_0(a_s) J_1(a_s)}{a_s} \, f^\perp_{0s} = 1 - \frac{3}{2} \alpha_s + \dots \\*
\Gamma^\perp_{11} (\alpha_s) & \doteq \frac{1}{\nsp} \int {\rm d}^3\bb{v} \, \frac{v^2_\perp}{\vthprp{s}^2} \frac{2J_0(a_s) J_1(a_s)}{a_s} \left(\frac{v_\parallel}{\vthprl{s}}\right) f^\perp_{0s} \times \left( \frac{\dupar}{\vthprl{s}} \right)^{-1} = 1 - \frac{3}{2} \alpha_s C_{11s} + \dots \\*
\Gamma^\perp_{20} (\alpha_s) & \doteq \frac{1}{\nsp} \int {\rm d}^3\bb{v} \left[ \frac{2v^2_\perp}{\vthprp{s}^2} \frac{J_1(a_s)}{a_s} \right]^2 f^\perp_{0s} = 2 \left( 1 - \frac{3}{2} \alpha_s C_{20s} + \dots \right),
\end{align}
\end{subequations}
where $\alpha_s \doteq k^2_\perp \rho^2_s / 2$. In doing so, we have introduced two additional coefficients, given by
\begin{equation}\label{eqn:c11c20}
C_{11s} \doteq \frac{1}{\nsp} \int {\rm d}^3\bb{v} \, \frac{v^2_\perp}{\vthprp{s}^2} \frac{v_\parallel}{\dupar} f_{0s}  \qquad {\rm and} \qquad C_{20s} \doteq \frac{1}{\nsp} \int {\rm d}^3\bb{v} \, \frac{1}{2} \frac{v^4_\perp}{\vthprp{s}^4} f_{0s},
\end{equation}
both of which equate to unity for a drifting bi-Maxwellian (Eq.~\ref{eqn:biMax}).\footnote{If there are no interspecies parallel drifts and if $f_{0s}$ and $f^\perp_{0s}$ are symmetric about $v_\parallel = \dupar$, then then $\dupar C_{11s} = \dupar\Gamma^\perp_{\ell 1}(\alpha_s) =0$.} Note the numbering scheme used for the $\Gamma_{\ell m}$ subscripts, which reflects the number of powers $\ell$ of $v^2_\perp$ and $m$ of $v_\parallel$ in the integrand.

With these definitions in hand, and working in the Fourier domain, we can express Equations (\ref{eqn:gkqn}), (\ref{eqn:gkprlamp}), and (\ref{eqn:gkpbalance}), which represent respectively the quasineutrality constraint and the parallel and perpendicular components of Amp\`{e}re's law, as follows:
\begin{align}\label{eqn:ggkqn}
\sum_s q_s \int {\rm d}^3\bb{v} \, J_0(a_s) g_{s\bs{k}} &= \sum_s \frac{q^2_s \nsp \varphi_{\bs{k}}}{\tprp{s}} \left[ \kzero{s} - \Gamma^\perp_{00}(\alpha_s) \right] - \sum_s q_s \nsp \Gamma^\perp_{10}(\alpha_s) \frac{\dBprlk}{B_0} \nonumber\\*\mbox{} &- \sum_s \frac{q^2_s \nsp \dupar \Aprlk}{c\tprp{s}} \left[ \kone{s} - \Gamma^\perp_{01}(\alpha_s) \right] ,
\end{align}
\begin{align}\label{eqn:ggkprlamp}
\sum_s q_s \int {\rm d}^3\bb{v} \, v_\parallel J_0(a_s) g_{s\bs{k}} &= \sum_s \frac{q^2_s\nsp\dupar\varphi_{\bs{k}}}{\tprp{s}} \left[ \kone{s} - \Gamma^\perp_{01}(\alpha_s)\right] - \sum_s q_s \nsp \dupar \Gamma^\perp_{11}(\alpha_s) \frac{\dBprlk}{B_0} \nonumber\\*\mbox{} &+ \left\{ \frac{c^2k^2_\perp}{4\pi} + \sum_s \frac{q^2_s\nsp}{m_s} \bigl[ 1 - \Gamma_{00}(\alpha_s) \bigr] \right. \nonumber\\*\mbox{} &- \left. \sum_s \frac{q^2_s\nsp}{m_s} \frac{\tprl{s}}{\tprp{s}} \left( 1 + \frac{2\duparsq}{\vthprl{s}^2} \right) \left[ \ktwo{s} - \Gamma^\perp_{02}(\alpha_s) \right] \right\} \frac{\Aprlk}{c} ,
\end{align}
\begin{align}\label{eqn:ggkprpamp}
\sum_s \frac{\betaprp{s}}{\nsp} \int {\rm d}^3\bb{v} \, \frac{2v^2_\perp}{\vthprp{s}^2} \frac{J_1(a_s)}{a_s} g_{s\bs{k}} &= - \sum_s \betaprp{s} \frac{q_s \varphi_{\bs{k}}}{\tprp{s}} \Gamma^\perp_{10}(\alpha_s) - \left[ 2 + \sum_s \betaprp{s} \Gamma^\perp_{20}(\alpha_s) \right] \frac{\dBprlk}{B_0} \nonumber\\*\mbox{} &+ \sum_s \betaprp{s} \frac{q_s\dupar \Aprlk}{c \tprp{s}} \Gamma^\perp_{11}(\alpha_s) .
\end{align}
Note that $\varphi_{\bs{k}}$ always enters the field equations in combination with $-\dupar \Aprlk / c$ (cf.~Eq.~\ref{eqn:varphiprime}).

\subsection{Massless Electron Fluid}\label{sec:gkelectrons}

In this Section, we carry out an expansion of the electron gyrokinetic equation in powers of $( m_e / m_i )^{1/2} \simeq 0.02$ (for hydrogen plasma). This expansion is done while still considering $\betaprl{s}$, $\betaprp{s}$, $\tauprl{s}$, $\tauprp{s}$, and $k_\perp \rho_i$ to be order unity, so that $k_\perp \rho_e \sim k_\perp \rho_i \, ( m_e / m_i )^{1/2} \ll 1$. Then we can expand the Bessel functions arising from averaging over the electron ring motion,
\begin{equation}
J_0(a_e) = 1 - \frac{1}{4} a^2_e + \dots , \quad \frac{2 J_1(a_e)}{a_e} = 1 - \frac{1}{8} a^2_e + \dots ,
\end{equation}
and evaluate the fields $\varphi$, $A_\parallel$, and $\dBprl$ at $\bb{r} = \bb{R}_e$. The electron kinetic equation, accurate up to and including the first order in $(m_e / m_i)^{1/2}$, then reads
\begin{align}\label{eqn:gkelectron}
\order{1}{\pD{t}{g_e}} + \order{0}{v_\parallel \pD{z}{g_e}} &+ \frac{c}{B_0} \biggl\{ \order{1}{\varphi} - \order{0}{\frac{v_\parallel A_\parallel}{c}} - \order{1}{\frac{\tprp{e}}{e} \frac{v^2_\perp}{v^2_{{\rm th}\perp e}} \frac{\dBprl}{B_0}} , \, g_e \biggr\}
\nonumber\\*
\mbox{} &= \frac{e}{\tprl{e}} \, v_\parallel \biggl( \order{0}{\frac{1}{c} \pD{t}{A_\parallel}} + \order{0}{\eb \bcdot \grad \varphi} - \order{0}{\frac{\tprp{e}}{e}\, \eb \bcdot \grad \frac{v^2_\perp}{\vthprp{e}^2} \frac{\dBprl}{B_0}} \biggr)f^\parallel_{0e} \nonumber\\*
\mbox{} &- \frac{e}{\tprl{e}} \, \dupare \biggl(  \order{1}{\frac{1}{c} \pD{t}{A_\parallel}} + \order{1}{\eb \bcdot \grad \varphi} - \order{1}{\frac{\tprp{e}}{e}\, \eb \bcdot \grad \frac{v^2_\perp}{\vthprp{e}^2} \frac{\dBprl}{B_0}} \biggr) f^\parallel_{0e} ,
\end{align}
where we have indicated underneath each term the lowest order to which that term enters when compared with $v_\parallel \partial \df{e} / \partial z$. We refer the reader to \S 4.1 and, in particular, equations (80)--(82) in S09 for details on obtaining this ordering. 

We expand $g_e = g^{(0)}_e + g^{(1)}_e + \dots$ in powers of $(m_e/m_i)^{1/2}$ and carry out the expansion to the first two orders.

\subsubsection{Zeroth Order}

To zeroth order, the electron kinetic equation is
\begin{equation}\label{eqn:gkelectron0}
\eb\bcdot\grad \left( g^{(0)}_e + \frac{v^2_\perp}{\vthprl{e}^2} \frac{\dBprl}{B_0} f^\parallel_{0e} \right) - \frac{e}{\tprl{e}} \left( \frac{1}{c} \pD{t}{A_\parallel} + \eb \bcdot \grad \varphi \right) f^\parallel_{0e} = 0.
\end{equation}
Equation (\ref{eqn:gkelectron0}) is identical to the electron kinetic equation obtained from KRMHD (Eq.~\ref{eqn:eleckinetic}), as it should be (since $a_e \ll 1$). Multiplying Equation (\ref{eqn:gkelectron0}) by $\tprl{e} / e\nem$ and integrating over the velocity space, we obtain
\begin{equation}\label{eqn:gkefield}
\frac{1}{c} \pD{t}{A_\parallel} + \eb \bcdot \grad \varphi = \frac{1}{\czero{e}} \, \eb \bcdot \grad\,  \frac{\tprl{e}}{e}\! \left( \frac{\dne}{\nem} + \Delta_{1e} \frac{\dBprl}{B_0} \right) ,
\end{equation}
which matches the expression for (minus) the parallel electric field given by Equation (\ref{eqn:efield}). Substituting Equation (\ref{eqn:gkefield}) back into Equation (\ref{eqn:gkelectron0}) gives
\begin{equation}\label{eqn:g0e}
g^{(0)}_e = \left[ \frac{1}{\czero{e}} \left( \frac{\dne}{\nem} + \Delta_{1e} \frac{\dBprl}{B_0} \right) - \frac{v^2_\perp}{\vthprl{e}^2} \frac{\dBprl}{B_0} \right] f^\parallel_{0e} ,
\end{equation}
from which follows the equations of state for the electrons, Equations (\ref{eqn:pprle}) and (\ref{eqn:pprpe}). 

\subsubsection{First Order}

At first order, Equation (\ref{eqn:gkelectron}) reads
\begin{align}
\pD{t}{g^{(0)}_e} + v_\parallel \eb \bcdot \grad g^{(1)}_e &+ \frac{c}{B_0} \left\{ \varphi - \frac{\tprp{e}}{e} \frac{v^2_\perp}{\vthprp{e}^2} \frac{\dBprl}{B_0} \, , g^{(0)}_e \right\} 
\\*&\mbox{} = - \frac{e}{\tprl{e}}  \,\dupare \left( \frac{1}{c} \pD{t}{A_\parallel} + \eb\bcdot\grad \varphi - \frac{\tprp{e}}{e} \eb\bcdot\grad \frac{v^2_\perp}{\vthprp{e}^2} \frac{\dBprl}{B_0} \right) f^\parallel_{0e} .\nonumber
\end{align}
Using Equations (\ref{eqn:gkefield}) and (\ref{eqn:g0e}) and integrating Equation (\ref{eqn:gkelectron}) over the velocity space, we find
\begin{align}\label{eqn:gkelectron1}
\left( \pD{t}{} + \dupare \eb\bcdot\grad \right) \left( \frac{\dne}{\nem} - \frac{\dBprl}{B_0} \right) + \frac{c}{B_0} \left\{ \varphi \, , \frac{\dne}{\nem} - \frac{\dBprl}{B_0} \right\} + \eb \bcdot \grad u_{\parallel e} \nonumber\\*\mbox{} + \frac{\cone{e}}{\czero{e}} \frac{c\tprp{e}}{eB_0} \left\{ \frac{\dne}{\nem} \, , \frac{\dBprl}{B_0} \right\} = 0 ,
\end{align}
where the parallel electron velocity is first order:
\begin{equation}\label{eqn:g1e}
u_{\parallel e} = u^{(1)}_{\parallel e} = \frac{1}{\nem} \int {\rm d}^3\bb{v} \, v_\parallel g^{(1)}_e .
\end{equation}
The first two terms in Equation (\ref{eqn:gkelectron1}) can be combined upon identifying
\begin{equation}\label{eqn:gkdbydt}
\D{t}{} = \pD{t}{} + \bb{u}_E \bcdot \grad = \pD{t}{} + \frac{c}{B_0} \{ \varphi \, , \dots \}
\end{equation}
as the Lagrangian time derivative measured in a frame transported at the $\bb{E} \btimes \bb{B}$ drift velocity, $\bb{u}_E = -c \grad_{\!\perp} \varphi \btimes \ez / B_0$ (cf.~Eq.~\ref{eqn:dbydt}):
\begin{equation}\label{eqn:gkinduction}
\left( \D{t}{} + \dupare \eb\bcdot\grad \right) \left( \frac{\dne}{\nem} - \frac{\dBprl}{B_0} \right) + \eb \bcdot \grad u_{\parallel e} + \frac{\cone{e}}{\czero{e}} \frac{c\tprp{e}}{eB_0} \left\{ \frac{\dne}{\nem} \, , \frac{\dBprl}{B_0} \right\} = 0 .
\end{equation}
As we will show in the following section, Equation (\ref{eqn:gkinduction}) indicates that, at first order in the mass-ratio expansion, the magnetic flux is {\em not} tied to the $\bb{E}\btimes\bb{B}$ flow. This marks a departure from KRMHD, where $k_\perp \rho_i$ is effectively zero at all relevant orders. However, Equation (\ref{eqn:gkinduction}) does not signal the breakdown of magnetic-flux conservation.

\subsubsection{Magnetic-Flux Conservation}

While it may not be readily apparent, Equations (\ref{eqn:gkefield}) and (\ref{eqn:gkinduction}) state that the magnetic flux is frozen into the electron flow velocity and is therefore exactly conserved in that frame. To see that this is indeed the case, we begin by combining Equation (\ref{eqn:gkefield}) for the parallel electric field with the lowest-order expression for the perpendicular electric field $-\grad_{\!\perp} \varphi'_e$ associated with the Alfv\'{e}nic fluctuations measured in the frame of the drifting electrons to obtain the total electric field,
\begin{equation}
\bb{E} = ( \msb{I} - \eb \eb ) \bcdot \left[ - \grad \varphi'_e + \frac{1}{\czero{e}} \grad \frac{\tprl{e}}{e} \left( \frac{\dne}{\nem} + \Delta_{1e} \frac{\dBprl}{B_0} \right) \right] - \frac{1}{\czero{e}} \grad \frac{\tprl{e}}{e} \left( \frac{\dne}{\nem} + \Delta_{1e} \frac{\dBprl}{B_0} \right) .
\end{equation}
Then Faraday's law becomes
\begin{equation}\label{eqn:fluxfreezinginelectrons}
\pD{t}{\bb{B}} = - c \grad \btimes \bb{E} = \grad \btimes \bigl( \bb{u}_{\rm eff} \btimes \bb{B} \bigr) ,
\end{equation}
with
\begin{equation}
\bb{u}_{\rm eff} = \eb \btimes \grad \frac{c}{B} \left[ \, \varphi'_e - \frac{1}{\czero{e}} \frac{\tprl{e}}{e} \left( \frac{\dne}{\nem} + \Delta_{1e} \frac{\dBprl}{B_0} \right) \right]
\end{equation}
being the effective velocity into which the magnetic flux is frozen. Therefore, there is a frame in which the magnetic flux is exactly conserved. We now show that this frame is associated with the electron flow velocity. 

We use Equation (\ref{eqn:gkbdotgrad}) to rewrite Equation (\ref{eqn:gkinduction}) in the following equivalent form:
\begin{equation}\label{eqn:fluxfreezing1}
\left( \pD{t}{} + \dupare \pD{z}{} \right) \left( \frac{\dBprl}{B_0} - \frac{\dne}{\nem} \right) + \frac{c}{B_0} \left\{ \varphi'_e - \frac{\cone{e}}{\czero{e}} \frac{\tprp{e}}{e} \frac{\dBprl}{B_0} , \frac{\dBprl}{B_0} - \frac{\dne}{\nem} \right\} =  \eb \bcdot \grad u_{\parallel e} .
\end{equation}
To the left-hand side of this equation we add zero, written in a rather auspicious guise:
\[
\frac{c}{B_0} \left\{ \frac{1}{\czero{e}} \frac{\tprl{e}}{e} \left( \frac{\dBprl}{B_0} - \frac{\dne}{\nem} \right) , \frac{\dBprl}{B_0} - \frac{\dne}{\nem} \right\} .
\]
Equation (\ref{eqn:fluxfreezing1}) then becomes
\begin{align}\label{eqn:fluxfreezing2}
\left( \pD{t}{} + \dupare \pD{z}{} \right) &\left( \frac{\dBprl}{B_0} - \frac{\dne}{\nem} \right) 
\\*&\mbox{} + \frac{c}{B_0} \left\{ \varphi'_e - \frac{1}{\czero{e}} \frac{\tprl{e}}{e} \left( \frac{\dne}{\nem} + \Delta_{1e} \frac{\dBprl}{B_0} \right) , \frac{\dBprl}{B_0} - \frac{\dne}{\nem} \right\} =  \eb \bcdot \grad u_{\parallel e} , \nonumber
\end{align}
and we may identify the first term in the Poisson bracket as the perpendicular component of the electron flow velocity,
\begin{equation}\label{eqn:uperpe}
\bb{u}_{\perp e} = \ez \btimes \grad_{\!\perp} \frac{c}{B_0} \left[ \, \varphi'_e - \frac{1}{\czero{e}} \frac{\tprl{e}}{e} \left( \frac{\dne}{\nem} + \Delta_{1e} \frac{\dBprl}{B_0} \right) \right] .
\end{equation}
It is clear that the leading-order contribution to $\bb{u}_{\rm eff}$ is precisely that given by Equation (\ref{eqn:uperpe}). This identification made, we can interpret Equation (\ref{eqn:gkinduction}) as the reduced electron continuity equation. To wit, Equation (\ref{eqn:fluxfreezing2}) expressed in terms of $\bb{u}_{\perp e}$,
\begin{equation}
\left( \pD{t}{} + \dupare \pD{z}{} + \bb{u}_{\perp e} \bcdot \grad_{\!\perp} \right) \left( \frac{\dBprl}{B_0} - \frac{\dne}{\nem} \right) = \eb\bcdot\grad u_{\parallel e}
\end{equation}
may be combined with the parallel component of the induction equation (\ref{eqn:fluxfreezinginelectrons}) expanded to $\mc{O}(\epsilon \omega B_0)$,
\begin{equation}\label{eqn:fluxfreezing3}
\left( \pD{t}{} + \dupare\pD{z}{} + \bb{u}_{\perp e} \bcdot \grad_{\!\perp} \right) \frac{\dBprl}{B_0} = \eb \bcdot \grad u_{\parallel e} - \grad \bcdot \bb{u}_e ,
\end{equation}
to find
\begin{equation}
\left( \pD{t}{} + \dupare \pD{z}{} + \bb{u}_{\perp e} \bcdot \grad_{\!\perp} \right) \frac{\dne}{\nem} = -\grad \bcdot \bb{u}_e ,
\end{equation}
the reduced electron continuity equation.

\subsubsection{Field Equations}

The two fluid-like Equations (\ref{eqn:gkefield}) and (\ref{eqn:gkinduction}) form the system that describes the electrons. These are closed by the gyrokinetic equation for $g_i$ (Eq.~\ref{eqn:ggyrokinetic} with $s=i$), and by the three integral relations derived from quasineutrality (Eq.~\ref{eqn:ggkqn}) and the parallel (Eq.~\ref{eqn:ggkprlamp}) and perpendicular (Eq.~\ref{eqn:ggkprpamp}) components of Amp\`{e}re's law. We now express the latter explicitly in terms of $\dne$, $u_{\parallel e}$, $\varphi$, $A_\parallel$, $\dBprl$, and $g_i$.

Expanding the summation over species and using Equations (\ref{eqn:g0e}) and (\ref{eqn:g1e}) to compute the velocity-space moments of $g_e$, Equations (\ref{eqn:ggkqn}), (\ref{eqn:ggkprlamp}), and (\ref{eqn:ggkprpamp}) become, respectively,
\begin{align}\label{eqn:gkqnk}
\order{0}{\frac{\delta n_{e\bs{k}}}{\nem}} &- \order{0}{\sum_i c_i \Gamma^\perp_{10}(\alpha_i) \frac{\dBprlk}{B_0}} + \order{1}{\sum_i c_i \left[ \kzero{i} - \Gamma^\perp_{00}(\alpha_i) \right] \frac{Z_i e \varphi_{\bs{k}}}{\tprp{i}}}
\nonumber\\*
\mbox{} &- \order{1}{\sum_i c_i \left[ \kone{i} - \Gamma^\perp_{01}(\alpha_i) \right] \frac{Z_i e}{\tprp{i}} \frac{\dupari \Aprlk}{c}} =  \order{0}{\sum_i \frac{c_i}{\nip} \int {\rm d}^3\bb{v}\, J_0(a_i) g_{i\bs{k}}} ,
\end{align}
\begin{align}\label{eqn:gkprlampk}
\Biggl\{ \order{1}{\frac{k^2_\perp c B_0}{4\pi e \nem}} &+ \order{1}{\sum_s c_s \Omega_s \bigl[ 1 - \Gamma_{00}(\alpha_s) \bigr] } - \order{1}{\sum_s c_s \Omega_s \frac{\tprl{s}}{\tprp{s}} \left( 1 + \frac{2\duparsq}{\vthprl{s}^2} \right) \left[ C^\perp_{2s} - \Gamma^\perp_{02}(\alpha_s) \right] } \Biggr\} \frac{\Aprlk}{B_0}
\nonumber\\*
&\mbox{} + \order{1}{\sum_i c_i \dupari \left[ C^\perp_{1i} - \Gamma^\perp_{01}(\alpha_i) \right] \frac{Z_i e \varphi_{\bs{k}}}{\tprp{i}}} + \order{2}{ \sum_i c_i \dupari \left[ 1 - \Gamma^\perp_{11}(\alpha_i) \right] \frac{\dBprlk}{B_0}} + \order{0}{u_{\parallel e\bs{k}}}
\nonumber\\*
&\mbox{}  = \order{0}{\sum_i \frac{c_i}{\nip} \int {\rm d}^3\bb{v}\, v_\parallel J_0(a_i) g_{i\bs{k}}} = \sum_i c_i u_{\parallel i\bs{k}} ,
\end{align}
\begin{align}\label{eqn:gkprpampk}
\order{0}{\frac{C^\parallel_{1e}}{C^\parallel_{0e}} \left( \frac{\delta n_{e\bs{k}}}{\nem} + \Delta_{1e} \frac{\dBprlk}{B_0} \right) }+ \order{0}{\left[ \sum_i c_i \frac{\tauprp{i}}{Z_i} \Gamma^\perp_{20}(\alpha_i) + \frac{2}{\betaprp{e}} - 2 \Delta_{2e} \right] \frac{\dBprlk}{B_0}} \nonumber\\*
\mbox{} - \order{1}{ \sum_i c_i \frac{\tauprp{i}}{Z_i} \left[ 1 - \Gamma^\perp_{10}(\alpha_i) \right] \frac{Z_i e \varphi_{\bs{k}}}{\tprp{i}} } + \order{1}{\sum_i c_i \frac{\tauprp{i}}{Z_i} \left[ 1 - \Gamma^\perp_{11}(\alpha_i) \right] \frac{Z_i e}{\tprp{i}} \frac{\dupari \Aprlk}{c} }
\nonumber\\*
\mbox{}  = - \order{0}{\sum_i \frac{\tauprp{i}}{Z_i} \frac{c_i}{\nip}  \int {\rm d}^3\bb{v} \, \frac{2v^2_\perp}{\vthprp{i}^2} \frac{J_1(a_i)}{a_i} g_{i\bs{k}}} ,
\end{align}
where we have used $\sum_i c_i \dupari = u'_{\parallel 0e} \simeq 0$ (which follows from the mass-ratio expansion). The lowest order in $k_\perp \rho_i$ at which each term enters is indicated underneath that term, following the subsidiary ordering discussed in \S 5.2 of S09. As promised, if we retain only the zeroth-order terms, Equations (\ref{eqn:gkqnk})--(\ref{eqn:gkprpampk}) reduce to their respective KRMHD Equations (\ref{eqn:neutrality2})--(\ref{eqn:pbalance2}).

\subsection{Inertial-Range Turbulence: Reduction to KRMHD}\label{app:gkreduction}

Thus far, we have constructed the theory for electrons, which determines the equations of state of the electron fluid, evolves the parallel component of the vector potential via a generalised Ohm's law, and demonstrates that the magnetic flux is convected by the perpendicular electron flow. In this Section, we proceed to derive the gyrokinetic theory for the ions, and show that it reproduces the KRMHD equations (\ref{sum:psi})--(\ref{sum:ionkinetic}) in the long-wavelength limit relevant to the inertial range.

\subsubsection{Compressive Fluctuations}

Substituting the expression for $\partial A_\parallel / \partial t$ that follows from Equation (\ref{eqn:gkefield}) into the gyrokinetic equation (Eq.~\ref{eqn:ggyrokinetic}), we find that $g_i$ satisfies
 \begin{eqnarray}\label{eqn:gkion}
\order{0}{ \pD{t}{g_i} + v_\parallel \pD{z}{g_i} + \frac{c}{B_0} \{ \langle \chi \rangle_{\gai} \, , g_i \} } = - \frac{Z_i e}{\tprl{i}} \bigl( v_\parallel - \dupari \bigr) \Biggl\langle \order{1}{ \frac{1}{B_0} \{ A_\parallel \, , \varphi - \langle \varphi \rangle_{\gai} \} }
\nonumber\\*
\mbox{} +  \order{0}{\eb\bcdot \grad \left[  \frac{1}{\czero{e}} \frac{T_{\parallel 0{\rm e}}}{e} \left( \frac{\dne}{\nem} + \Delta_{1{\rm e}} \frac{\dBprl}{B_0} \right) - \left\langle \frac{\bb{v}_\perp \bcdot \bb{A}_\perp}{c} \right\rangle_{\gga} \right] }\Biggr\rangle_{\gga} f^\parallel_{0i} .
\end{eqnarray}
We have indicated underneath each term the lowest order in $k_\perp \rho_i$ at which that term enters. The zeroth-order terms should reduce to the ion drift-kinetic equation (Eq.~\ref{sum:ionkinetic}). Using Equations (\ref{eqn:gkdbydt}) and (\ref{eqn:gkbdotgrad}) to group terms on the left-hand side of Equation (\ref{eqn:gkion}) into the Lagrangian operators ${\rm d}/{\rm d}t$ and $\eb\bcdot\grad$, and replacing $(Z_i e / \tprl{i}) \langle \bb{v}_\perp \bcdot \bb{A}_\perp / c \rangle_{\gai}$ with the lowest-order expression for it, $-(v^2_\perp/\vthprl{i}^2)(\dBprl/B_0)$, we find
\begin{equation}
\left( \D{t}{} + v_\parallel \eb \bcdot \grad \right) g_i + \bigl( v_\parallel - \dupari \bigr) \eb\bcdot\grad \left[ \frac{1}{\czero{e}} \frac{Z_i}{\tauprl{i}} \left( \frac{\dne}{\nem} + \Delta_{1e} \frac{\dBprl}{B_0} \right) + \frac{v^2_\perp}{\vthprl{i}^2} \frac{\dBprl}{B_0} \right] f^\parallel_{0i} = 0 ,
\end{equation}
as promised.

\subsubsection{Alfv\'{e}nic Fluctuations}\label{app:gkalfven}

If we now multiply Equation (\ref{eqn:gkion}) by $c_i / \nip$, sum over the ionic species, integrate over the velocity space (keeping $\bb{r}$ constant), and use Equations (\ref{eqn:gkqnk}) and (\ref{eqn:gkprlampk}) to express the velocity-space integrals of $g_i$, we obtain
\begin{align}\label{eqn:gkvorticity}
\pD{t}{} \sum_i  c_i  \frac{Z_i e}{\tprp{i}} \order{1}{ \left\{ \left[ \kzero{i} - \Gamma^\perp_{00} ( \alpha_i ) \right] \varphi_{\bs{k}} - \left[ \kone{i} - \Gamma^\perp_{01}(\alpha_i) \right] \frac{\dupari \Aprlk}{c} \right\} }
\nonumber\\*
\mbox{} + \pD{t}{} \!\!\order{0}{ \left[ \frac{\delta n_{e\bs{k}}}{\nem} - \sum_i c_i \Gamma^\perp_{10} (\alpha_i) \frac{\dBprlk}{B_0} \right] } + \pD{z}{} \biggl( \order{0}{u_{\parallel e\bs{k}}} + \order{1}{\frac{k^2_\perp c B_0}{4\pi e\nem} \frac{\Aprlk}{B_0}} \biggr)
\nonumber\\*
\mbox{} + \pD{z}{} \sum_i c_i \dupari \, \biggl\{ \order{2}{\left[ 1 - \Gamma^\perp_{11}(\alpha_i) \right] \frac{\dBprlk}{B_0}} + \order{1}{\left[ \kone{i} - \Gamma^\perp_{01}(\alpha_i) \right] \frac{Z_i e \varphi_{\bs{k}}}{\tprp{i}} } \biggr\}
\nonumber\\*
\mbox{} + \pD{z}{} \sum_s c_s \Omega_s \order{1}{\left\{ 1 - \Gamma_{00}(\alpha_s) - \frac{\tprl{s}}{\tprp{s}} \left( 1 + \frac{2\duparsq}{\vthprl{s}^2} \right) \left[ \ktwo{s} - \Gamma^\perp_{02} (\alpha_s) \right] \right\} \frac{\Aprlk}{B_0} }
\nonumber\\*
\mbox{} + \order{0}{\frac{c}{B_0} \sum_i \frac{c_i}{\nip} \int {\rm d}^3\bb{v} \, J_0(a_i) \, \bigl\{ \langle \chi \rangle_{\gai} \, , g_{i} \bigr\}_{\bs{k}} } = 0
\end{align}
It is straightforward to show that the zeroth-order component of this equation is identical to Equation (\ref{eqn:gkelectron1}) and so vanishes. Next we consider the first-order terms. Noting that
\begin{equation*}
\frac{k^2_\perp c B_0}{4\pi e\nem} = \sum_s \frac{c_s}{\Omega_s} k^2_\perp \vasq \qquad {\rm and} \qquad \sum_s \frac{c_s}{\Omega_s} = \frac{\rho_0}{e\nem} \frac{c}{B_0} ,
\end{equation*}
we multiply Equation (\ref{eqn:gkvorticity}) by $-e \nem / \rho_0$, inverse Fourier transform back into real space, and use Equation (\ref{eqn:gkstreamflux}) to get
\begin{align}
\pD{t}{} \nabla^2_\perp \Phi& + \left\{ \Phi \, , \nabla^2_\perp \Phi \right\} - \left[ 1 + \sum_s \frac{\betaprl{s}}{2} \left( \Delta_s - \frac{2\duparsq}{\vthprl{s}^2} \right) \right] \left(  \valf  \pD{z}{} \nabla^2_\perp \Psi + \left\{ \Psi \, , \nabla^2_\perp \Psi \right\} \right) 
\nonumber\\*
& \mbox{} + \left( \sum_i \frac{m_i \nip \dupari}{\rho_0 \valf} \right)\nabla^2_\perp  \left( \pD{t}{} \Psi + \{ \Phi \, , \Psi \} + \valf \pD{z}{} \Phi \right) = 0 .
\end{align}
The final term in parentheses vanishes by the reduced induction equation (\ref{eqn:redinduction}), leaving the reduced vorticity equation (\ref{eqn:redforce}).

This completes our derivation of KRMHD from the $k_\perp \rho_i \ll 1$ limit of the non-Maxwellian gyrokinetic theory.

\section{Coefficients for a Bi-Kappa Distribution Function}\label{app:bikappa}

A bi-kappa distribution function is often used to describe the non-thermal electron population in the solar wind and, in particular, its suprathermal ($T_e \sim 60~{\rm eV}$) halo \citep[e.g.,][]{vasyliunas68,mpl97,mpr97,maksimovic05}. In this Appendix, we evaluate all of the dimensionless $C_{\ell s}$ coefficients introduced in the main text and in Appendix \ref{app:gk} for the bi-kappa distribution function
\begin{equation}\label{eqn:bikappa}
f_{\textrm{bi-}\kappa , s}(v_\parallel , v_\perp ) \doteq \frac{\nsp}{\sqrt{\pi\kappa} \thetaprl{s}} \frac{1}{\pi\kappa\thetaprp{s}^2} \frac{\Gamma(\kappa+1)}{\Gamma(\kappa-1/2)} \left[ 1 + \frac{( v_\parallel - \dupar )^2}{\kappa\thetaprl{s}^2} + \frac{v^2_\perp}{\kappa\thetaprp{s}^2} \right]^{-(\kappa+1)} ,
\end{equation}
where $\Gamma$ is the Gamma function, $\kappa > 3/2$ is the spectral index, and
\begin{equation}\label{eqn:thetaprl}
\thetaprl{s} \doteq \vthprl{s} \sqrt{1 - \frac{3}{2\kappa}} \qquad{\rm and}\qquad \thetaprp{s} \doteq \vthprp{s} \sqrt{1 - \frac{3}{2\kappa}} ,
\end{equation}
are the effective parallel and perpendicular thermal speeds, respectively; $\vthprl{s}$ and $\vthprp{s}$ are defined as in Equation (\ref{eqn:vths}). At low and thermal energies, the bi-kappa distribution approaches a Maxwellian distribution, whereas at high energies it exhibits a non-thermal tail that can be described as a decreasing power law. Note that Equation (\ref{eqn:bikappa}) tends to the bi-Maxwellian distribution (Eq.~\ref{eqn:biMax}) as $\kappa \rightarrow \infty$.

Evaluation of the $C_{\ell s}$ coefficients is eased by rewriting $f_{\textrm{bi-}\kappa , s}$ in integral form:
\[
f_{\textrm{bi-}\kappa , s} = \int_0^\infty {\rm d}t \, \frac{t^\kappa \, {\rm e}^{-t}}{{\Gamma(\kappa-1/2)}} \frac{\nsp}{\sqrt{\pi\kappa} \thetaprl{s}} \exp\left[ - \frac{( v_\parallel - \dupar )^2}{\kappa\thetaprl{s}^2} \,t \right]  \frac{1}{\pi\kappa\thetaprp{s}^2} \exp\left( - \frac{v^2_\perp}{\kappa\thetaprp{s}^2} \,t \right) .
\]
We then have from Equation (\ref{eqn:ecoeff}):
\begin{equation}
\czero{e} = \left( 1 - \frac{1}{2\kappa} \right) \left(1 - \frac{3}{2\kappa}\right)^{-1} , \quad \cone{e} = \ctwo{e} = 1 ;
\end{equation}
from Equation (\ref{eqn:kcoeffs}):
\begin{equation}
\kzero{s} = \kone{s} = \left( 1 - \frac{1}{2\kappa} \right) \left(1 - \frac{3}{2\kappa}\right)^{-1} , \quad \ktwo{s} = \left( 1 + \frac{2\duparsq}{\vthprl{s}^2} \kzero{s} \right) \left( 1 + \frac{2\duparsq}{\vthprl{s}^2} \right)^{-1} ;
\end{equation}
and from Equation (\ref{eqn:c11c20}):
\begin{equation}
C_{11i} = 1 , \quad C_{20i} = \left( 1 - \frac{3}{2\kappa} \right) \left( 1 - \frac{1}{2\kappa} \right)^{-1} .
\end{equation}
The $C^\parallel_{\ell i}$ coefficients defined by Equation (\ref{eqn:icoeff}) involve Landau-like integrals, which may be written in terms of the modified plasma dispersion function
\begin{equation}\label{eqn:kappaZ}
Z_\kappa (\xi) \doteq \frac{\Gamma(\kappa+1)}{\Gamma(\kappa+1/2)} \frac{1}{\sqrt{\pi\kappa}} \int^{\infty}_{-\infty} {\rm d}x \, \frac{1}{x-\xi} \left( 1 + \frac{x^2}{\kappa} \right)^{-(\kappa+1)}
\end{equation}
introduced by \citet{st91,st92}:
\begin{subequations}
\begin{equation}
\czero{i} = \left( 1 - \frac{1}{2\kappa} \right) \left( 1 - \frac{3}{2\kappa} \right)^{-1} \bigl[ 1 + \xi_i Z_\kappa(\xi_i) \bigr] ,
\end{equation}
\begin{equation}
\cone{i} = \frac{1}{2\kappa} + \left( 1 - \frac{1}{2\kappa} \right) \left( 1 + \frac{\xi^2_i}{\kappa} \right) \bigl[ 1 + \xi_i Z_\kappa(\xi_i) \bigr] ,
\end{equation}
\begin{align}
\ctwo{i} &= \frac{1}{\kappa} \left[ 1 - \frac{3}{4\kappa} + \left( 1 - \frac{3}{2\kappa} \right) \frac{\xi^2_i}{2\kappa} \right] \left( 1 - \frac{1}{\kappa} \right)^{-1}
\nonumber\\*
\mbox{} &+ \left( 1 + \frac{\xi^2_i}{\kappa} \right)^2 \left( 1 - \frac{3}{2\kappa} \right) \left( 1 - \frac{1}{2\kappa} \right) \left( 1 - \frac{1}{\kappa} \right)^{-1} \bigl[ 1 + \xi_i Z_\kappa(\xi_i) \bigr] ,
\end{align}
\end{subequations}
where $\xi_i \doteq ( \omega - k_\parallel \dupari ) / |k_\parallel| \vthprl{i}$ is the dimensionless Doppler-shifted phase speed.

\section{Nomenclature}\label{app:nomenclature}

In this Appendix, for the reader's benefit we provide a glossary of frequently used symbols in our formulations of non-Maxwellian KRMHD and gyrokinetics. Each symbol is accompanied by a textual description and a numerical reference to either the section(s) in which the symbol was introduced or, if available, the equation by which the symbol was defined (given in parentheses). Throughout the manuscript, the subscript ``0'' appended to any of the following symbols denotes an equilibrium value; the pre-factor ``$\delta$'' denotes a fluctuation. The species index $s = i$ (for ion), $e$ (for electron), or $\alpha$ (for alpha).

\renewcommand\arraystretch{1.5}
\setlength{\tabcolsep}{0.03\textwidth}
\xentrystretch{-0.14}
\vspace{0.5em}
\begin{xtabular}{p{0.06\textwidth} p{0.64\textwidth} p{0.15\textwidth}}
\multicolumn{3}{l}{\textbf{Miscellaneous}} \\
$m_s$ & mass of species $s$ & \S\ref{sec:KMHD} \\
$q_s$ & charge of species $s$ $\bigl(  = Z_s e \bigr)$ & \S\ref{sec:KMHD} \\
$e$ & electric charge magnitude & \S\ref{sec:KMHD} \\
$c$ & speed of light & \S\ref{sec:KMHD} \\
$\msb{I}$ & unit dyadic & \S\ref{sec:KMHD} \\
$\epsilon$ & expansion parameter $\bigl( \ll 1 \bigr)$ & \S\ref{sec:ordering}, \S\ref{app:gkordering} \\
$J_0,~J_1$ & zeroth- and first-order Bessel functions & \S\ref{sec:fourier} \\
[0.5em]
\multicolumn{3}{l}{\textbf{Lengthscales}} \\
$L$ & fiducial macroscale & \S\ref{sec:introduction}, \S\ref{app:gk} \\
$\lambda_{\rm mfp}$ & collisional mean free path $\bigl( = \vthprl{i}/\nu_{ii} \bigr)$ & \S\ref{sec:introduction} \\
$\rho_s$ & gyroradius of species $s$ $\bigl( = \vthprp{s} / \Omega_s \bigr)$ & \S\ref{sec:KMHD}, \S\ref{app:gk}\\
$k^{-1}_\parallel$ & (inverse) parallel wavenumber & \S\ref{sec:ordering}, \S\ref{app:gkordering} \\
$k^{-1}_\perp$ & (inverse) perpendicular wavenumber & \S\ref{sec:ordering}, \S\ref{app:gkordering} \\ 
[0.5em]
\multicolumn{3}{l}{\textbf{Frequencies}} \\
$\omega$ & frequency of the fluctuations $\bigl( = \omega_{\rm r} + \imag \gamma \bigr)$ & \S\ref{sec:ordering}, \S\ref{app:gkordering} \\
$\nu_{ii}$ & ion--ion collision frequency & \S\ref{sec:introduction}, \S\ref{app:gkordering} \\
$\xi_s$ & dimensionless phase speed $\bigl( = \omega / k_\parallel \vthprl{s} - \dupar / \vthprl{s} \bigr)$ & \S\ref{sec:mirror}, \S\ref{app:disprel} \\
$\Omega_s$ & gyrofrequency of species $s$ $\bigl(  = q_s B_0 / m_s c \bigr)$ & \S\ref{sec:introduction}, \S\ref{app:gk} \\ 
[0.5em]
\multicolumn{3}{l}{\textbf{Phase-Space Coordinates}} \\
$v_\parallel$ & velocity-space coordinate parallel to the magnetic field & \S\ref{sec:KMHD}, (\ref{eqn:velocity}) \\
$v_\perp$ & velocity-space coordinate perpendicular to the magnetic field & \S\ref{sec:KMHD}, (\ref{eqn:velocity}) \\
$w_\perp$ & velocity-space coordinate perpendicular to the magnetic field and peculiar to the mean perpendicular flow of species $s$ & \S\ref{sec:KMHD} \\
$\bb{r}$ & real-space coordinate & \S\ref{sec:KMHD} \\
$\varepsilon_s$ & kinetic energy of a particle of species $s$ as measured in the frame of the Alfv\'{e}nic fluctuations and the equilibrium species drift & (\ref{eqn:energy}), (\ref{eqn:gkenergy}) \\
$\mu_s$ & first adiabatic invariant of a particle of species $s$ & (\ref{eqn:mu}), (\ref{eqn:gkmu}) \\
$\vartheta$ & gyrophase angle & (\ref{eqn:velocity}) \\
$\bb{R}_s$ & guiding-centre position of species $s$ & (\ref{eqn:guidingcentre}) \\ 
$\overline{\varepsilon}_s$ & total energy of a particle of species $s$ in the frame of the equilibrium species drift & (\ref{eqn:ebar}) \\
$\varepsilon_{0s}$ & kinetic energy of a particle of species $s$ in the frame of the equilibrium species drift & (\ref{eqn:ebar}) \\
$\varepsilon_{1s}$ & first-order correction to the kinetic energy of species $s$ & (\ref{eqn:ebar}) \\
$\overline{\mu}_s$ & gyrophase-dependent part of the first adiabatic invariant of a particle of species $s$ & (\ref{eqn:mubar}) \\
$\mu_{0s}$ & lowest-order magnetic moment of a particle of species $s$ $\bigl( = m_s v^2_\perp / 2 B_0 \bigr)$ & (\ref{eqn:mubar}) \\
$\mu_{1s}$ & first-order gyrophase-dependent correction to the magnetic moment & (\ref{eqn:mubar}) \\
$a_s$ & dimensionless velocity-space coordinate perpendicular to the magnetic field $\bigl( = k_\perp v_\perp / \Omega_s \bigr)$ & \S\ref{sec:fourier} \\
[0.5em]
\multicolumn{3}{l}{\textbf{Distribution Functions}} \\
$f_s$ & distribution function of species $s$ & \S\ref{sec:KMHD} \\
$f^\parallel_{0s}$ & dimensionless derivative of the equilibrium distribution of species $s$ with respect to the square of the parallel velocity peculiar to the equilibrium species drift & (\ref{eqn:fprlfprp}), (\ref{eqn:gkDf0s}) \\
$f^\perp_{0s}$ & dimensionless derivative of the equilibrium distribution of species $s$ with respect to the square of the perpendicular velocity  & (\ref{eqn:fprlfprp}), (\ref{eqn:gkDf0s}) \\
$f_{\textrm{bi-M},s}$ & bi-Maxwellian distribution function of species $s$ & (\ref{eqn:biMax}) \\
$g_s$ & perturbed distribution function if $f_s$ is taken to be a function of $v_\parallel$ and $\mu_s$ (in KRMHD); gyrocentre distribution function (in gyrokinetics) & (\ref{eqn:gs}), (\ref{eqn:gkgs}) \\
$\widetilde{f}_s$ & distribution function of species $s$ as a function of $(\varepsilon_s,\mu_s)$ & \S\ref{sec:eandmu} \\
$\widetilde{g}_i$ & passively mixed, undamped, ballistic component of the perturbed ion distribution function $g_i$ & (\ref{eqn:gtilde}) \\
$h_s$ & gyrokinetic response & (\ref{eqn:hs}) \\
$\delta f_{1s,{\rm Boltz}}$ & leading-order Boltzmann response & (\ref{eqn:boltzmann}) \\
$\widetilde{h}_s$ & gyrokinetic response corrected for $\mu_s$ conservation & (\ref{eqn:hstilde}) \\
$f_{\textrm{bi-}\kappa,s}$ & bi-kappa distribution function of species $s$ & (\ref{eqn:bikappa}) \\
[0.5em]
\multicolumn{3}{l}{\textbf{Moments of the Zeroth-Order Distribution Function}} \\
$\cell{s}$ & dimensionless coefficients related to perpendicular moments of the parallel-differentiated equilibrium distribution function; includes Landau resonance for ions & (\ref{eqn:ecoeff}), (\ref{eqn:icoeff}) \\
$Z_{\rm M}$ & Maxwellian plasma dispersion function & (\ref{eqn:maxwZ}) \\
$C^\perp_{\ell s}$ & dimensionless coefficients related to parallel moments of the perpendicular-differentiated equilibrium distribution function & (\ref{eqn:kcoeffs}) \\
$\Gamma^{(\perp)}_{\ell m}$ & several dimensionless $v^{2\ell}_\perp$-, $v^m_\parallel$-, and Bessel-function--weighted integrals over the (perpendicular-differentiated) equilibrium distribution function & (\ref{eqn:gammas}) \\
$Z_\kappa$ & kappa plasma dispersion function & (\ref{eqn:kappaZ}) \\
[0.5em]
\multicolumn{3}{l}{\textbf{Densities}} \\
$n_s$ & number density of species $s$ $\bigl(  = \int {\rm d}^3\bb{v} \, f_s \bigr)$ & \S\ref{sec:KMHD} \\
$\rho$ & volume density of plasma $\bigl(  = \sum_s m_s n_s \bigr)$ & \S\ref{sec:drifts} \\
$c_s$ & charge-weighted ratio of number densities $\bigl(  = Z_s \nsp / \nem\bigr)$ & \S\ref{sec:ordering} \\
[0.5em]
\multicolumn{3}{l}{\textbf{Velocities}} \\
$\bb{u}_s$ & mean velocity of species $s$ $\bigl(  = n^{-1}_s \int{\rm d}^3\bb{v} \, \bb{v} f_s \bigr)$ & \S\ref{sec:KMHD} \\
$\bb{u}_{\perp s}$ & mean perpendicular velocity of species & \S\ref{sec:KMHD} \\
$\bb{u}$ & centre-of-mass velocity $\bigl( =\sum_s m_s n_s \bb{u}_s / \sum_s m_s n_s \bigr)$ & \S\ref{sec:drifts} \\
$\bb{u}_\perp$ & perpendicular centre-of-mass velocity $\bigl( = c\bb{E}\btimes\bb{B}/B^2 \bigr)$ & \S\ref{sec:drifts} \\
$u'_{\parallel s}$ & mean parallel velocity of species $s$ measured in a frame comoving with the centre-of-mass velocity & (\ref{eqn:uprls}) \\
$\valf$ & Alfv\'{e}n speed & (\ref{eqn:valf}) \\
$\vthprl{s}$ & parallel thermal speed of species $s$ & (\ref{eqn:vths}) \\
$\vthprp{s}$ & perpendicular thermal speed of species $s$ &(\ref{eqn:vths}) \\
$\valfeff$ & effective Alfv\'{e}n speed & (\ref{eqn:vaeff}) \\
$\theta_{\parallel s}$ & effective parallel thermal speed of species $s$ for a bi-kappa distribution function & (\ref{eqn:thetaprl}) \\
$\theta_{\perp s}$ & effective perpendicular thermal speed of species $s$ for a bi-kappa distribution function & (\ref{eqn:thetaprl}) \\
[0.5em]
\multicolumn{3}{l}{\textbf{Pressures}} \\
$\msb{P}_s$ & pressure tensor of species $s$ & (\ref{eqn:ptensor}) \\
$p_{\parallel s}$ & parallel pressure of species $s$ & (\ref{eqn:pprls}) \\
$p_{\perp s}$ & perpendicular pressure of species $s$ & (\ref{eqn:pprps}) \\
$p_\parallel$ & parallel pressure of plasma $\bigl(  = \sum_s p_{\parallel s} \bigr)$ & \S\ref{sec:drifts} \\
$p_\perp$ & perpendicular pressure of plasma $\bigl(  = \sum_s p_{\perp s} \bigr)$ & \S\ref{sec:drifts} \\
$\Delta_s$ & dimensionless pressure anisotropy of species $s$ & (\ref{eqn:paniso}) \\
$\betaprl{s}$ & ratio of parallel pressure of species $s$ to the magnetic pressure & (\ref{eqn:betas}) \\
$\betaprp{s}$ & ratio of perpendicular pressure of species $s$ to the magnetic pressure & (\ref{eqn:betas}) \\
$\beta_\parallel$ & ratio of parallel pressure of plasma to the magnetic pressure $\bigl( = \sum_s \betaprl{s}\bigr)$ & \S\ref{sec:ordering} \\
$\beta_\perp$ & ratio of perpendicular pressure of plasma to the magnetic pressure $\bigl( = \sum_s \betaprp{s}\bigr)$ & \S\ref{sec:ordering} \\
$\Delta_{\ell s}$ & dimensionless pressure anisotropy of the electrons weighted by $\cell{s}$ & (\ref{eqn:epaniso}), (\ref{eqn:ipaniso}) \\
$\widetilde{\Delta}_s$ & dimensionless pressure anisotropy of species $s$ augmented by the parallel ram pressure from equilibrium parallel drifts & (\ref{eqn:tildedelta}) \\
[0.5em]
\multicolumn{3}{l}{\textbf{Temperatures}} \\
$T_{\parallel s}$ & parallel temperature of species $s$ & (\ref{eqn:pprls}) \\
$T_{\perp s}$ & perpendicular temperature of species $s$ & (\ref{eqn:pprps}) \\
$\tauprl{s}$ & ratio of parallel temperature of species $s$ to the parallel electron temperature & (\ref{eqn:taus}) \\
$\tauprp{s}$ & ratio of perpendicular temperature of species $s$ to the perpendicular electron temperature & (\ref{eqn:taus}) \\
[0.5em]
\multicolumn{3}{l}{\textbf{Electromagnetic Fields and Potentials}} \\
$\bb{E}$ & electric field & \S\ref{sec:KMHD} \\
$\bb{B}$ & magnetic field & \S\ref{sec:KMHD} \\
$\eb$ & unit vector in the magnetic-field direction $\bigl( = \bb{B} / B \bigr)$ & \S\ref{sec:KMHD} \\
$E_\parallel$ & parallel electric field & (\ref{eqn:eprl}), (\ref{eqn:efield}) \\
$\bb{E}_\perp$ & perpendicular electric field $\bigl( = - \bb{u}_\perp \btimes \bb{B} / c \bigr)$ & \S\ref{sec:drifts} \\
$\Phi$ & velocity stream function & (\ref{eqn:streamflux}) \\
$\Psi$ & magnetic flux function & (\ref{eqn:streamflux}) \\
$\zeta^\pm$ & generalised Elsasser potentials & (\ref{eqn:elsasserpotentials}) \\
$\bb{z}^\pm$ & generalised Elsasser fields & (\ref{eqn:elsasserfields}) \\ 
$\varphi$ & electrostatic scalar potential & (\ref{eqn:BcurlA}) \\
$\bb{A}$ & magnetic vector potential & (\ref{eqn:BcurlA}) \\
$\bb{j}$ & current density & (\ref{eqn:ampere}) \\
$\varphi'_s$ & electrostatic scalar potential in the frame of the parallel-drifting species $s$ & (\ref{eqn:varphiprime}) \\
$\chi$ & gyrokinetic potential & (\ref{eqn:gkpotential}) \\
[0.5em]
\multicolumn{3}{l}{\textbf{Functions Defined for Parallel Kinetics (\S\ref{sec:parallelkinetics})}} \\
$G_n$ & integral over the perpendicular velocity space in Eq.~(\ref{sum:dne}), which relates the fluctuating electron number density to the lower-order moments of $g_i$ & (\ref{eqn:GnBdef}) \\
$G_B$ & integral over the perpendicular velocity space in Eq.~(\ref{sum:dBprl}), which relates the fluctuating magnetic-field strength to the lower-order moments of $g_i$ & (\ref{eqn:GnBdef}) \\
$F^\parallel_{\ell i}$ & various perpendicular moments of the parallel-differentiated equilibrium ion distribution function & (\ref{eqn:Felli}) \\
$\lambda^{(\,\dots)}_i$ & coupling coefficients quantifying the influence of the density and magnetic-field-strength fluctuations on the kinetic fluctuations; superscripts are $nn$, $nB$, $Bn$, and $BB$ & (\ref{eqn:GnB2}), \S\ref{app:coefficients} \\
$G^\pm$ & eigenvectors resulting from diagonalising the reduced ion kinetic equation for a bi-Maxwellian plasma & (\ref{eqn:Gpm}) \\
$\Lambda^\pm$ & (inverse) eigenvalues resulting from diagonalising the reduced ion kinetic equation for a bi-Maxwellian plasma & (\ref{eqn:lambdapm}) \\
$\sigma_i$ & useful ion coefficient & (\ref{eqn:sigma}) \\
$\varsigma_i$ & useful ion coefficient & (\ref{eqn:varsigma}) \\
$\varpi_i$ & useful ion coefficient & (\ref{eqn:varpi}) \\
$\kappa_i$ & useful ion coefficient & (\ref{eqn:kappa}) \\
[0.5em]
\multicolumn{3}{l}{\textbf{Invariants}} \\
$W^\pm_{\rm AW}$ & Alfv\'{e}n-wave invariants associated with $\zeta^\pm$ & (\ref{eqn:WAWpm}) \\
$W_{\rm AW}$ & Alfv\'{e}n-wave invariant & (\ref{eqn:WAW}) \\
$W_{\rm compr}$ & compressive invariant & (\ref{eqn:Wcompr}) \\
$W^\pm_{\rm compr}$ & compressive invariants associated with $G^\pm$ & (\ref{eqn:invariants2}) \\
$W_{\widetilde{g}_i}$ & compressive invariant associated with $\widetilde{g}_i$ & (\ref{eqn:Wg}) \\
$W$ & generalised free energy & (\ref{eqn:W}) \\ 
[0.5em]
\multicolumn{3}{l}{\textbf{Differential and Integral Operators}} \\
${\rm D}/{\rm D}t$ & time derivative measured in a frame co-moving with the Alfv\'{e}nic fluctuations and streaming along the magnetic field at velocity $v_\parallel$ $\bigl( = \partial/\partial t + \bb{u}_{\perp s}\bcdot\grad + v_\parallel \eb \bcdot \grad \bigr)$ & \S\ref{sec:KMHD} \\
$\mf{D}$ & differential operator measuring the velocity-space anisotropy of a distribution function & (\ref{eqn:Df0s}), (\ref{eqn:gkDf0s}) \\
$\{ ~\cdot\; , ~\cdot~ \}$ & Poisson bracket & (\ref{eqn:rmhdbracket}), (\ref{eqn:gkbracket}) \\
${\rm d}/{\rm d}t$ & time derivative measured in a frame co-moving with the Alfv\'{e}nic fluctuations & (\ref{eqn:dbydt}), (\ref{eqn:gkdbydt}) \\
$\eb\bcdot\grad$ & space derivative measured along the exact magnetic-field direction & (\ref{eqn:bdotgrad}), (\ref{eqn:gkbdotgrad}) \\
$\dot{f}_s$ & time derivative of the function $f_s(t,\bb{r},\bb{v})$ taken along the full phase-space trajectory of a particle of species $s$ & (\ref{eqn:vlasov}) \\
$\langle \,\dots \rangle_{\gas}$ & ring average over $\vartheta$ at fixed $\bb{R}_s$ & (\ref{eqn:ringaverage}) \\
$\langle \,\dots \rangle_{\bs{r}}$ & gyro-average over $\vartheta$ at fixed $\bb{r}$ & (\ref{eqn:gyroaverage}) \\
\end{xtabular}

\bibliographystyle{jpp}
\bibliography{kscac15}

\end{document}